\documentclass[11pt]{article}

\usepackage[T1]{fontenc}
\usepackage[utf8]{inputenc}

\usepackage[letterpaper,margin=1in]{geometry}

\usepackage{amsmath,amssymb}

\usepackage{lmodern}
\usepackage{microtype}

\usepackage{graphicx}
\graphicspath{{./figures/}}

\DeclareMathOperator{\sgn}{sgn}

\usepackage{authblk}

\usepackage[hidelinks]{hyperref}

\usepackage[style=authoryear,natbib=true,uniquename=false]{biblatex}
\DefineBibliographyStrings{english}{url= \textsc{url} ,  }
\DeclareDelimFormat{nonameyeardelim}{\addcomma\space}
\usepackage{booktabs}
\usepackage{array}
\usepackage{tabularray}
\UseTblrLibrary{booktabs}
\UseTblrLibrary{siunitx}
\usepackage{siunitx}
\usepackage{dcolumn}
\usepackage{rotating}

\usepackage{xstring}

\newcommand{\vectitletextdefinitions}[1]{%
    \protect\IfStrEqCase{#1}{%
        {I}{Newtonian Incorrect}%
        {II}{Applied Force Exceeds Resistance for Constant Speed}%
        {III}{Impetus Force, Mostly on Circular Path}%
        {IV}{Impetus Plus Centrifugal or Centripetal Force on Circular Path}%
        {V}{Active Dominates Massive Dominates Passive}%
        {VI}{Impetus Force, Mostly Along Linear Path}%
        {VII}{Impetus Force on Both Linear and Circular Paths}%
        {VIII}{Passive Object Pushed ``Because in the Way''}%
        {IX}{After Rocket Starts or Stops, Goes in Direction of Latest Force or Earlier Motion}%
        {X}{Omission of Reaction Force (Normal or From Passive Object), Usually With Gravity}%
        {XI}{Straight Path Preferred vs. Curved Paths}%
        {XII}{Eliminated (Similar to XI but Weaker)}%
        {XIII}{Decelerating After Impulse From Rocket Firing}%
        {XIV}{2M and M Balls Differ by Factor of 2}%
        {XV}{Latest Force Dominates After Sudden Force}%
        {XVI}{Force Change Increases Speed, Sometimes With Initial Acceleration}%
        {XVII}{Continues Curving After Inward Force Removed}%
        {XVIII}{Air Exerts Significant Drag and Downward Force}%
        {XIX}{No Acceleration While Rocket Firing, Then Speed Decreases}%
        {XX}{Eliminated (Similar to XIX but Less Sparse)}%
        {XXI}{2M Ball Falls Significantly Faster, but Not by Factor of 2}%
        {XXII}{Eliminated (Low Bootstrap Correlations)}%
        {XXIII}{Missing Centripetal Force}%
        {XXIV}{Eliminated (Appears To Be a Residual)}%
        {XXV}{Eliminated (Low Bootstrap Correlations)}%
        {XXVI}{Motion Diagrams, v and a Confused}%
        {XXVII}{Gravity Stronger Closer to Ground}%
    }%
{}}
\newcommand{\vecname}[1]{\textit{\vectitletextdefinitions{#1}} (#1)}
\newcommand{\vecnameonly}[1]{\textit{\vectitletextdefinitions{#1}}}

\newcommand{\samplenametextdefinitions}[1]{%
    \protect\IfStrEqCase{#1}{%
        {SU}{High School Modeling}%
        {U}{Large Public 3}%
    }%
{}}
\newcommand{\sample}[1]{\textit{\samplenametextdefinitions{#1}}}

\title{Discovering Misconceptions and Misunderstandings From Administrations
       of Research-Designed Multiple Choice Instruments}

\author[1]{Martin Segado}
\author[2]{Aaron Adair}
\author[3]{John Stewart}
\author[2]{David Pritchard}

\small\affil[1]{Department of Mechanical Engineering, Massachusetts Institute of Technology, Cambridge, MA}
\small\affil[2]{Department of Physics, Massachusetts Institute of Technology, Cambridge, MA}
\small\affil[3]{Department of Physics and Astronomy, West Virginia University, Morgantown, WV}

\date{}   

\begin{document}

\maketitle

\begin{abstract}
Misconceptions are ``alternate hypotheses'' that are incorrect according to established theories of how the world works.  Often held with confidence by students, they are relatively insensitive to context, often seem like common-sense views, and noted for being resistant to remediation using traditional instruction methods.  To find misconceptions we analyze $\sim$34,000 administrations of a pioneering concept test in Newtonian mechanics---the Force Concept Inventory---using our robust Bayesian implementation of a flexible multidimensional item-response model for multiple-choice data. In contrast to most earlier work, this model permits answer choices within a single question to have different directions in the multidimensional space of student knowledge, making it an excellent fit for concept inventories in which different incorrect responses may reflect distinct misconceptions. Our method extracts 21 unique dimensions from our data; we explore and contrast various factor rotation methods for transforming these to maximize their interpretability and consistency. These yield several alternative ``sparse'' solutions in which most dimensions load strongly on just a few incorrect answer choices (distractors). Examining the content of these reveals a total of 22 robust dimensions for which the strongest distractors reflect a coherent idea identifiable as a misconception or misunderstanding.

Prior research summarizing open-ended responses to concept questions and performing interviews has uncovered some misconceptions similar to ours, or for which ours are elaborations. Motivated by the realization that many of our discovered misconceptions are surprisingly similar to previously-accepted theories of mechanics, we broadly sort them by historical era: \textit{Ancient} (learned by infants but codified and extended by Greeks), \textit{Medieval} (reactions and extensions of Aristotelian ideas), and \textit{Post-Newtonian} (including known modern misconceptions as well as two which appear novel).

Given our collection of identified misconceptions and their associated distractor loadings, we also present a simple method for computing a set of ``misconception scores'' for each student. We then examine the overall prevalence of the different misconceptions in our sample, both prior to and subsequent to instruction, as a function of raw pre-test score. Surprisingly varied patterns of remediation emerge: some misconceptions persist largely unchanged by instruction, while for others instruction proves significantly more effective for either above- or below-average students. In general, we find that many misconceptions are poorly remediated for students of average or lower ability. Our hope is that our work will serve as a guide for developing, evaluating, and improving specific interventions for each of these while providing Physics instructors with a new and useful tool for class-level formative assessment.
\end{abstract}

\section{Introduction}

\textit{Misconceptions} present major barriers to deep learning for students, especially in STEM.  A misconception is a stable, theory-like alternative conception to the current paradigm of how the world works.  An individual holding a misconception applies it in a relatively context-independent manner and associates it with other ideas with the result that it is robust against most forms of traditional instruction \citep{Docktor2010c,etkina2005}, principally  because it requires belief revision, mental model transformation, or categorical shifts by the student \citep{Chi2008}.  On a research-designed multiple choice instrument, such misconceptions manifest as an intellectually-coherent set of wrong answer choices (distractors) that are preferentially selected by students holding the misconception (to greater or lesser degrees depending on the extent to which they hold it).

A significant fraction of misconceptions in introductory mechanics arise from the fundamental misconception that (as Aristotle argued) force \textit{causes} motion rather than (as Newton's second law proclaims) force \textit{changes} motion.  Supported by everyday observations and common sense, this misbelief could lead one to the misconception that, whatever an object's recent history of motion, the current direction of motion is determined by the current force or the last force to act (if there is some belief in inertia)---or alternatively, that an object that is moving without an obvious external force on it must possess some sort of internal ``impetus force'' acting in the direction of motion.  Holders of such misconceptions would be expected to select distractors containing wording like ``a force in the direction of motion'' on a mechanics concept test, and to do so with some consistency across several questions probing straight-line motion—whether that motion is horizontal, vertical, results from a previous force (whether steady or impulsive), or has gravity acting obliquely to the line of motion.

Misconceptions involve domain-relevant concepts (unlike the more general phenomenological primitives viewpoint \citep{disessa1993} or dual process theories \citep{gette2018}), are less context-dependent than the resources viewpoint \citep{scherr2007}, may involve ontological confusions \citep{chi1993}, and are broader than an unlearned knowledge component \citep{Koedinger2012}.  Our misconceptions each involve a \textit{set} of wrong answers and not just a single wrong response as some authors have assumed \citep{bani-salameh2017}.

The expectation that students with a certain misconception will preferentially select a subset of distractors on different questions based on the same alternative hypothesis motivates our strategy to discover misconceptions and measure how strongly each misconception is held.  We find such patterns using our implementation \citep{segado2025} of a very general model from item response theory (IRT): the Multidimensional Nominal Categories Model or MNCM \citep{revuelta2020}. This model extracts multiple dimensions of knowledge from multiple-choice data (up to 21 in our case) and does so at the level of individual distractors rather than of entire questions. This is followed by the application of exploratory factor rotation methods to search for patterns in the extracted dimensions that are most likely to correspond to actual underlying student thought processes. Judicious use of Bayesian methods with hierarchical priors lends the needed robustness to our approach, while variational inference techniques borrowed from machine learning make the problem tractable even for large datasets.

In this paper we apply these methods to $\sim$34,000 administrations of the Force Concept Inventory \citep{FCI,halloun1995},  a pioneering research-designed multiple choice instrument whose distractors were found from research on common misunderstandings and misconceptions or by distilling responses when the questions were administered open response. We additionally extend our procedures to ensure that the discovered misconceptions are consistent across different educational institutions, highly prevalent in both pre- and post-instruction data, robust across different sub-samples of students as determined by non-parametric bootstrapping, and easy to interpret. The result is 22 robust patterns of distractors identifiable with misconceptions.  The majority of these are consistent with various historical conceptions of mechanics—though two appear to be novel—so we present them in a framework that includes key concepts from ancient, medieval, and post-Newtonian eras.

We further extend our analysis by developing procedures to measure the amount of each robust misconception held by a student or a class.  Since our dataset contains both pre- and post-instruction administrations of the FCI for every student, we can measure the decrease in the prevalence of each misconception after instruction.  We find dramatic differences on the effect of instruction, with numerous cases where the instruction is effective mainly for students of high initial ability but relatively ineffective for most students (in accord with previous work on misconceptions), but also a significant number that are held primarily by students of average initial ability and that are remediated effectively by current instruction.

The ability to measure the change in the prevalence of each misconception for low-, medium-, and high-ability students offers unprecedented opportunities to evaluate and improve instruction.  It also promises to measure the efficacy of experimental interventions that mediate the various misconceptions we have discovered—an important task given the robustness of most misconceptions.

\section{Background: the MNCM-Bayes method}

The MNCM is a highly flexible IRT model designed to explain categorical data (such as student responses to the multiple-choice questions on the FCI). In its most general form, it is distinct from many other multidimensional IRT models in that it treats multidimensionality as a property of the individual answer options; that is, it allows each possible answer in a multiple-choice question to have its own `direction' in the multidimensional space of misconceptions \citep{revuelta2020}. This property makes the model a good fit for analyzing the FCI and other research-based concept tests, in which the answer choices in a given question often represent different misconceptions (or different combinations of misconceptions).

As the model and our Bayesian implementation of the same have already been described in a previous paper by our group \citep{segado2025}, we will limit this section to the minimum background necessary for understanding the rest of this work.

\subsection{Mathematical description of the MNCM}

As with other multidimensional IRT models, the MNCM posits that student responses to questions on a test arise from the interaction of various student traits (which may correspond to different dimensions of knowledge, for example) and various properties of the questions themselves---neither of which is directly observable. Specifically, the probability $p_{sic}$ of some student $s$ responding to some question (or ``item'') $i$ by selecting some answer choice $c$ is given by
\begin{equation}
    p_{sic} = \frac{\exp t_{sic}}{\sum\limits_{c'=1}^C \exp t_{sic'}}
    \label{eq:newprobs}
\end{equation}
where $C$ is the number of answer choices for the question and $t_{sic}$ is a \textit{response tendency} which depends linearly on the student's traits ${\theta_{s1}, \theta_{s2}, \dots, \theta_{sD}}$:
\begin{equation}
    t_{sic} \overset{\text{def}}{=} \sum\limits_{d=1}^D \theta_{sd} a_{icd} + b_{ic}.
    \label{eq:tendencydef}
\end{equation}

The parameters $a_{icd}$ and $b_{ic}$ in the above equation are, as their subscripts suggest, properties of the various answer choices on a given question. In this work we are primarily concerned with the slope parameters $a_{icd}$, which encode the strength of the relationship between the student traits and the response tendencies; there are $I \times C$ such slopes in each dimension $d$.

\subsection{Model invariances and parameter interpretability}

The parameters of the MNCM are not uniquely defined: for a given dataset, there are many sets of parameter values which explain the data equally well. This fact is crucial when attempting to interpret the outputs of the model, as some ways of constraining these free degrees of freedom may yield results which are considerably easier to understand (or more likely to match the true underlying thought processes of students).

While we will not cover all invariances of the MNCM here, there are two which we view as necessary for understanding the present work. First, the slope parameters (or ``loadings'') are \textit{translation invariant}: shifting the $C$ slopes in any given question and dimension by an additive constant leaves the response probabilities unchanged due to the normalization in Eq.~\eqref{eq:newprobs}. We make use of this invariance to set the slope for the correct answer to zero in each question and dimension, which we have found effective in highlighting misconceptions (which represent deviations from Newtonian-correctness).

Second, the model is \textit{rotation invariant}: since Eq.~\eqref{eq:tendencydef} contains the equivalent of an inner product between $\theta$ and $a$, we may rotate both variables in $D$-dimensional space without changing the response tendencies. Rotating the variables (or equivalently, the coordinate system) allows us to search for simpler, more interpretable explanations of the data, and is standard practice in both exploratory factor analysis and its IRT equivalent \citep[``item factor analysis'';][]{bock1988}. Further details as well as an illustrative example may be found in Section 6.1 of our first paper \citep{segado2025}.

Standard methods for factor rotation exist in many statistical software packages; we use the GPArotation package for the R language~\citep{bernaards2005,rcoreteam2021}. GPArotation implements a gradient projection algorithm which, given a randomly-initialized rotation matrix as a starting point, numerically updates this matrix to minimize some differentiable measure of complexity in the rotated slopes. This in turn leads to much sparser solutions, in which most dimensions load heavily on only a subset of individual distractors. Many different measures of complexity exist, however, and selecting a suitable one which yields robust, interpretable results is a key goal of the present work.

\subsection{Our hierarchical Bayesian implementation}

Our implementation of the MNCM (which we call the MNCM-Bayes method) has several key properties which are important to understand this work. We cover these briefly below while referring readers to our earlier work for a more comprehensive treatment of the material including pointers to relevant literature.

First, our method takes a hierarchical Bayesian approach to inferring the parameters of the MNCM from data. Without rehashing the details here, this means that it seeks to find \emph{probability distributions} over the possible values of each model parameter (all $\theta_{sd}$'s, $a_{icd}$'s, and $b_{ic}$'s) given the observed data and some reasonable prior expectations of how these parameters will be distributed. This principled approach to managing uncertainty lends considerable robustness to parameter estimates, especially when observations are limited (e.g.\ for infrequently-selected answer choices).

Second, the method possesses an emergent self-selecting dimensionality property: given the flexibility to fit a large number of dimensions to a dataset, it chooses to use only as many as are justifiable from the amount of information present. Larger sample sizes typically result in the method returning a higher-dimensional fit, while smaller sample sizes are treated more conservatively to avoid overfitting. We leverage this property in the present work to recover the most dimensions possible from our data while remaining confident that we are not simply fitting random noise (though we do also take steps to verify this by comparing results obtained on different subsets of our data as we will discuss in section~\ref{sec:methods}).

\section{Methods to Find Misconceptions }
\label{sec:methods}

The central objective of this paper is to discover as many student misconceptions as possible from student responses to the FCI.  We build on the foundation of our group's recent work on fitting the MNCM using variational Bayesian inference with hierarchical priors \citep{segado2025}, which analyzes testee response data to determine a set of $D$ orthogonal \textit{principal vectors} each indicating the loading of that dimension onto \textit{each} possible response (question and choice) on the instrument.  In that work we exploited the indeterminacy of the MNCM model vis-à-vis rotations of those principal vectors to find \textit{sparse distractor vectors}, each of which loaded heavily (i.e. had large components) on just a small set of distractors. We exhibited two such sparse distractor vectors in that work whose most heavily-loaded distractors appeared consistent with particular misconceptions, and identified these by appropriate descriptive names.

Our approach is one solution to a key issue that arises in applying multi-dimensional psychometric models to real data---namely interpreting the meaning of the extra dimensions recovered from the data.

\subsection{Data and cleaning}

The dataset used for our study includes responses to both pre- and post-instruction administrations of the FCI (revised version ``v95''; \cite{halloun1995}) collected from $\sim$17,000 students. The bulk of these ($\sim$10,000) are high school students taught by teachers who attended a modeling workshop run by the ASU Modeling Instruction Program. The remainder were collected from seven North American colleges and universities. A full description of the data may be found in earlier work by our group \citep{Perez-Lemonche2019}, though we note that this misidentified the source of the high school data and erroneously labeled it ``Large Public 1''; the present work will instead refer to this subset of our data as \sample{SU}.

The original data contains a small fraction of tests with missing responses. While it is possible to model these within an IRT paradigm, we chose instead to discard all data from students who left any question unanswered (whether on the pre-test or the post-test) based on our expectation that students not completing all the questions were less likely to provide carefully-considered responses to those which they did complete. We also carefully checked for duplication in the data and discarded all entries from students with more than exactly one pre-test and one post-test submission.

\subsection{Initial comparisons of rotation methods}
\label{sec:sparse-vec-similarity}

We begin our analyses by exploring the consistency of different rotation methods across data from different schools as well as between pre- and post-instruction administrations of the FCI. The central concept of the misconceptions approach to student cognition is that misconceptions should be stable constructs across students, institutions, or levels of instruction, although the \emph{extent} to which they are held may of course differ across these conditions. This means that the set of distractors that probe a particular misconception—and ideally, the extent to which they probe it as indicated by their loadings—should remain similar when the method is applied to different datasets.

Of course, consistency across datasets isn't the only important criterion: it is also necessary for the content of each dimension's distractors to share one or more common themes which are identifiable as misconceptions or other cognitive constructs.

In our exploration of rotation methods, we therefore look for those which produce the most universal and interpretable distractor vectors by the following criteria:
\begin{enumerate}
    \item Distractor vectors should be similar across different educational institutions, suggesting that the misconceptions represent patterns that are widely shared by students independently of admission biases or instruction.
    \item Distractor vectors should be similar when obtained using pre- and post-instruction data; i.e., the misconceptions themselves should be stable, even if their overall prevalence decreases with instruction.
    \item Distractor vectors should each load primarily on a few distractors that are consistent with one or two identifiable misconceptions.
\end{enumerate}

\subsubsection{Illustrating similarity across educational institutions}

\begin{figure}
    \centering
    \includegraphics[width = 5.5in]{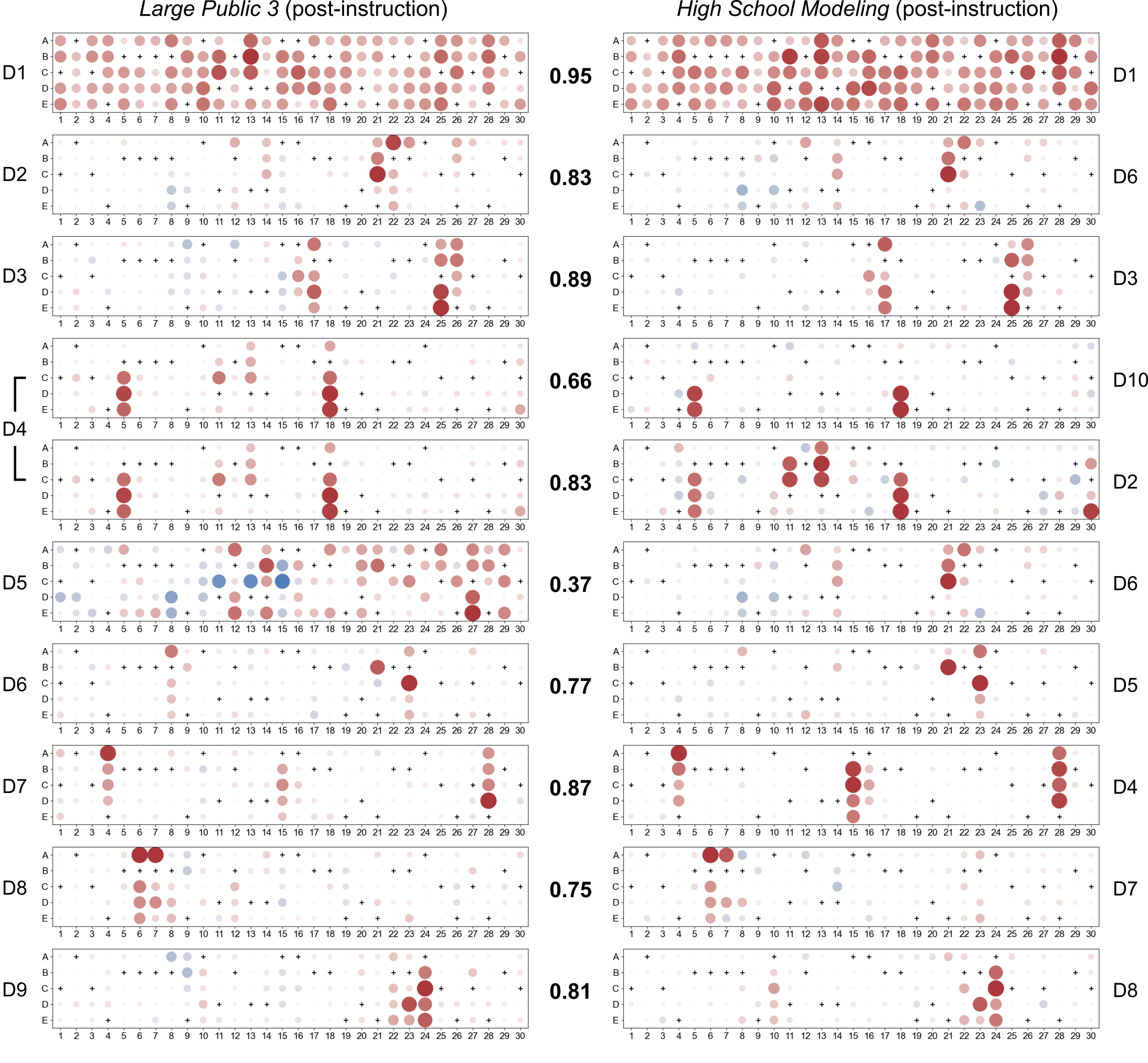}
    \caption[BigeominQ-rotated distractor vectors from post-instruction \sample{SU} matched with those from \sample{U}]{\textbf{BigeominQ-rotated distractor vectors from post-instruction \sample{SU} matched with those from \sample{U}.} The top 9 post-instruction vectors from \sample{U} are on the left.  Dimension 4 is doubled since it ``matches'' with two \sample{SU} vectors---typical of comparisons with results where more dimensions are recovered. Correlation coefficients (uncentered Pearson) are shown in bold between each pair of matched vectors.}
    \label{U vs SU BigeominQ}
\end{figure}

\begin{figure}   
    \centering
    \includegraphics[width = 4in]{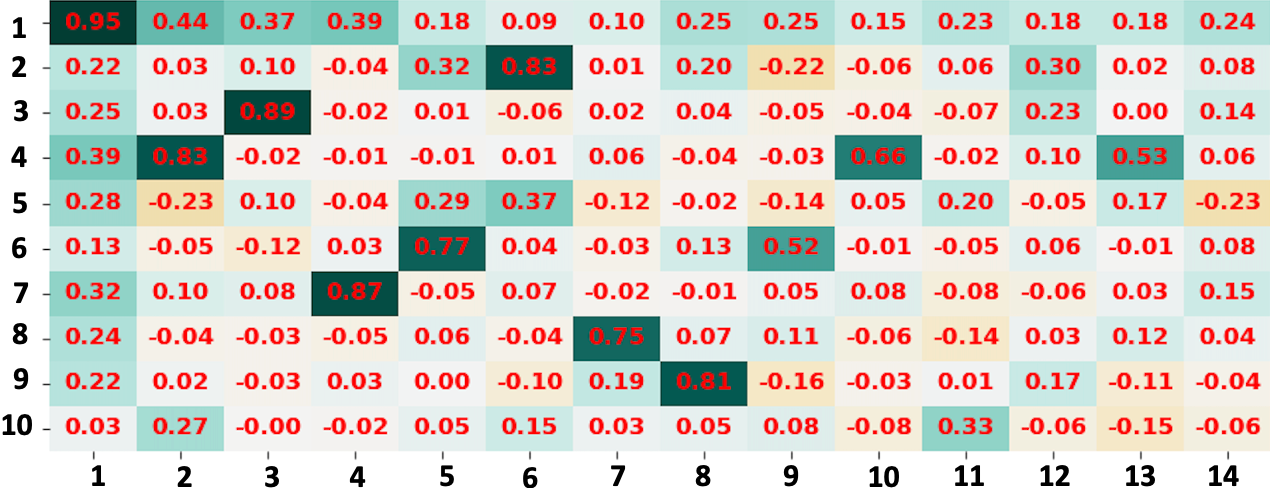}
    \caption[Uncentered Pearson correlations of post-instruction \sample{U} vs \sample{SU} distractor vectors]{\textbf{Uncentered Pearson correlations of post-instruction \sample{U} vs \sample{SU} distractor vectors.} The 10 vectors from \sample{U} (rows) are correlated with the 14 vectors from \sample{SU} (columns); both used the BigeominQ rotation method.  Dark shading indicates coefficients $>0.75$ and highlights that 8 of the 10 \sample{U} vectors correlate with one and only one ``similar'' vector from \sample{SU}. Vector 5 obtained from \sample{U} (in Fig.~\ref{U vs SU BigeominQ}, this is the sixth vector from the top in the left column) is relatively complex and does not have a similar vector in \sample{SU}.}
    \label{U vs SU triangle}
\end{figure}

Figures \ref{U vs SU BigeominQ} and \ref{U vs SU triangle} illustrate the issue of similarity between the two largest samples of post-test data from different educational institutions.  The key finding is that 8 of the 10 sparse vectors from the smaller sample, \sample{U}, can be paired with highly similar sparse distractor vectors from the larger sample, \sample{SU}, with correlation $> 0.75$, but not in the same rank order. In addition, Fig. \ref{U vs SU triangle} shows that the matrix of correlations between the sparse vectors obtained from the two data subsets is not diagonal. The difference in ordering between these results may reflect differences in the instruction, or possibly differences in the entering student bodies; clearly looking at pre- and post-instruction data together would help answer this question.  (Note: we used uncentered correlation coefficients to highlight components of distractors furthest from the Newtonian-correct answers.)

We frequently observed that the sample from which we extracted more dimensions had two dimensions whose response patterns matched different parts of the pattern from a single dimension in a sample where fewer dimensions were recovered.  Fig.~\ref{U vs SU triangle} shows an example of this where vector 4 of \sample{U} has a correlation of 0.83, 0.66, and 0.53 with ``similar'' dimensions of \sample{SU}. Such cases suggest that the dimensions in the larger sample may represent related concepts and, more broadly, that the structure of the dimensions overall might be hierarchical (a topic we hope to explore further in future work).

Occasionally we observed sparse vectors that were fairly complex, and that generally didn't correlate with dimensions from any other educational institutions---for example, vector 5 of \sample{U}. Sometimes, particularly when analyzing smaller samples, we could see noticeable correlations with \emph{several} vectors obtained from larger samples. Even in these cases, we could not see any overarching physics themes represented by them, suggesting that they originated from more complex concatenations or simply from students lacking a coherent overview of force and motion.  This last explanation is consistent with the fact that more of these were detected from pre-instruction data sets.

\subsubsection{Our similarity criterion}

To quantify the degree of ``similarity'', we developed the following \textit{similarity index} to quantify the similarity of two sets of distractor dimensions:
\begin{equation}
    \label{similarity}
    \mathcal{S}=4 \left( \sum_{d1=1}^{D_1}  \sum_{d2=1}^{D_2} (C_{d1,d2} - 0.5)^2 \right)/{D_1}
\end{equation} 
where the sums ignore terms when the uncentered Pearson correlation coefficient of the relevant vectors ($C_{d1,d2}$) is less than 0.5.  

This expression applies to two sets of distractor vectors with number of dimensions $D_1 \leq D_2$. If every dimension of the $D_1$ set of vectors correlated perfectly ($C=1$) with one of the $D_2$ set of vectors, this expression would evaluate to unity.  

We applied our similarity index to a large number of rotated solutions obtained by different methods in the GPArotation package. For each method, we compare the three pairs formed from only the three largest samples in our dataset. The remaining five samples were excluded from these comparisons as their lower sample sizes did not support the recovery of as many sparse vectors, and those that were recovered were often combinations of two or more recovered from the larger samples. This analysis was repeated for both pre- and post-instruction data.

Our results in Table \ref{tab:instruction_comparison} show that rotations that constrain student traits to be orthogonal (those with names ending in `T') outperform their oblique counterparts (ending in `Q') in similarity index. We believe that this result stems from the greater number of degrees of freedom inherent in oblique rotations: this added flexibility allows them to achieve lower complexity scores (and often easier-to-interpret solutions), albeit at the expense of some reduction in consistency across datasets.

\begin{table}
    \centering
    \footnotesize
    \begin{tabular}{ll|ll}
        \toprule
        \multicolumn{2}{c|}{Pre-Instruction} & \multicolumn{2}{c}{Post-Instruction} \\
        \midrule
        bentlerQ  & 0.010170  & bentlerQ  & 0.012305        \\
        oblimin       & 0.014716       & tandemII        & 0.013579        \\
        cfQ           & 0.015112       & tandemI        & 0.014040        \\
        quartimin     & 0.015112       & oblimin         & 0.015433        \\
        bentlerT      & 0.015351       & bentlerT        & 0.019421        \\
        bifactorQ     & 0.016199       & quartimin       & 0.020169        \\
        cfT           & 0.016271       & cfQ             & 0.020169        \\
        quartimax     & 0.016271       & cfT             & 0.020660        \\
        geominQ       & 0.018052       & quartimax       & 0.021044        \\
        tandemII      & 0.018276       & geominQ         & 0.022104        \\
        tandemI       & 0.019035       & bifactorQ       & 0.022207        \\
        bigeominQ     & 0.023413       & bigeominQ       & 0.027025        \\
        bifactorT     & 0.024274       & entropy         & 0.027852        \\
        geominT       & 0.024501       & bigeominT       & 0.028198        \\
        bigeominT     & 0.024666       & geominT         & 0.028271        \\
        entropy       & 0.024853       & bifactorT       & 0.029768        \\
        \bottomrule
    \end{tabular}
    \caption[Cross-school similarity index for various transformations]{\textbf{Cross-school similarity index for various transformations.} The post-instruction similarities are larger than the pre-instruction similarities---consistent with our previous suggestion that the student views are more consistent after instruction.  Note also that the T (orthogonal) transformations always have more similarity than the corresponding Q (oblique) transformations.}
    \label{tab:instruction_comparison}
\end{table}

\subsubsection{Similarity across pre- and post-tests}

We now examine the similarity of the discovered sparse vectors between separate pre- and post-instruction analyses of combined data from \emph{all} educational institutions in our sample.  This is a more stringent test of the invariance of sparse vectors for two reasons.  Firstly, the overall student perspective should be strongly modified by instruction, so given a particular misconception (e.g.\ that gravity is appreciably stronger near the earth's surface than a short distance above it) it might be activated differently in different contexts due to other learnings about these contexts.  Secondarily, combining data from all institutions increases the size of the datasets, which in turn allows us to recover more dimensions than is possible from each source in isolation.

\begin{figure}   
    \centering
    \includegraphics[width = 5in]{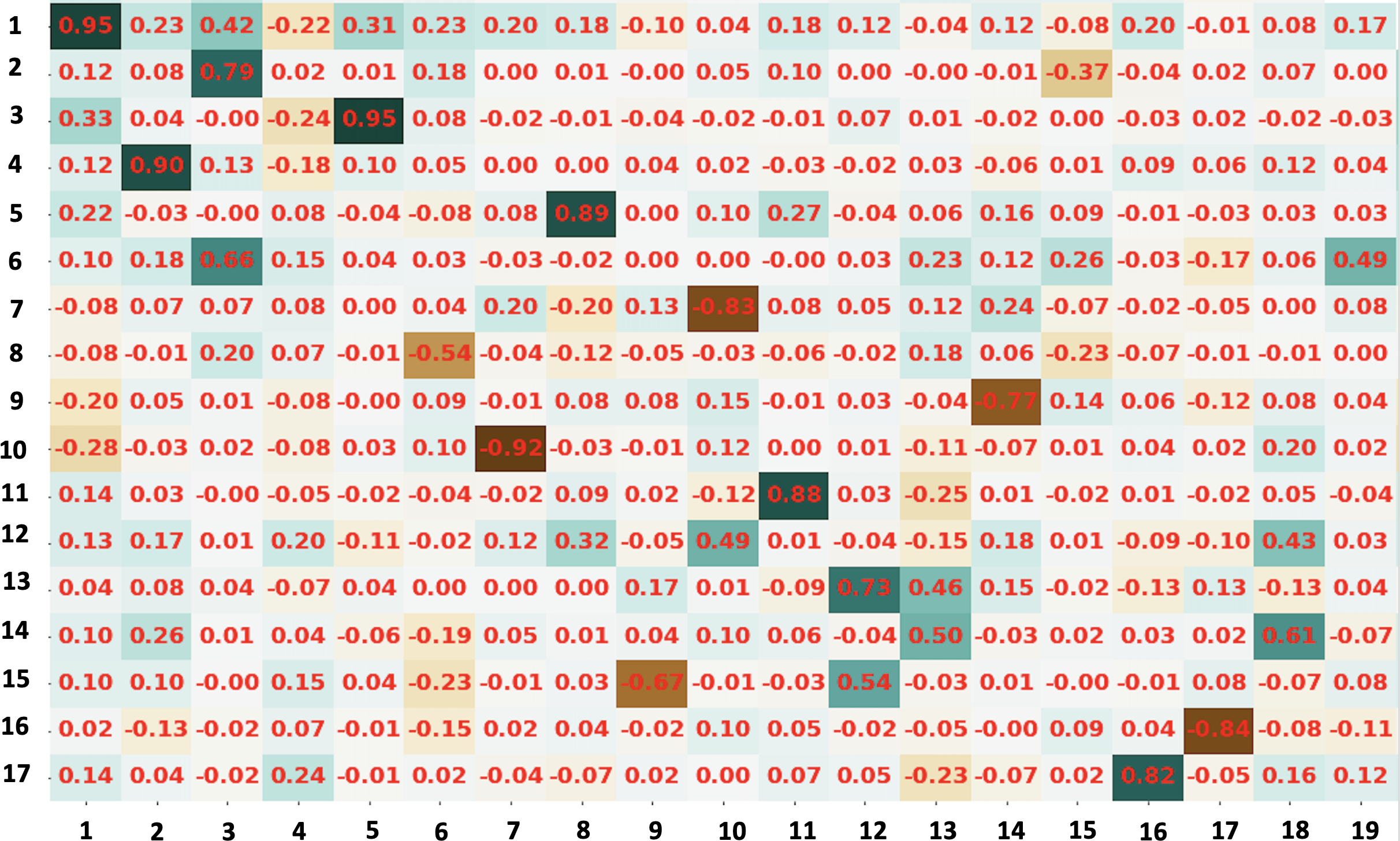}
    \caption[Uncentered Pearson correlations of pre- and post-instruction distractor vectors]{\textbf{Uncentered Pearson correlations of pre- and post-instruction distractor vectors} (columns and rows respectively) after BigeominQ rotation. Twelve of these vectors correlate at above 0.72, showing good overlap between the two sets of discovered sparse vectors.}
    \label{pre-post correlation}
\end{figure}

We find considerable similarity between pre- and post-instruction distractor vectors as shown in Figure~\ref{pre-post correlation}.  Twelve of these distractor vectors correlate above 0.72. In contrast, the typical correlation due to noise can be judged by off-diagonal correlations of distractor vectors with different dimensions from either the pre- or post-instruction; their magnitude is always less than 0.33 and only $\sim$10\% exceed 0.20. This comparison shows that most of the misconceptions associated with the sparse vectors on the pre- and post-test are highly similar, and therefore represent coherent beliefs that resist modification by instruction. Note, however, that the \textit{prevalence} of these various misconceptions is significantly diminished by instruction as we shall show in the following section.

As with our earlier comparisons across schools, we report overall similarity scores for several rotation methods in Table~\ref{tab:prepost_comparison}. Interestingly, the best score here was obtained by an oblique method (bigeominQ) rather than an orthogonal one.

\begin{table}
    \centering
    \footnotesize
    \begin{tabular}{l l}
    \toprule
    BigeominQ & 0.68 \\
    BifactorT & 0.65 \\
    BifactorQ & 0.62 \\
    Entropy & 0.57 \\
    Principal & 0.42 \\
    BigeominT & 0.36 \\
    \bottomrule
    \end{tabular}
    \caption[Cross-instruction similarity index (select transformations only)]{\textbf{Cross-instruction similarity index (select transformations only).} Pooled data from all schools was used to generate both pre- and post-instruction results, which were then compared using our similarity index after various rotations.}
    \label{tab:prepost_comparison}
\end{table}

Overall, we were pleasantly surprised by the striking similarities of the sparse distractor vectors found by the various methods, both across educational institutions and before/after instruction. These strongly suggest that many of our sparse distractor vectors represent educationally-important clusters of distractors which are valid in multiple contexts for our diverse sample of testees. 

\subsection{The Best Transformations}
\label{sec:besttransformations}

We now consider differences in the nature of the transformations alongside the universality and interpretability criteria listed in Section~\ref{sec:sparse-vec-similarity}.

All the transformations we tried promote sparsity in the set of distractor loadings; however, they differ in several important ways that permit their categorization into four classes along two different axes. The first axis is whether the transformation treats all dimensions equally when attempting to induce sparsity, or whether one dimension is treated as a `special case' and allowed to load on all distractors without penalty. The latter type promote results with so-called \textit{bifactor structure}, which have one general and several specific dimensions; in GPArotation, only four methods (bifactorT, bifactorQ, bigeominT, and bigeominQ) are of this type. The second axis is whether the different dimensions of student ability must remain orthogonal following the transformation or whether they can be correlated, a distinction often indicated by appending ``T'' or ``Q'' (respectively) to the names of transformations offering a choice between these alternatives.

Examining Table~\ref{similarity}, the wide range of similarity values combined with the fact that the rank ordering of the various rotation methods is quite similar for both pre- and post-test data suggests that the transformations bifactor, bigeomin, to some extent entropy should generally produce more universal sparse distractor vectors.

We note that although the entropy rotation did not explicitly enforce a bifactor structure, it nevertheless tended to converge on one: as for the bifactor and bigeomin rotations, this solution included a non-sparse ``Newtonian incorrect'' dimension and numerous sparse dimensions (for an example, see Figure~\ref{8solutions} later in this paper). This finding lends weight to the idea that our FCI data has an inherently-bifactor structure, with most of the variability in student response behavior explained by a single general factor and the remainder explained by a long tail of specific misconceptions.

Addressing the question of oblique vs.\ orthogonal rotations requires examination of the sparse vectors with an eye to interpretability. Inspection of loading vectors from a variety of transformations shows that orthogonal transformations are generally less sparse, having a larger number of distractors with non-negligible loadings across many dimensions. This is of course not surprising given our cognitive context: we find it unlikely that real misconceptions are unrelated to the extent that they may \emph{all} be considered orthogonal. In an orthogonal rotation, therefore, some dimensions will necessarily contain projections of distractor loadings from other correlated or anti-correlated misconceptions.

In summary, we found that the distractor vectors determined from the same data by the various methods appear highly similar (see Figure~\ref{8solutions}), with correlations above 0.8 when similar sparse vectors are matched up (to compensate for the dimensions appearing in different orders).  In addition, we found that many sparse vectors match up well across data obtained from different educational institutions.  However, we found several cases where a sparse distractor vector from one source is clearly a combination of two sparse distractor vectors from a source with a larger sample size (for which more and sparser vectors were discovered).

\subsection{Finding the Most Robust Distractor Vectors}
\label{sec:eightsolprocedure}

In addition to illustrating robustness and informing our choice of rotation methods, the results above demonstrate a key property of our method: it extracts more dimensions as the sample size increases. (This behavior was explored in more detail in our previous paper; see \cite{segado2025}.) In order to find the largest possible number of distractor vectors it therefore makes sense to analyze the largest possible dataset, here made by combining data from all educational institutions both before and after instruction. Combining multiple sources in this way also has the advantage of averaging out source-specific features, hopefully leading to more generalizable results.

When we combined all pre-instruction data or all post-instruction data, the MNCM-Bayes method recovered 19 or 17 dimensions, respectively. Combining all pre- and post-instruction data from all educational institutions into a single $\sim$34k-administration dataset yielded 21 recovered dimensions. Guided by our earlier explorations of rotation methods, we transformed our complete-data results using the bifactorQ, bigeominQ, and entropy rotations. We also included the commonly-used quartimin rotation for comparison; this rotation allows correlations in the abilities but does \emph{not} accommodate a global non-sparse dimension.

As alluded to earlier, some rotation methods may get stuck in local minima, an issue which is usually dealt with by repeating the procedure multiple times with different random initial rotation matrices. In many cases, the simplicity of the local rotation minima is similar to that of the global minimum, and these alternate solutions may be advantageous because they provide a `menu' of psychologically-plausible candidates for further examination \citep{rozeboom_glory_1992}. In the end, the most interpretable sparse solution---and ideally, the one closest to students' underlying thought processes---may prove to be one of these local minima rather than the one that minimizes the rotation objective function \citep{nguyen2023}. We therefore report not only the global minimum attained by each rotation, but also all salient local minima found when running each method multiple times starting from different randomly-initialized rotation matrices. We use 100 such starts for each method as suggested by \citeauthor{hattori2017} \citeyearpar{hattori2017} in the context of Geomin rotations (which tend to yield numerous local minima).

The complete procedure for obtaining our candidate set of sparse solutions from the 21-dimensional fit on our combined dataset, then, is as follows:
\begin{enumerate}
    \item Collect all slope parameters for each distractor of each item into a $120 \times 21$-dimensional matrix. Note that we exclude the correct answer categories, which are used as a reference and therefore always have a slope of zero; this leaves $30 \times (5 - 1) = 120$ free slopes for each dimension.
    \item Generate 100 random $21 \times 21$ rotation matrices to use as starting points for the rotation methods.
    \item For each of the four selected rotation methods (bifactorQ, bigeominQ, entropy, and quartimin):
    \begin{enumerate}
        \item Run the method 100 times—once for each of the 100 random initial starting rotations generated in the previous step—on the $120 \times 21$-dimensional matrix of distractor slopes, yielding 100 rotated slope-matrix solutions.
        \item \label{item:signed-permutation-step} Align the resulting solutions by permuting and sign-reversing the dimensions of solutions 2--100 as necessary to best match those of solution 1 (chosen as an arbitrary reference). This step is necessary because the rotation methods are insensitive to any signed permutation of the sparse dimensions and may return these in arbitrary order and with arbitrary signs. We define a ``best match'' permutation as one that maximizes the sum of uncentered Pearson correlations between the aligned dimensions in a given pair of solutions; this may be treated as a standard linear sum assignment problem and solved using off-the-shelf software routines.
        \item Apply hierarchical agglomerative clustering to the 100 re-ordered and sign-corrected slope matrices to identify sets of identical solutions. The metric used for clustering is the cosine distance (equal to one minus the uncentered Pearson correlation), and a maximum distance of $10^{-4}$ is permitted between any two solutions in a cluster. 
        \item Retain only clusters containing 10 or more solutions. Each retained cluster represents a commonly-encountered minimum of the rotation function, and is summarized by the cluster centroid (i.e. the mean of the slope matrices in the cluster).
    \end{enumerate}
    \item For all commonly-encountered minima found in the previous step, permute or sign-reverse the dimensions (using the same procedure described in step~\ref{item:signed-permutation-step}) to align the results across the different rotation methods.
\end{enumerate}

\begin{sidewaysfigure}
    \includegraphics[width=\textheight]{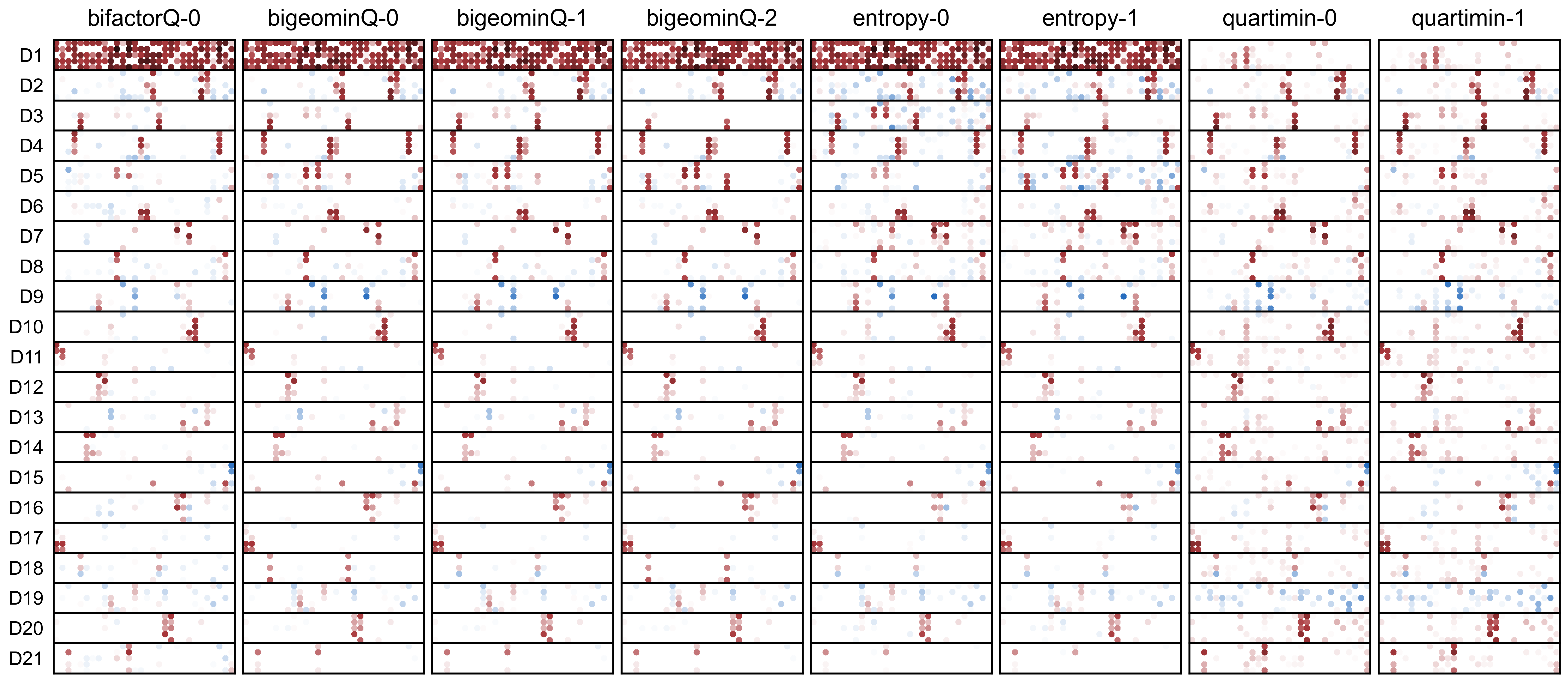}
    \caption[Eight candidate sparse solutions]{\textbf{Eight candidate sparse solutions}. These were obtained by applying our method to the combined FCI dataset including both pre- and post-instruction results from multiple schools. The eight solutions are largely similar, with the notable exception being the first dimension of the quartimin-rotated solutions. Further differences are discussed in the text.}
    \label{8solutions}
\end{sidewaysfigure}

Applying the procedure above yielded a total of eight candidate solutions: one from the bifactorQ method, three from the bigeominQ method, and two each from the entropy and quartimin methods. A high-level visual summary of these eight solutions is shown in Figure~\ref{8solutions}.

\subsection{Selection of Final Sparse Vectors for Interpretation}
\label{sec:selection-of-best-vectors}

The various rotation solutions obtained in the preceding section represent eight alternative ``lenses'' through which we can view our data. Statistically, all fit the observed data equally well due to the rotational invariance of the MNCM (i.e., the computed response probabilities are identical for all rotated solutions); consequently, metrics beyond statistical likelihood are needed in order to evaluate these further.

While there is no substitute for examining the content of the distractors in each sparse vector to determine whether they are consistent with an interpretable construct (which we will do in section~\ref{contentanalysis}), there are several quantitative metrics we can apply to evaluate the overall quality of our solutions. We will use these to select a subset of the most promising sparse vectors for further interpretation.

\subsubsection{A simple metric of overall solution quality}

One of the simplest quality metrics for a factor-rotated solution is the \emph{hyperplane count}: the total number of slopes that are equal (or nearly equal) to zero \citep{cattell1963}. Larger hyperplane counts imply a smaller number of non-ignorable relationships between distractors and the rotated dimensions; that is, they serve as another measure of overall simplicity. Prior work has found that solutions with a higher hyperplane count tend to be the ones that best correspond to the true underlying structure of the data in cases where this structure was available for comparison \citep{nguyen2023}. As the slopes for the correct answer categories are fixed at zero, we compute the hyperplane counts of each solution based only on the distractor slopes using an inclusion criterion of $|a| < 0.2$. 

\begin{figure}
    \centering
    \includegraphics[width=4in]{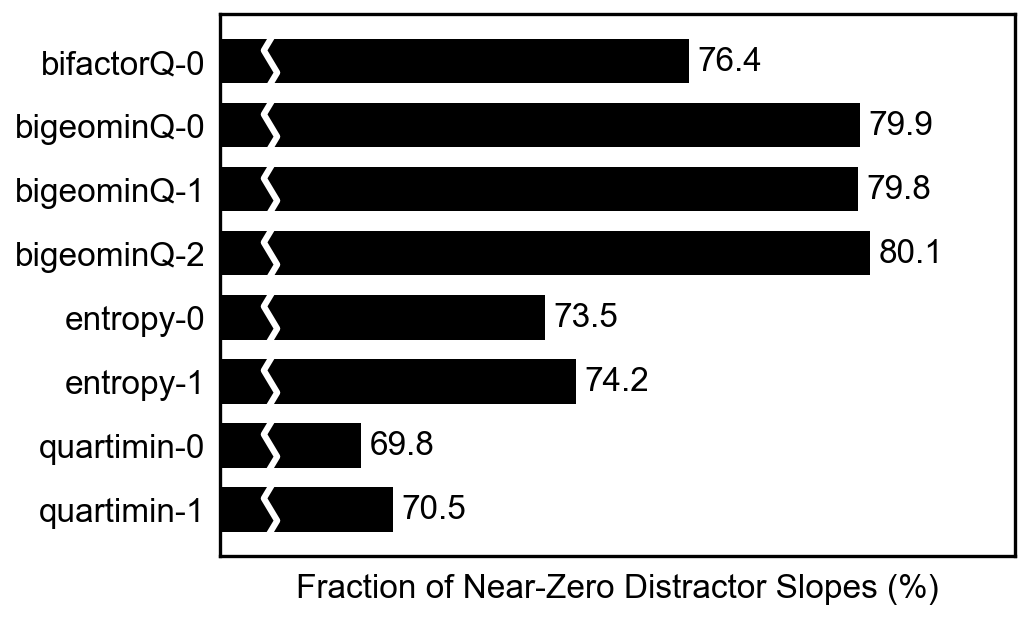}
    \caption[Hyperplane fractions for eight candidate solutions]{\textbf{Hyperplane fractions for eight candidate solutions}, here defined as the fraction of distractor slopes with magnitude less than 0.2. Solutions with more nearly-zero distractor slopes (i.e., greater hyperplane count) are typically easier to interpret and perhaps more likely to match real psychological processes.}
    \label{simplemetrics}
\end{figure}

We report our findings as a fraction of the number of free slopes in all the sparse vectors of that rotation to facilitate interpretation; these are shown in Figure~\ref{simplemetrics}, and unambiguously favor the bigeominQ-rotated solutions. While not shown in the figure, these results are also robust to the choice of threshold value used for the hyperplane count, with one of the bigeominQ solutions (typically bigeominQ-2) achieving the highest score for any choice of threshold between 0.002 and 0.3.

\subsubsection{Evaluating robustness via non-parametric bootstrapping}
\label{sec:bootstrap}

In addition to the simple metrics above, we evaluate the robustness of our results across different subsamples of our data using a non-parametric bootstrap approach. We generate 500 bootstrap samples in total; each of these is formed by drawing students at random, with replacement\footnote{i.e., allowing a student to be randomly selected multiple times or not at all}, up to the size of the original 34k-administration dataset. In this initial bootstrap evaluation, both pre- and post-instruction responses are included for every sampled student: that is, if a given student's pre-test data appears in a bootstrap dataset, so does their post-test data. (We will later explore pre- and post-instruction trends separately.) For each of the 500 bootstrap samples, we fit the MNCM using our MNCM-Bayes method, permitting up to 25 dimensions as when fitting the parent dataset and identifying the results to have orthogonal $a$ vectors and orthonormal $\theta$s.

Following extraction of the MNCM coefficients, we then align the results of all bootstrap replications to a common coordinate system using orthogonal generalized Procrustes rotation \citep{gower1975}. This step is necessary for two reasons, one theoretical and the other practical. The theoretical reason is that our identification constraints cannot fully resolve rotational indeterminacy: if two dimensions have equal variance, the solution retains a rotational degree of freedom within the plane of these dimensions. The sign indeterminacy of the MNCM is also not resolved by our identification constraints, and so an alignment step is needed to ensure that the signs of each dimension are consistent across replications. The practical reason is that even when orthogonalizing the slopes \emph{can} fully constrain rotation, a principal coordinate system may still be highly sensitive to minor fluctuations in coefficient values and therefore vary widely across different replications. This occurs when the variances of any two dimensions in the parent dataset are similar, even if not exactly equal.

\begin{figure}
    \centering
    \includegraphics[width=4in]{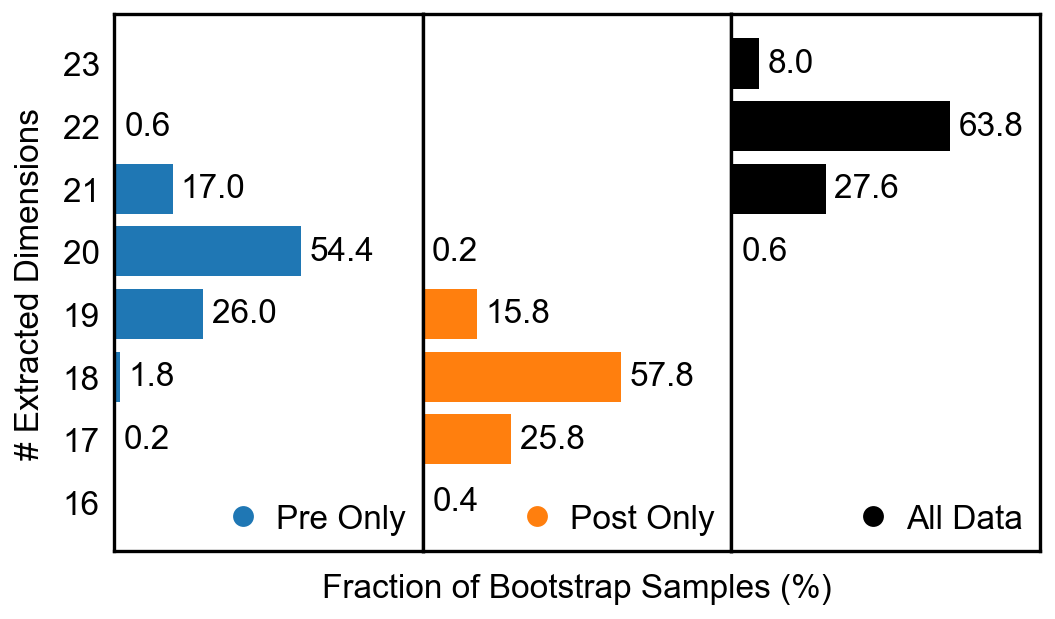}
    \caption[Histograms of extracted number of dimension in bootstrap evaluations]{\textbf{Histograms of extracted number of dimension in bootstrap evaluations}. The three panes show the distributions, over 500 bootstrap samples, of the number of dimensions extracted from only pre-instruction data, only post-instruction data, and all data combined.}
    \label{fig:bootstrap_hist_all}
\end{figure}

As the MNCM-Bayes method exhibits self-selecting dimensionality, the number of dimensions extracted differs across bootstrap replications. We retain only the first 21 dimensions of the aligned solutions; this matches the dimensionality of the solution obtained on the parent dataset and also represents the number of dimensions shared across all but 0.6\% of the bootstrap replications (see Figure~\ref{fig:bootstrap_hist_all}, ``All Data'').

We use oblique target rotations to map the ensemble of aligned and truncated bootstrap solutions into the coordinate systems of each of the eight sparse solutions under evaluation. These rotations are computed analytically using the approach described by \citeauthor{korth1976} \citeyearpar{korth1976}, which simultaneously maximizes the uncentered Pearson correlation\footnote{``Uncentered'' because the Newtonian correct slopes are constrained to zero. These are alternately known as the coefficient of factor congruence, Tucker's congruence coefficient, the cosine similarity, and the congruence coefficient} between each individual dimension of the rotated and target $A$ matrices. We apply a deadband of $\pm0.2$ to both the rotated and target vectors prior to computing these correlations since these small coefficients are below our estimate of the error—i.e., they act like noise. Mathematically, we transform the slopes in each according to
\begin{equation}
    \label{eq:deadband}
    x \mapsto \sgn(x) \max(|x| - 0.2, 0).
\end{equation}
This transformation leaves the larger slopes largely unchanged.

\begin{sidewaysfigure}   
    \centering
    \includegraphics[width=0.9\textheight]{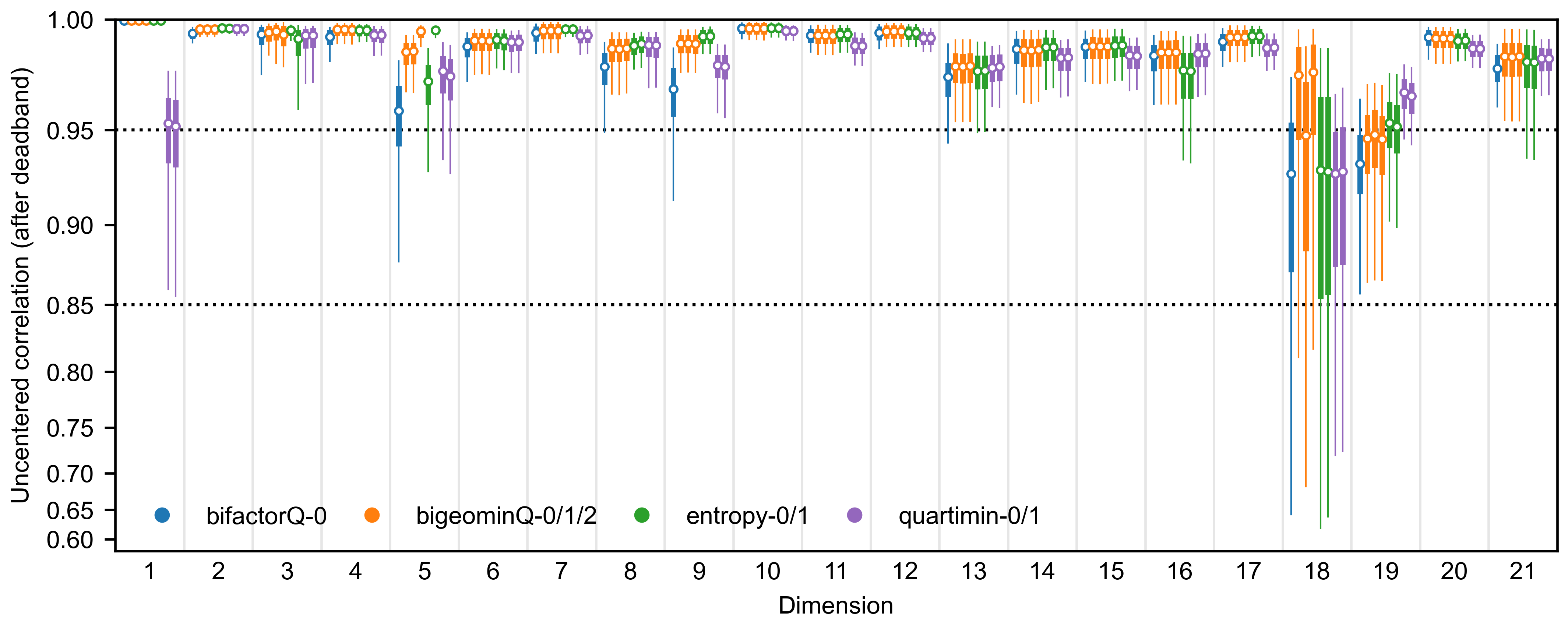}
    \caption[Bootstrapped correlation coefficients for eight candidate solutions]{\textbf{Bootstrapped correlation coefficients for eight candidate solutions}, computed per-dimension after applying a deadband of $\pm0.2$ as described in the text. White points indicate median values across bootstrap samples, while thick and thin lines indicate interquartile ranges and central 95\% percentile intervals, respectively.}
    \label{eightsolcongruences}
\end{sidewaysfigure}

The bootstrapped correlation coefficients for each dimension of our eight candidate solutions are summarized in Figure~\ref{eightsolcongruences}. As is clear from the figure, the majority of dimensions in all rotations have modified correlation coefficients exceeding 0.95, indicating excellent reproducibility across subsamples of our data. The most conspicuous exception is dimension 18, for which only bigeominQ-0 and bigeominQ-2 
achieved an acceptable range of correlation coefficients (though even for these many samples fell below 0.85, which we adopt as a minimum threshold for good agreement between factor patterns following \cite{lorenzo-seva2006}).

\subsubsection{Choice of robust vectors for further analysis}
\label{sec:choice-of-robust-vectors}

Based on the metrics computed above as well as our qualitative judgment, we now choose a subset of \textit{robust distractor vectors} from our candidate solution set to examine further. For dimensions which appear qualitatively identical between different rotation methods (most of the dimensions in this set), we select the bigeominQ-0 version as our exemplar. This solution represents the lowest minimum attained by any of the bigeominQ results, and is nearly tied for high hyperplane count and good bootstrap reproducibility. For dimensions in which the differences appear more substantive, we typically choose two exemplars from either the bigeominQ-1, bigeominQ-2, or bifactorQ-0 solutions. The final set of vectors selected for additional analysis are assigned roman numerals for ease of identification and are listed in Table~\ref{tab:bestvector}, with the full details of each (including coefficient values, distractor descriptions, labels according to existing misconception taxonomies, and other metrics) listed in the supplemental materials for this paper.

\begin{table}
    \centering
    \footnotesize
    \begin{tabular}{llp{0.8in}|llp{0.8in}}
        \toprule
        Rom. & Dim. & Solution & Rom. & Dim. & Solution \\
        \midrule
        I & 1 & \textasteriskcentered & XV & 12 & \textasteriskcentered \\
        II & 2 & \textasteriskcentered & XVI & 13 & \textasteriskcentered \\
        III & 3 & bigeominQ-1 & XVII & 14 & \textasteriskcentered \\
        IV & 3 & bigeominQ-2 & XVIII & 15 & \textasteriskcentered \\
        V & 4 & \textasteriskcentered & XIX & 16 & \textasteriskcentered \\
        VI & 5 & bigeominQ-1 & XX & 16 & bifactorQ-0 \\
        VII & 5 & bigeominQ-2 & XXI & 17 & \textasteriskcentered \\
        VIII & 6 & \textasteriskcentered & XXII & 18 & bigeominQ-1 \\
        IX & 7 & \textasteriskcentered & XXIII & 18 & bigeominQ-2 \\
        X & 8 & \textasteriskcentered & XXIV & 19 & \textasteriskcentered \\
        XI & 9 & \textasteriskcentered & XXV & 19 & bifactorQ-0 \\
        XII & 9 & bifactorQ-0 & XXVI & 20 & \textasteriskcentered \\
        XIII & 10 & \textasteriskcentered & XXVII & 21 & \textasteriskcentered \\
        XIV & 11 & \textasteriskcentered &  &  &  \\
        \bottomrule
    \end{tabular}
    \caption[Robust distractor vectors chosen for further analysis]{\textbf{Robust distractor vectors chosen for further analysis.} Each vector is uniquely identified with a roman numeral. Vectors selected from the bigeominQ-0 solution are marked with an asterisk (\textasteriskcentered) for clarity, with all others explicitly labeled in the table.}
    \label{tab:bestvector}
\end{table}

\subsubsection{Additional pre/post-test bootstrap evaluation}
\label{sec:pre-post-bootstrap}

Finally, we also compute the median correlation coefficients obtained from additional bootstrap analyses on only pre-instruction or post-instruction data and include these in our discussion of discovered misconceptions (see Table~\ref{tab:all-misconceptions-table}), with the complete set for all robust sparse vectors again appearing in our supplemental materials. These results serve as a preliminary indicator of the extent to which pre- or post-instruction data alone support the sparse vectors obtained from the combined dataset; we hope to explore such differences in more detail using a true multi-group extension of our method in future work.

The procedure used to generate pre- and post-data bootstrap results was identical to that used for the combined-data analyses: students are randomly sub-sampled, the model is fit using our Bayesian method, results are identified and aligned to a common coordinate system, and the set of aligned bootstrap results is then rotated to maximize the uncentered correlation with each of the eight candidate solutions obtained from the combined data (Section~\ref{sec:eightsolprocedure}).

An important caveat of this approach is that the lower effective sample size of the pre-only and post-only datasets leads to fewer dimensions being extracted by our Bayesian method on average. This is evident in Figure~\ref{fig:bootstrap_hist_all}, where we observe that only 19- and 17-dimensional fits are supported by the majority of pre- and post-test bootstrap replications, respectively. This places a ceiling on the bootstrap correlations obtained: even if the psychological processes governing the pre- and post-test data were in fact identical to those underlying the complete data, it is not possible to transform these results (which both have fewer than 21 dimensions) such that they simultaneously achieve perfect correlations with all 21 linearly-independent dimensions obtained from the combined dataset. When interpreting these results, therefore, it is important to remember that any observed reductions in the correlation coefficients stem partly from the smaller sample size of the pre- and post-test only datasets and not solely to underlying differences in the data-generating processes.

\section{Robust Distractor Vectors Represent Misconceptions and Misunderstandings}
\label{contentanalysis}

The result of all of the above processes is a carefully-considered set of distractor vectors and supporting bootstrap metrics suitable for further interpretation; these are documented in full in the supplemental materials to this paper. In this section we examine these to see whether they in fact represent identifiable misconceptions.  This analysis tests the key assumption motivating this research: that the robust subsets of distractors identified using a purely-exploratory combination of Bayesian IRT and factor rotation methods will be readily identifiable as cognitive misconceptions or misunderstandings. We find that this assumption is strongly confirmed because examination of the text of the most heavily weighted distractors in each sparse dimension by domain experts shows that they nearly always pertain to an obvious misconception.

To interpret each distractor vector, we first examine the distractors with the largest slopes in the context of their associated questions to determine which beliefs or known misconceptions would be consistent with their selection. (When making this determination, it is occasionally helpful to contrast these with the response categories in the same questions that do \emph{not} have sizable loadings, including the correct answers.) This set of consistent beliefs and misconceptions is examined to see if these distractors are consistent with one (or two) clear-cut hypotheses, in which case we assign that distractor vector a descriptive name. In most cases we find that this is a known misconception or a particular refinement of one that is known, although we find two which are novel.

Close examination of the discovered misconceptions (and also of their remediation by instruction in the next section) has not revealed any obvious distinguishing characteristics that would enable us to classify them into intellectually distinct groups. However, noting that many of our misconceptions were accepted beliefs at some pre-Newtonian point in time, we have elected to organize them using historical eras: Ancient, Medieval, and finally Post-Newtonian (with suitable references to prior research).  Ancient is used to connote conceptions acquired by infants and toddlers which were codified and extended largely by Greek philosophers, Medieval refers to the time from $\sim$500~A.D.\ to 1600~A.D., and Post-Newtonian refers to the era since Galileo and Newton.

Our 22 named and categorized distractor vectors are listed in Table~\ref{tab:all-misconceptions-table}; we discuss these by historical era below.

\subsection{Ancient Views of Force and Motion}
\label{sec:ancient}

Before they can talk, human infants learn to parse the visual and haptic (sense of touch) world into individual objects and to classify them as inanimate (passive) or animate (having agency). Animate objects are able to move (generally toward some objective), whereas inanimate objects are naturally at rest with respect to their surroundings unless pushed or pulled with some minimum force \citep{spelke2022}. The idea that maintaining motion requires more force than the resistance is our strongest specific misconception, \vecname{II}.  The Greeks formalized these notions, adding a tendency to move towards the center of the earth (which they knew was spherical) due to the ``natural'' force of gravity.  

Once moving, Aristotle's golden rule of motion applies: ``Force is to force as time is to time inversely'' \citep{dijksterhuis1969}. This formulation is based on the Greek conception of motion as a transition: from one place and time to another place and time (with both variables separated by fixed intervals).  Thus this rule says that if the force is doubled, the time is halved.  Invoking the more modern concept of speed, this rule says that force causes speed in proportion.  Our misconception \vecname{XIV} is consistent with this rule, while \vecname{XXI} and \vecname{XVI} are closely related. Furthermore,  \vecname{XXVII} is a logical consequence of Aristotle's rule if the universal increase in speed observed for falling objects is combined with the belief that force is proportional to motion. 

\begin{table*}   
    \includegraphics[width=\textwidth]{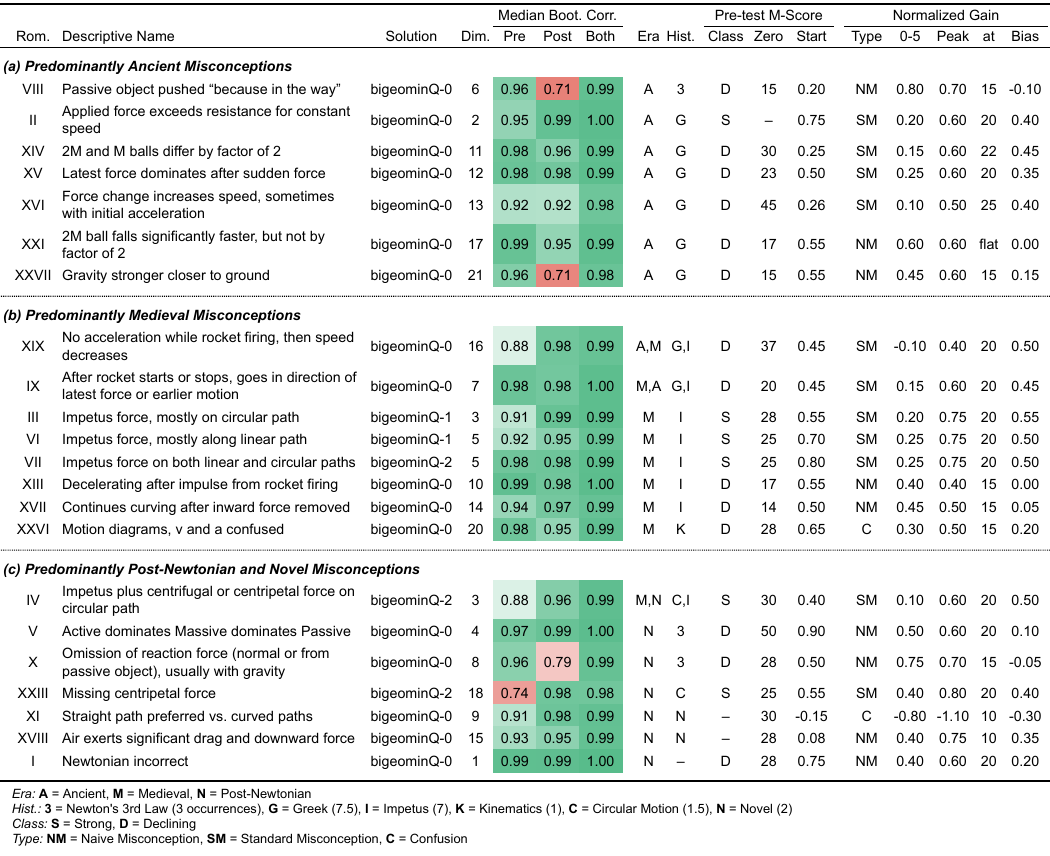}
    \caption[Table of named misconceptions]{\textbf{Table of named misconceptions}, broadly organized by historical eras. \textit{Rom.}\ is the roman numeral that indexes our robust distractor vectors (cf. Table~\ref{tab:bestvector}). The \textit{Median Boot.\ Corr.}\ columns show the medians of the bootstrap correlations described in Section~\ref{sec:bootstrap}, with green and red shading used for values above and below 0.85, respectively. \textit{Hist.} contains an alternate categorization schema based primarily on historical views of our misconceptions. The various columns grouped under the headings \textit{Pre-test M-Score} and \textit{Normalized Gain} describe additional features related to the prevalence of each misconception in our data and will be discussed later in the text (Section~\ref{sec:student-class-prevalence}).}
    \label{tab:all-misconceptions-table}
\end{table*}

The robust distractor vectors that reflect predominantly-Ancient ideas of force and motion are listed in Table~\ref{tab:all-misconceptions-table}(a). The pervasiveness of the unity of force and motion in these misconceptions causes us to consider it the central pre-Newtonian misconception: that \textbf{force causes motion}---and consequently that motion when the force is constant is motion with constant velocity, and hence no acceleration---in contrast to Newton's \textbf{force changes motion}, i.e. gives acceleration and changes velocity.  A corollary of ``force causes motion'' is that there is no inertia or other memory of motion previous to the latest force or impulse as in \vecname{XV} although here there is continuing motion after the impulse stops. 

Two of our discovered misconceptions related to questions involving Newton's 3rd Law arise from the ancient fundamental belief that animate objects possess the ability to initiate their own motion and to exert forces on other objects, and the logical follow-up belief that inanimate objects are incapable of initiating motion or of exerting forces. This suggests one-way interactions like \vecname{VIII}. Closely related is \vecname{X}; we have chosen to classify this as predominantly post-Newtonian rather than predominantly ancient because it directly contradicts Newton's 3rd Law.

\subsection{Medieval Views of Force and Motion}
\label{sec:medieval}

Medieval scholars improved on ancient ideas of force and motion in two major ways.  The Merton College school, among others, created an understanding of kinematics, differentiating between velocity and acceleration and inventing graphical ways to display and interrelate them.  Thus we assign \vecname{XXVI} as a Medieval misunderstanding since the Merton school did not straighten these out until late in this period. 

The major concern in medieval mechanics was to originate and elaborate on impetus theory \citep{drake_impetus_1975} which addresses the most obvious challenge to ``force causes motion in proportion''---why does the motion of an inanimate body continue after the initial motive force has ceased? Examples include projectiles that keep moving after the force of the thrower no longer acts, planets continuing in their orbits, and wheels rotating after being set spinning.

Based on the idea that the motion of animate bodies is due to some property or internal thing that allows them to move, impetus theory posits by analogy that the force that initially sets an inanimate object in motion also instills some kind of substance or thing---called \textit{impetus}---within that body that acts like an internal force to maintain the body's motion. The FCI was designed for students with little formal physics training and no idea of historical concepts like impetus, so our categorization of ``impetus force'' is inferred from the presence of wording such as ``a force in the direction of motion'' for a body moving at constant speed with no real external force acting along its path.

\begin{figure*}   
    \centering
    \begin{tblr}{
        stretch = 0,
        colspec = {Q[h,3.8in]p{1.7in}},
        cells = {font = \sffamily\raggedright\footnotesize}
    }
        \includegraphics[width=\linewidth]{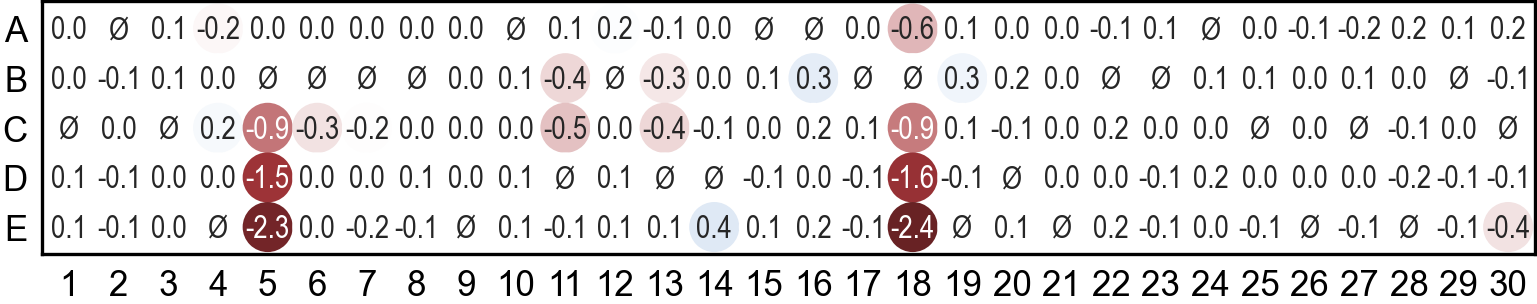} & \vecname{III} \\
        \includegraphics[width=\linewidth]{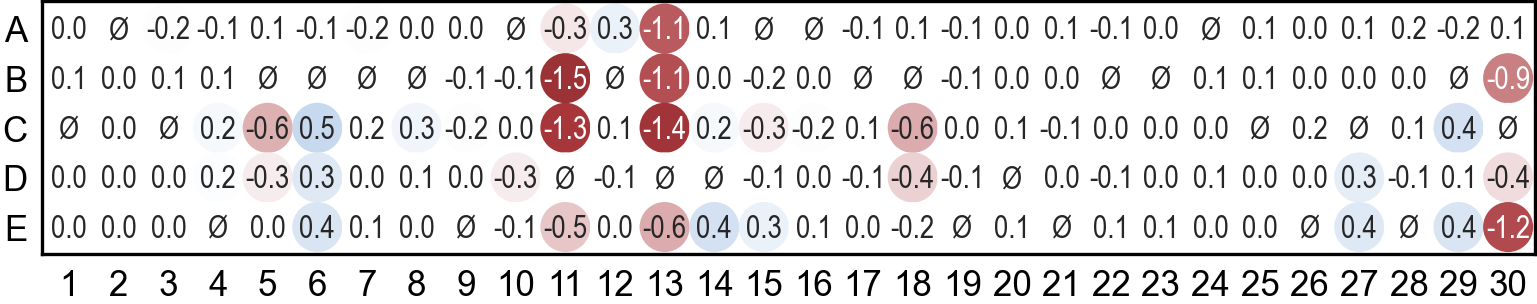} & \vecname{VI} \\
        \includegraphics[width=\linewidth]{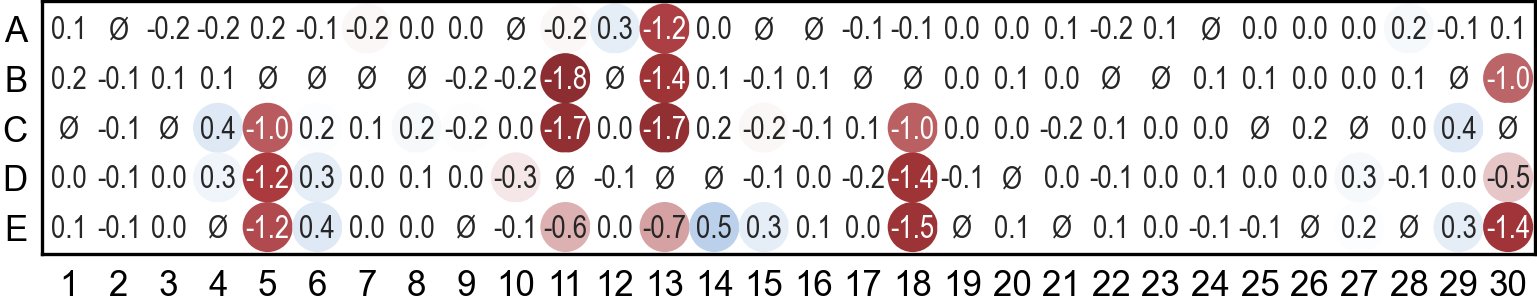} & \vecname{VII} \\
        \includegraphics[width=\linewidth]{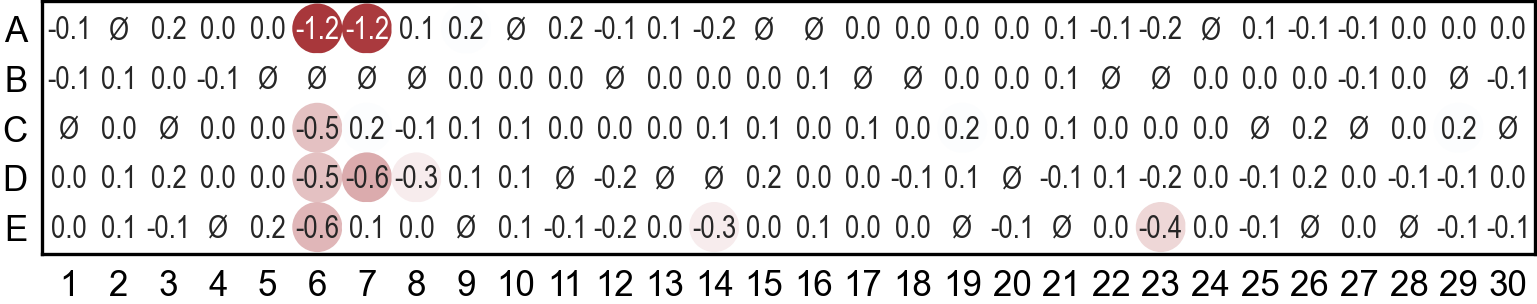} & \vecname{XVII}
    \end{tblr}
    \caption[Four impetus misconception vectors]{\textbf{Four impetus misconception vectors.} The first loads primarily on Q5 and Q18 which involve impetus force in circular motion, the second loads strongly on six distractors all involving an impetus force along the (straight-line) motion, and the third loads strongly on all distractors in the preceding two. The fourth describes a different but related concept: the continuation of a circular trajectory in the absence of radial forces. The work of the Merton College school on position, velocity, and acceleration towards the end of the Medieval period justifies including failure to distinguish amongst these variables as a Medieval misconception.}
    \label{4impetus}
\end{figure*}

The impetus vectors III, VI, and VII in Table~\ref{tab:all-misconceptions-table}(b) all involve an impetus force acting along the direction of motion; they differ primarily in whether the object is traversing a circular or a straight path. \vecname{III} loads heavily on circular impetus, especially when combined with a radial force; it loads weakly on impetus forces for linear paths. \vecname{VI} loads decisively on impetus force on linear paths. \vecname{VII} has slopes larger than 1.0 on both linear and circular paths. These overlapping misconceptions reflect that questions 5, 11, 13, and 18 on the FCI load on several misconceptions whose sparse vectors are not orthogonal and will therefore produce correlated misconception scores (to be discussed later) as is evident from the several dot plots shown in Figure~\ref{4impetus}.

Although various proponents of impetus theory agree on the existence of an impetus force, they espoused many alternatives for the motion that it causes.  For example, most impetus theorists posited that the impetus would eventually ``wear out'', causing the object to slow down, especially if external forces act (either against or perpendicular to the direction of motion). This concept is reflected in \vecname{XIII} where the rocket decelerates (but occasionally accelerates) with no external force present. Decreasing impetus is combined with ``force causes proportional motion'' in \vecname{XIX}, which is jointly assigned to the ancient and medieval ages.  

The last and best codification of impetus ideas is due to Buridan and his school in the 13th century.  He argued that in the absence of external forces an object containing impetus would continue in motion indefinitely and that the impetus was proportional to the weight times the speed, foreshadowing Galileo and Newton's ideas about inertia and momentum. However, they attributed continuous motion to inertia and asserted that motion was a state variable rather than some sort of internal substance or property \citep{jammer}.

Buridan also believed that an object in circular motion would continue in circular motion in the absence of any applied force, a common current misconception as evidenced by \vecname{XVII}.  

\subsubsection{Combined Ancient and Impetus Misconceptions}

Several of the questions on the FCI ask students to select the path of an object with a given motional history starting when the force on it changes.  It is then our problem to categorize the path, which often starts in one direction and curves towards another so that the simplest explanation involves two concepts. The dependence of present motion on past motion and forces is an area of variation in various impetus theories, for example \vecname{IX} which has some force-causes-motion ideas mixed in and is categorized accordingly.

In summary, these variations are not surprising, since impetus theory was an ad hoc theory based on the intuitions of period Greek, Arabic, and medieval natural philosophers---rather than on experiment or a precise mathematical formulation like Newton's laws.  

\subsection{Post-Newtonian and Novel Misconceptions}
\label{sec:post-newtonian-novel}

Newton's laws introduce or imply several new conceptions, notably that uniform circular motion is accelerated motion and that physical interactions require pairs of equal forces.  We shall discuss circular motion misconceptions first and then those following directly from Newton's 3rd Law.  Finally we will introduce two misconceptions that are new (and hence post-Newtonian) as well as an overall ``Newtonian Incorrect'' dimension. The relevant distractor vectors are listed in Table~\ref{tab:all-misconceptions-table}(c).

\subsubsection{Circular Motion}

Newton's laws imply that uniform circular motion involves (radial) acceleration.  This is a challenging concept for post-Newtonian students as the concept of acceleration perpendicular to the direction of travel and in the absence of a change of speed is hard to digest; for example, circular motion spontaneously emerged as a category on a concept test designed to test reasoning about momentum and energy in mechanics \citep{lee2017}.

On the FCI, confusions about centripetal and centrifugal force appear in conjunction with impetus force on a circular path in \vecname{IV}, which we classify as both medieval and post-Newtonian.  However, \vecname{XXIII} involves \textit{never invoking an inward (centripetal) force} when asked to identify forces acting on an object in circular motion and is clearly a post-Newtonian misconception; our bootstrap evaluations suggest that it is much more clearly defined after instruction than before (once students have had some formal exposure to the relevant concepts).

\subsubsection{Third Law Misconceptions}

One of the strongest \emph{sparse} distractor vectors we have found is \vecname{V}, which involves clear violations of Newton's 3rd Law in interactions between objects of different mass. In general, previous research has found that the active or more massive agent in such interactions is perceived by students as exerting the greater force. In support of the status quo, we observe that in a collision between a truck and a small car (Q4), distractors in which the truck exerts a greater force receive the strongest loadings; similarly, in a scenario involving a student on a rolling chair pushing another student of lesser mass (Q28), the strongest loadings are on distractors in which the heavier student exerts the greater force.

Questions 15 and 16 (a car pushing a disabled truck) present a more nuanced situation in which the active agent is clearly the \emph{less} massive of the two. The conflict between ``active wins'' and ``massive wins'' appears to be reflected in the distractor loadings, which do not clearly favor the car or the truck when both are accelerating (Q15) and only weakly favor the car once both reach a constant speed (Q16). However, we note that the \emph{overall fraction} of submissions favoring the truck is far smaller, and so the loadings for these distractors are far less consequential than those for the more common wrong answers (which strongly favor the car---hence our choice of name, \vecnameonly{V}).

Differences in the overall response fractions also help make sense of the non-zero distractor loadings favoring the passive, less-massive agents in Q4 and Q28. While the observed likelihood of selecting such answers \emph{does} in fact seem to vary with the amount of this misconception, they are still selected very infrequently overall.

Other misconceptions pertinent to the 3rd Law include \vecname{X} and the previously-mentioned \vecname{XXIII}.

\subsubsection{Novel Misconceptions}

We found two misconceptions that do not appear to have been observed previously.  The first concerns the idea that \vecname{XVIII}: Air is seen as pushing downwards on an empty chair (Q29), on an upward-moving elevator (Q17), and on a falling stone.  On Q30 (Tennis Ball Hit Into Wind) the two responses that \textit{do not} adduce ``a force exerted by the air'' acting on the ball have strongly positive weights (i.e., selecting these responses indicates that the student \textit{does not} have this misconception) confirming that students possessing this misconception believe that air always exerts substantial forces.  

The second novel distractor vector we found is \vecname{XL} which weighs heavily on all five of the seven questions that ask about the path that an object takes given various scenarios of prior motion and force (the other two concern circular motion).  In questions 8 and 23 the object is subject to no forces in the plane of the figure and all of the distractors with strongly negative slopes ($a \le -0.8$) are curved paths.  In questions 12, 14, and 21 the object is subject to one force in the plane of the figure, and all of the distractors with strongly positive slopes ($a \geq 0.8$) are straight paths, implying that holders of this misconception favor curved paths.  Thus the misconception (corresponding to the negative slopes) is preference for curved paths for an object with no force on it and avoidance of straight paths for objects with a single force on them, again suggesting a preference for curved paths.  We can find no consistent description of this misconception involving any dynamical law; for example, both ``constant force causes uniform motion'' and ``always gets path wrong'' are inconsistent with the observed universal preference for curved or straight paths.

\subsubsection{Newtonian Incorrect}

By far the strongest distractor vector we have found is \vecname{I}. This loads significantly ($|a|\ge 0.5$) on \textit{every} distractor and heavily ($|a|\ge 1.0$) all but 13 (out of 120 total distractors). Of its six ``most incorrect'' responses ($|a|\ge 3.0$), three involve impetus on a linear path, two involve the belief that passive objects exert no force, and one the Greek idea that a thrown ball falls because of a natural tendency to rest on the surface of the earth.

\subsection{Connections with prior work}

This current work produces an enhanced statistical model of the FCI that is related to and builds upon three substantial strands of physics education research: factor analysis, IRT, and network analysis.

Work in factor analysis has examined only whether the items were correctly answered; the items are dichotomously scored. Studies applying traditional Exploratory Factor Analysis have reported models with from 5 to 6 factors as optimal \citep{scott2012exploratory,semak2017}. Studies performing factor analysis using multidimensional IRT (MIRT) have identified 5 to 9 factors as optimal \citep{scott2015,stewart2018}. These studies, which considered only the correctness of each question and did not distinguish between different distractors, support the identification of many more dimensions in the present work when incorrect knowledge evidenced by specific distractors is also considered.

Many studies have used one of the numerous IRT models to explore the FCI. Most have only considered a single latent ability trait and do not particularly inform this study. The MIRT factor analysis studies above used a multidimensional version of the two-parameter logistic (2PL) IRT model to allow multiple ability traits under the assumption of dichotomous grading. A multidimensional version of Bock's Nominal Response Model \citep[NRM;][]{Bock1972} has also been used in some studies, and represents a limiting case of the MNCM in which all distractors for a question are constrained to share a single ``direction'' in the multidimensional space. Using a dataset related to the one examined here, the multidimensional NRM was applied by \citet{Stewart2021} to determine that 8 to 9 factors (dimensions) produced optimal fit statistics for the three largest samples examined in the present work. The study only explored rotations of the first two dimensions: the first was rotated to be parallel to Newtonian correctness (similar to ours), while the second was related to incorrect understanding of Newton's 3rd Law in the two largest samples. 

A few studies have applied more sophisticated IRT models, albeit in only one dimension. A model called the 2PL-NRM was used by \citet{eaton:2019} to build a partial credit scoring model of the FCI and by \citet{smith2020} to characterize the distractors of the Force and Motion Conceptual Evaluation, another popular conceptual mechanics instrument. The 2PL-NRM is a type of nested logit model which combines the 2PL (to model question correctness) with the NRM (to model the choice between particular distractors for incorrect responses). Multidimensional extensions of such models may provide an interesting avenue for future work.

Network analytic techniques have also been used to understand the FCI and other conceptual instruments. \citet{brewe2016} pioneered network analysis studies for multiple choice responses and called the technique module analysis. These methods form a network where the responses to the instrument are nodes; for example, FCI response 1A would be a node. Edges in the network are created when one student selects two responses. For example, if a student selected 1A and 2B there would be an edge between nodes 1A and 2B. The edge is given a weight that is a measure of the association of the two responses. Early work used either the number of times the two responses were chosen together \citep{brewe2016} or the correlation between the responses \citep{wells2019, wells2020}. These studies could only examine the correct response network. \citet{yang2020} extended module analysis to include incorrect responses by changing the edge weight to the partial correlation between responses correcting for overall instrument score. Module analysis analyzes the resulting network to find clusters of nodes (called communities or modules) which are more related to each other than to nodes outside the cluster.

\citet{wheatley2022} applied module analysis to the sample examined in the current work and found that the modules identified were fairly consistent across institutions. The most common incorrect modules were FCI items: Newton's 3rd law misconceptions \{4A, 15C, 28D\}, motion implies active forces \{5D, 18D\}, motion implies active forces with a centrifugal force \{5E, 18E\}, circular impetus \{6A, 7A\}, largest force determines motion \{17A, 25D\}, last force to act determines motion \{21B, 23C\}, and impetus dissipation \{23D, 24C\}. These correspond well to our sparse vectors \vecname{V}, \vecname{IV}, \vecname{XVII}, \vecname{II}, \vecname{IX}, and \vecname{XIII}, providing additional support that the analysis of this work is identifying stable patterns in student responses.

\subsection{Differences between Misconceptions: Pre- vs. Post-instruction}

As explained earlier, Table~\ref{tab:all-misconceptions-table} includes the median correlation coefficients obtained by comparing fits on numerous sub-samples of our data to each of our robust sparse vectors. This serves as a measure of consistency, with high values indicating that the robust vector was well supported across different sub-samples of students. Separate values are computed using pre-test data only, post-test data only, and data from both pre- and post-tests combined, providing some evidence for the universality (or lack thereof) of our misconceptions before and after instruction. (Note however that these values do \emph{not} indicate the prevalence of the misconceptions---only how well they serve to explain student response behavior in our sample. The prevalence before and after instruction will be discussed in section~\ref{sec:student-class-prevalence}.)

The foremost observation is that our discovered misconceptions or misunderstandings generally appear strongly in both pre- and post-instruction data, testifying to their universality.  Considering the 22 robust distractor vectors included in the aforementioned table, only four out of the 44 pre- and post- median bootstrap correlations fell below 0.85. The first three of these appear to be more strongly supported by pre-instruction data than by post-instruction data: \vecname{VIII}, \vecname{XXVII}, and \vecname{X}. The fourth, \vecname{XXIII}, appears to be far better defined in the post-instruction data compared to the pre-instruction data; one possible explanation is that formal instruction primes students to distinguish between centrifugal and centripetal forces, whereas this distinction may not be as obviously or relevant to them prior to instruction.

\section{Misconception Prevalence in Students and Classes}
\label{sec:student-class-prevalence}

Having found a robust set of misconceptions and misunderstandings\footnote{In the interest of readability, we will shorten ``misconceptions and misunderstandings'' to simply ``misconceptions'' for the remainder of this section, even if this isn't quite an appropriate description for all dimensions}, we now seek ways to quantify the amount of each misconception held by each student or class. This offers two routes for significantly improving pedagogy.  Firstly, it enables concept inventories such as the FCI to serve as multidimensional formative assessments, identifying where remedial instruction might be best directed at the classroom level or pointing individual students to instructional resources most likely to help them. Secondly, by comparing the prevalence of misconceptions before and after instruction, it allows us to assess the effectiveness of new instructional procedures developed to address them.

In the following subsections we first describe our approach for determining the amount of each misconception that a particular student has from their raw responses on the FCI.  We then apply this method to quantify the pre- and post-instruction prevalence of each misconception as a function of raw pre-test score (a proxy for students' initial state of knowledge), averaged over all educational institutions in our sample. These results allow us to investigate the effectiveness of instruction for both high- and low-performing students. We observe large differences across misconceptions, both in their overall pre-instruction prevalence as well as in their reduction after instruction.  The qualitatively-different behavior that we find again demonstrates the uniqueness of the various misconceptions.  

\subsection{Quantifying the amount of each misconception}

We begin by addressing an obvious question: why not simply use the latent traits $\theta_{sd}$ obtained while fitting the MNCM to give a \textit{misconception score} (m-score) for each student on each misconception? Fixing all item parameters $a_{icd}$ and $b_{ic}$ of the model equal to those of the robust misconception vectors found in the analysis already presented, it would be relatively straightforward to compute $\theta_{sd}$ estimates for additional students from their responses to the FCI\footnote{For example, by using the variational Bayesian method described in our earlier paper, but with the item parameters fixed instead of being treated as outputs of the method.}. However, abilities computed in this way have several undesirable attributes---most notably, that abilities in the sparse dimensions are strongly influenced by the selection of distractors in the non-sparse dimension, and hence subject to change by student responses not directly evidencing the particular misconception under study.

Therefore, we have decided to use a simple method that bases the m-score for a particular misconception dimension on only those distractors that the associated misconception loads heavily on. This simple method involves summing the loadings (modified slightly using a deadband of $\pm0.2$ to reduce small loadings on irrelevant distractors; see Eq.~\eqref{eq:deadband}) for all selected distractors and then dividing this by the sum of the most negative loadings (also after deadband) in each question. It is essentially the weighted fraction of distractors selected by a student which are relevant to a given misconception, with weights derived from the MNCM loadings determined earlier. These scores are easy to interpret: a student who selects only correct answers receives a zero (i.e., no misconception) and one who selects the most-heavily-weighted distractors for that misconception is normalized to one (i.e., maximum possible amount of misconception). This is close to the ``misconception scores'' used by \citet{wells2019} to describe a nearly-identical quantity, one that uses equal weight for all distractors (determined using the network analysis method).

The principal advantage of this method is that the misconception scores are determined solely by students' selection of distractors pertinent to each particular misconception (more or less so depending on their relative loadings). In contrast, MNCM trait estimates for one dimension are dependent on a student's responses to \emph{all} questions, including those which do not load on a given dimension, especially because they must be orthogonal to the general (non-sparse) dimension (although not to each other). This means that even if a student selects \textit{exactly the same distractors} relevant to a particular misconception before and after instruction, they would generally \emph{not} receive the same pre- and post-test trait estimate for that dimension as we think they should.

Our procedure for determining a given student's m-score for a given misconception may be summarized as follows:
\begin{enumerate}
    \item\label{itm:distractorweights} Generate distractor weights by applying a deadband of $\pm0.2$ to the corresponding MNCM loadings.
    \item\label{itm:scalingfactor} Generate a scaling factor by summing the most negative distractor weight in each question, across all questions.
    \item Sum all weights corresponding the student's selected distractors (either before or after instruction), then divide by the scaling factor to yield the student's misconception score.
\end{enumerate}
This procedure is repeated to yield pre- and post-instruction m-scores for all students in our dataset across all identified misconceptions, and may be easily applied to new student responses to yield misconception scores for future testees.

\subsection{Misconceptions scores of student groups}
\label{sec:mscoresofgroups}

While the m-scores computed above may be a helpful diagnostic tool for individual students, the mean m-scores for groups of students are more helpful for observing or quantifying trends. Such means serve as a measure of the overall prevalence of a misconception in a given group and could be used to examine how this differs across educational institutions or student demographics. In the present work we limit our exploration to (1) pre- and post-instruction misconception prevalence and (2) instructional effectiveness, both as functions of raw pre-test score.

One challenge in grouping students by pre-test score is that relatively small sample sizes can result if the score range is divided evenly (i.e., with a separate group for each numerical score); for example, only 24 students in our dataset achieved a raw pre-test score of 30. Small sample sizes increase the uncertainty of our mean m-scores, which in turn greatly increases the noise in relative measures of instructional effectiveness such as normalized gains. To address this, we bin some pre-test scores together in our analysis. These bins are calculated in two steps. First, we compute raw pre-test score quantiles for 30 equally-sized student groups, dropping any duplicate scores\footnote{The number of initial bins in this step is subjective, but we have found 30 to yield a suitable balance between noise (which favors larger bins) and pre-test score resolution (which favors smaller bins).}. These score quantiles are then used as cutoffs to group the students, with each score bin including the lower edge and excluding the upper edge. The resulting groups range in size from 456 to 1659 students.

For each group, we compute the mean pre- and post-test m-scores as well as their standard errors. We also compute our preferred measure of instructional effectiveness: the \textit{normalized gain} popularized by \citet{hake1998}. It represents the reduction of what students don't know (typically measured by number of incorrect responses, but here applied to the amount of a given misconception) as a fraction of what was initially unknown prior to instruction. Values of 1.0 correspond to the complete elimination of a misconception, whereas a value of zero indicates no reduction in the overall prevalence of a misconception. Normalized gains based on m-scores are computed as follows (recalling that m-scores are already a measure of what is \emph{not} known):
\begin{equation}
    \left< g\right> \overset{\text{def}}{=} \frac{\left<m_\text{pre}\right> - \left< m_\text{post}\right>}{\left<m_\text{pre}\right>}
     = 1 - \frac{\left< m_\text{post}\right>}{\left<m_\text{pre}\right>}.
\end{equation}
Asymptotic standard errors for $\left<g\right>$ may be derived using standard statistical approaches \cite[e.g.][]{lee2006} and are given by
\begin{equation}
    SE_{\left<g\right>} \approx \frac{1}{\sqrt{n}} \left| \frac{\left<m_\text{post}\right>}{\left<m_\text{pre}\right>} \right| \sqrt{
        \frac{\mathrm{Var}[m_\text{pre}]}{\left<m_\text{pre}\right>^2}
        + \frac{\mathrm{Var}[m_\text{post}]}{\left<m_\text{post}\right>^2} 
        - 2 \frac
            {\mathrm{Cov}[m_\text{pre}, m_\text{post}]}
            {\left<m_\text{pre}\right> \left<m_\text{post}\right>}
    },
\end{equation}
where $n$ is the sample size of the group.\footnote{Note that the error has a covariance term to account for the observed correlation between pre- and post-test results for the students.  In extremis, if the post-test misconception score of every student were exactly half of their pre-test misconception score, then \(\left< g\right> = 0.5\) and \(SE_{\left<g\right>} = 0\).}

For each misconception, the m-score is proportional to the loadings on various distractors and the number of wrong answers selected.  To give a zero-knowledge assessment of these effects we have calculated a ``random guessing'' baseline m-score in each score bin, for each misconception, and for both pre- and post-test results. We do so by considering a scenario in which each student selects among the four available distractors in each question with equal probability---that is, in which the probability of selecting the correct answer is $\text{RawScore}/30$ and that of selecting each distractor is $(1-\text{RawScore}/30)/4$. This gives a baseline estimate of whether that particular misconception is frequent or rare.

Misconception scores (pre- and post-) are illustrated in Figures~\ref{fig:mscores-ancient}, \ref{fig:mscores-medieval}, and \ref{fig:mscores-post-newtonian-novel}, corresponding to predominantly ancient, medieval, and post-Newtonian or novel misconceptions, respectively. Each binned metric (mean pre- and post-instruction m-scores and their normalized gain) is illustrated with a single point on the graphs; the area of this point is proportional to the number of students in the corresponding raw-score bin, and its abscissa is simply the mean raw pre-test score of the binned students. Asymptotic standard errors are shown with vertical bars.

For perspective, the misconception scores and normalized gains for our ``random guessing'' baseline are overlaid on the plots using dashed lines (with colors matching those of the corresponding metrics). Note: there is a large amount of evidence that students in our sample answer the FCI with conviction, even if at the same rate one would expect from random guessing; these random guessing curves should be viewed only as a basis for judging the prevalence of a particular misconception.

\newcommand{\mscorecommoncaptiontext}[1]{\textbf{Binned m-scores and pre-post gains for predominantly #1 misconceptions}, shown as functions of pre-test raw score. Dot areas are proportional to the number of students in each bin and error bars show linearized standard errors. Dashed lines represent ``random guessing'' baselines (see Section~\ref{sec:mscoresofgroups}).}

\newcommand{\mscoreshortcommoncaptiontext}[1]{Binned m-scores and pre-post gains for predominantly #1 misconceptions}

\begin{figure*}
    \centering
    \includegraphics[width=0.5\textwidth]{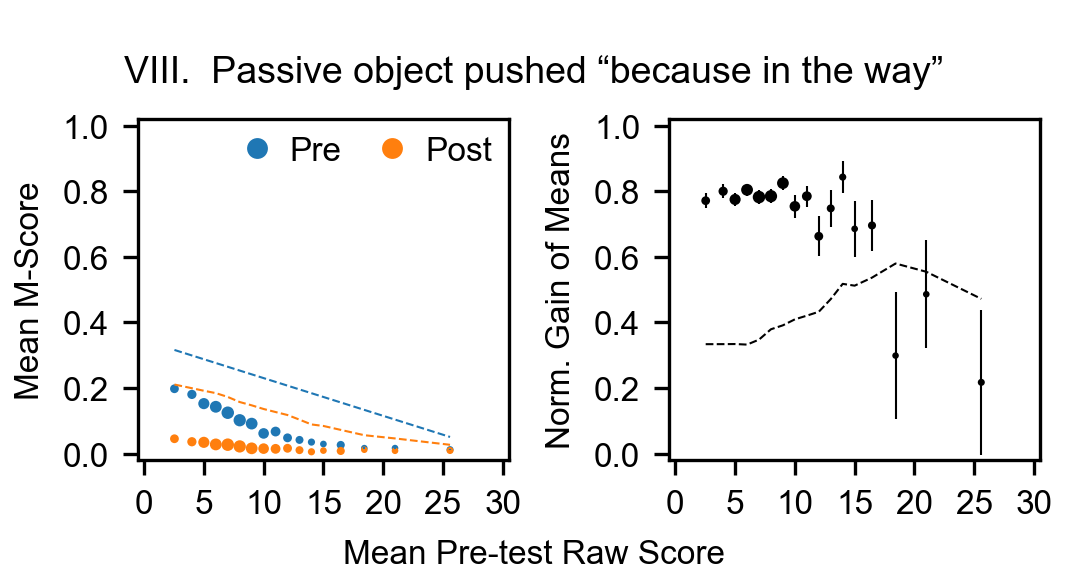}%
    \includegraphics[width=0.5\textwidth]{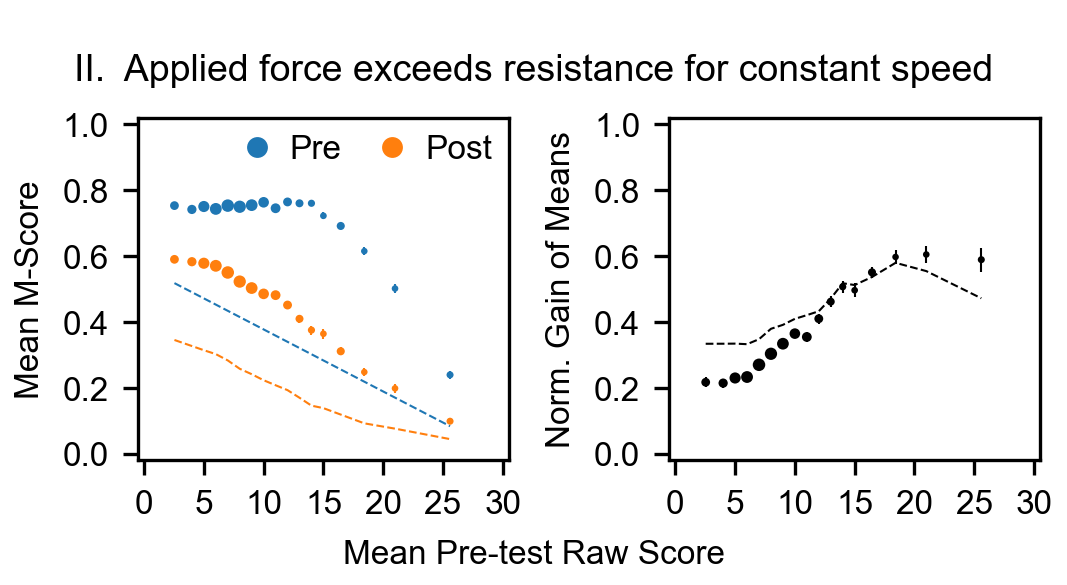}\\
    \includegraphics[width=0.5\textwidth]{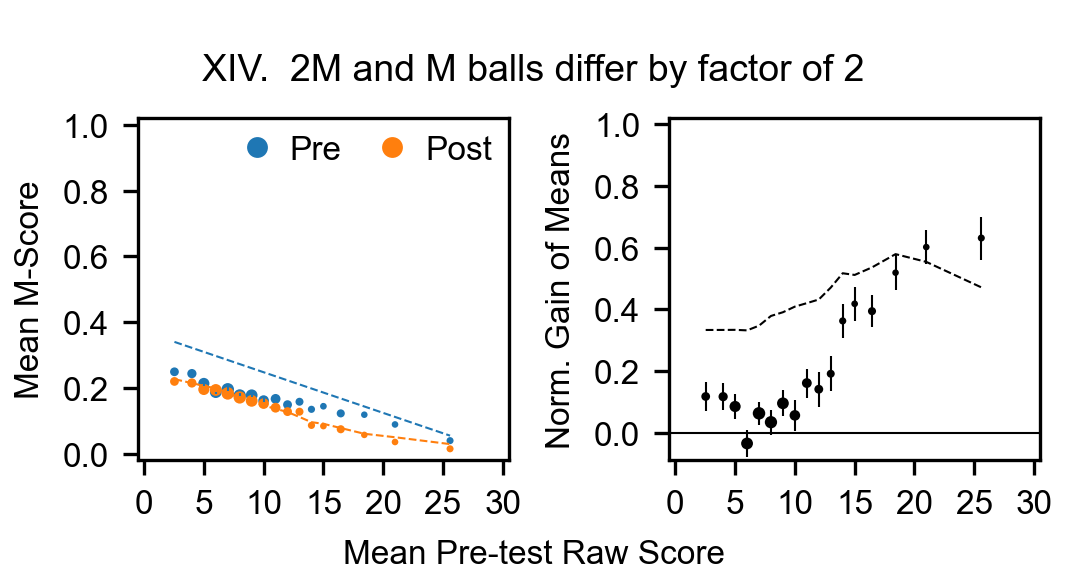}%
    \includegraphics[width=0.5\textwidth]{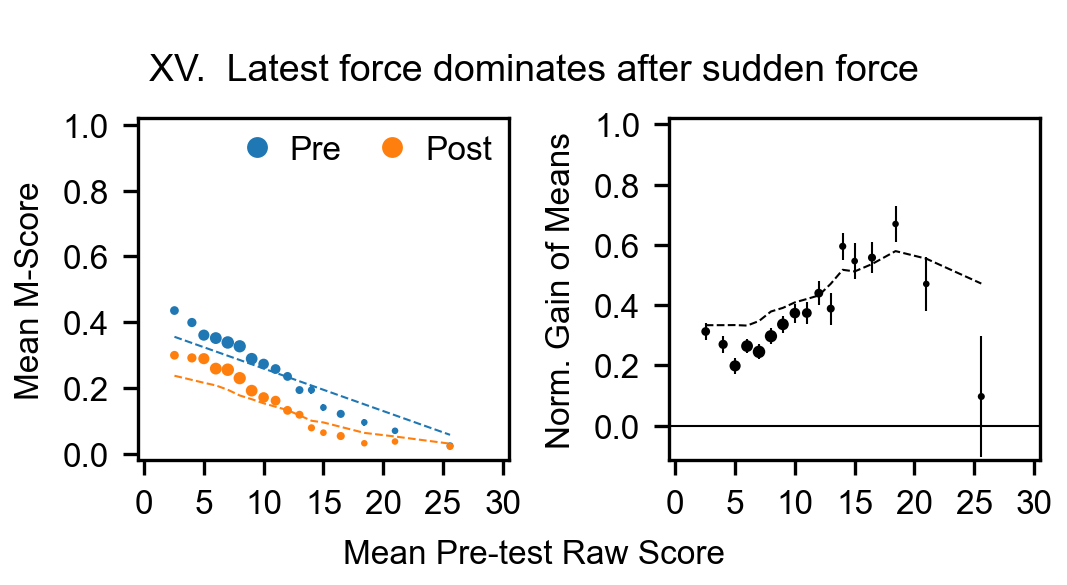}\\
    \includegraphics[width=0.5\textwidth]{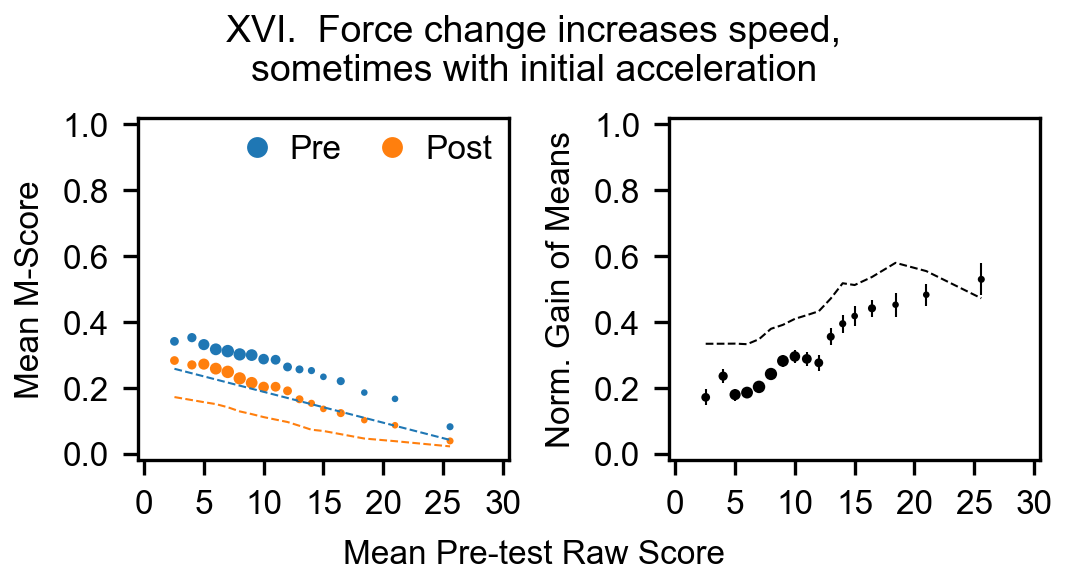}%
    \includegraphics[width=0.5\textwidth]{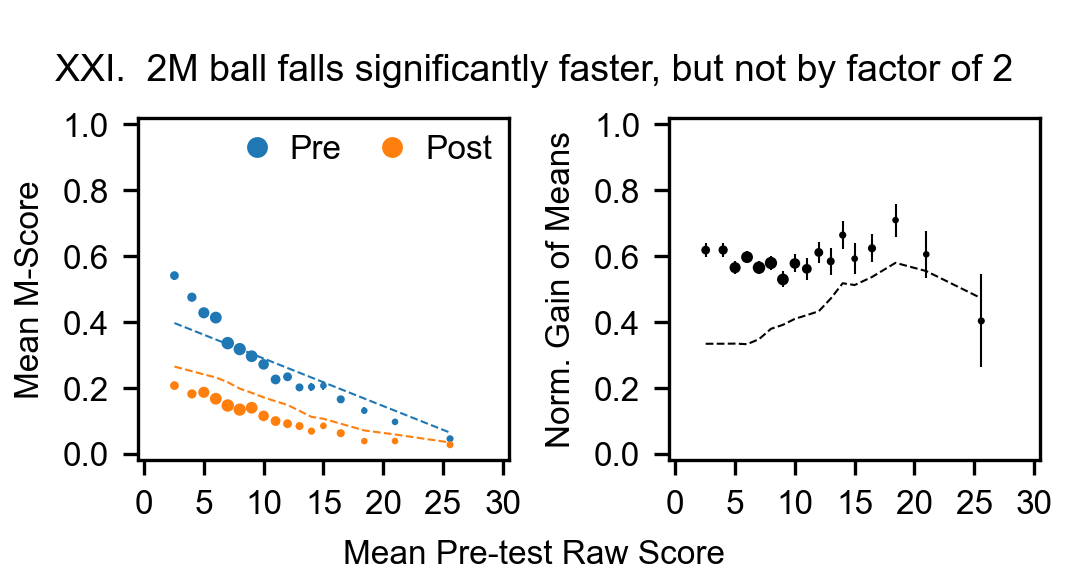}\\
    \includegraphics[width=0.5\textwidth]{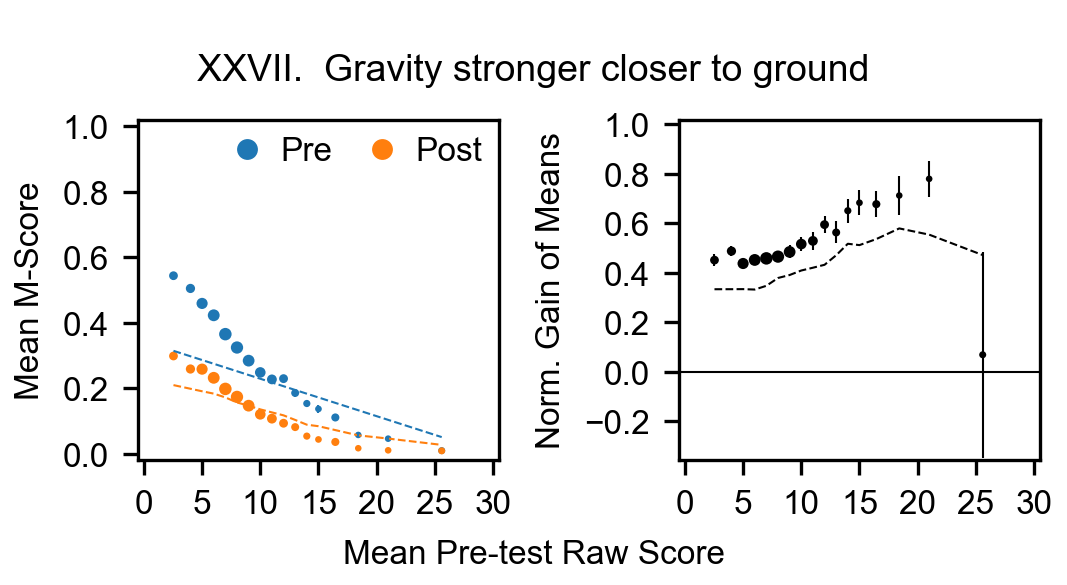}%
    \caption[\mscoreshortcommoncaptiontext{\emph{ancient}}]{\mscorecommoncaptiontext{\emph{ancient}}}
    \label{fig:mscores-ancient}
\end{figure*}

\begin{figure*}
    \centering
    \includegraphics[width=0.5\textwidth]{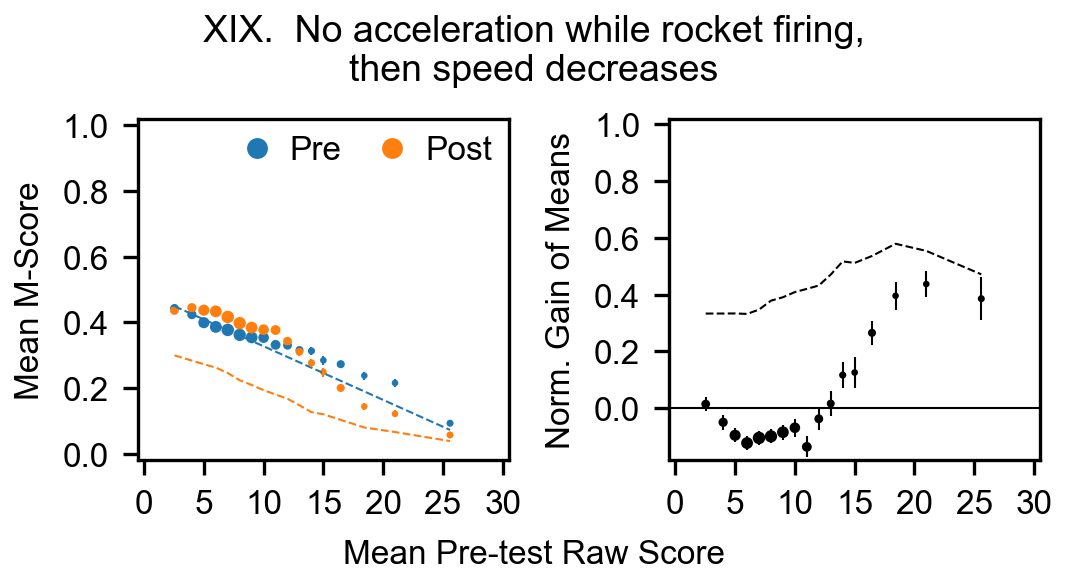}%
    \includegraphics[width=0.5\textwidth]{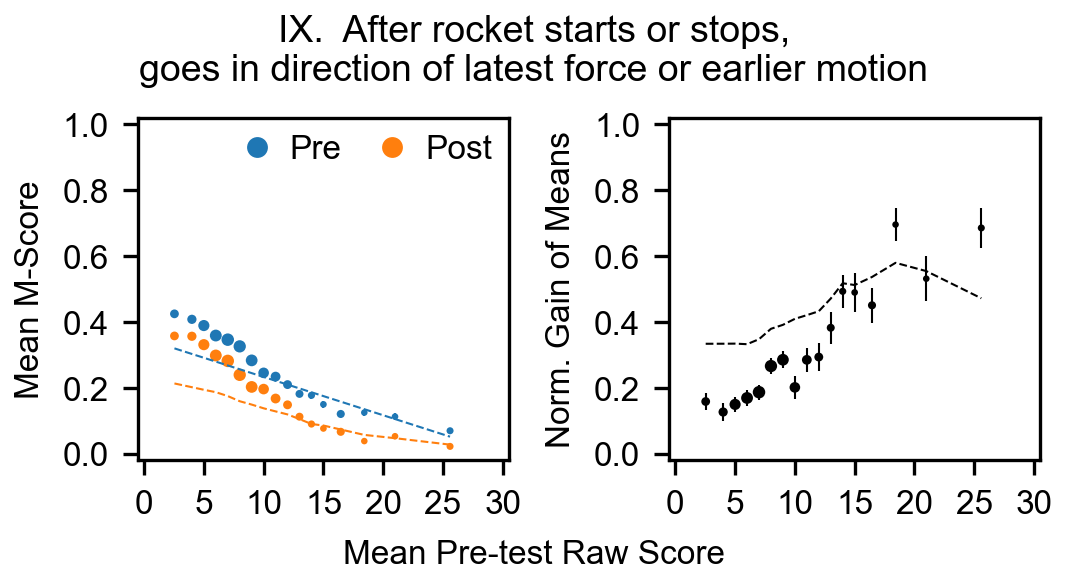}\\
    \includegraphics[width=0.5\textwidth]{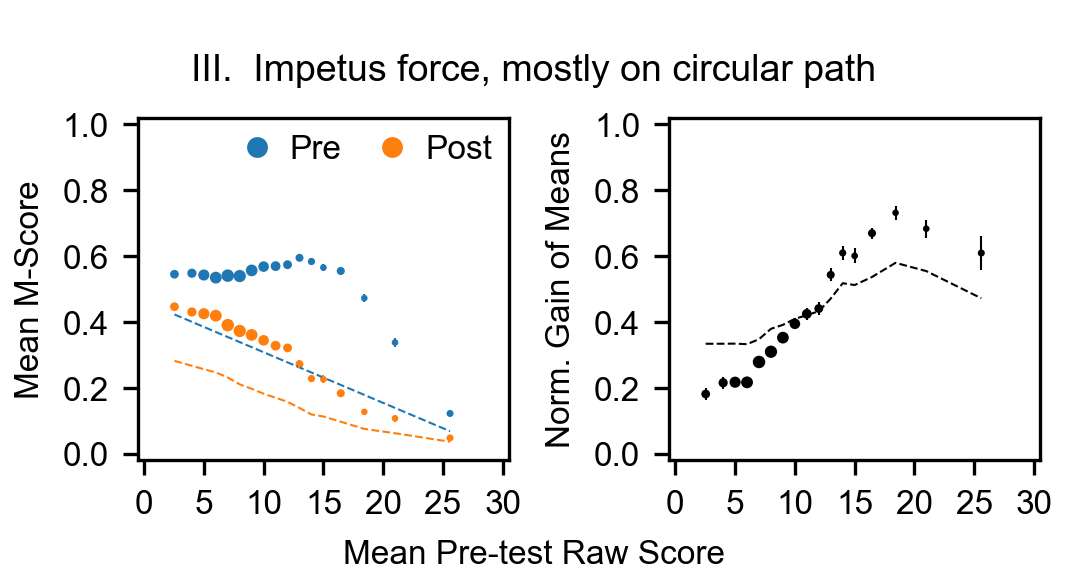}%
    \includegraphics[width=0.5\textwidth]{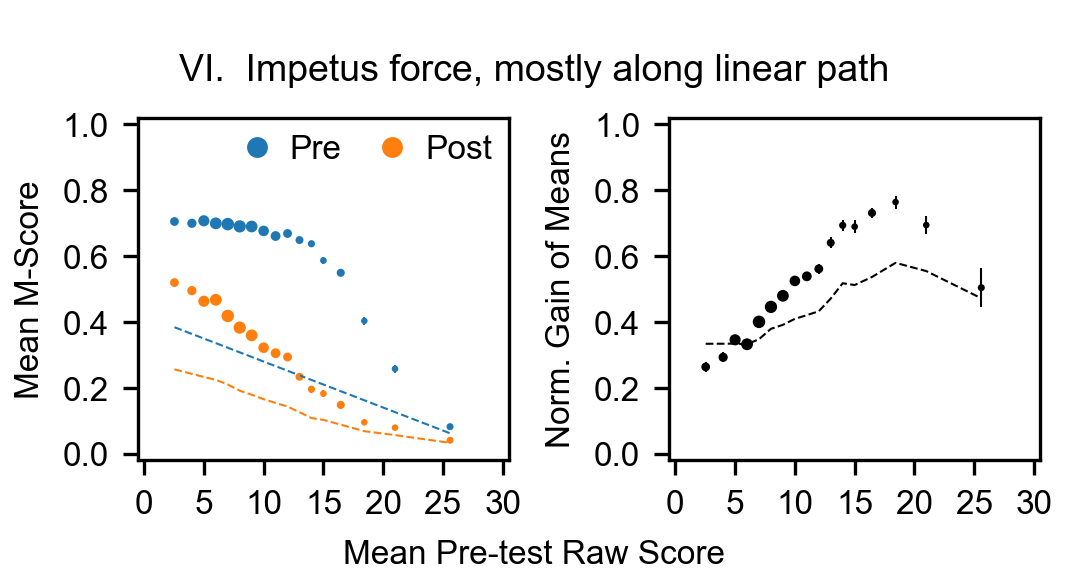}\\
    \includegraphics[width=0.5\textwidth]{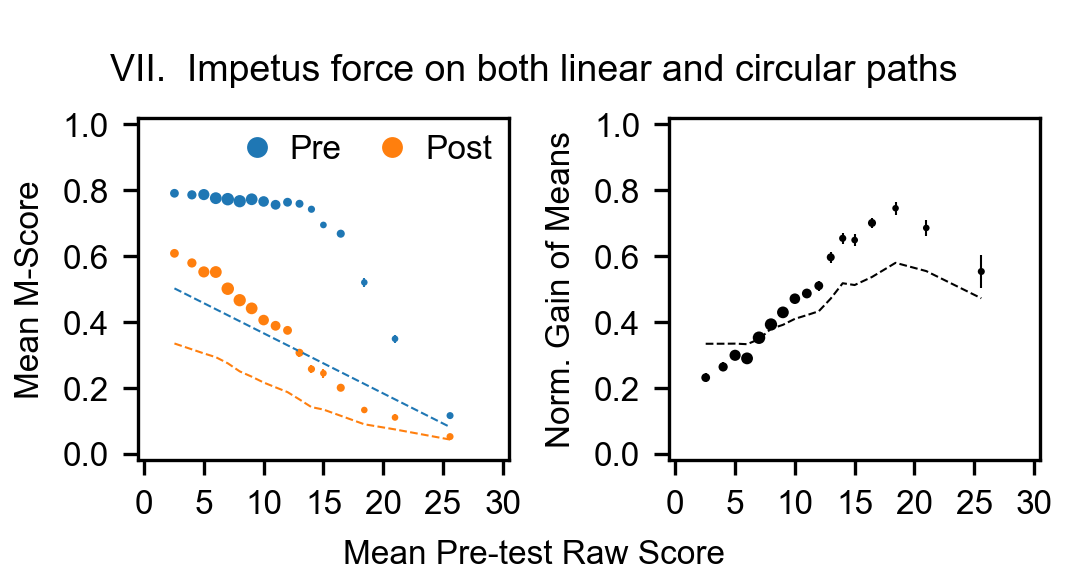}%
    \includegraphics[width=0.5\textwidth]{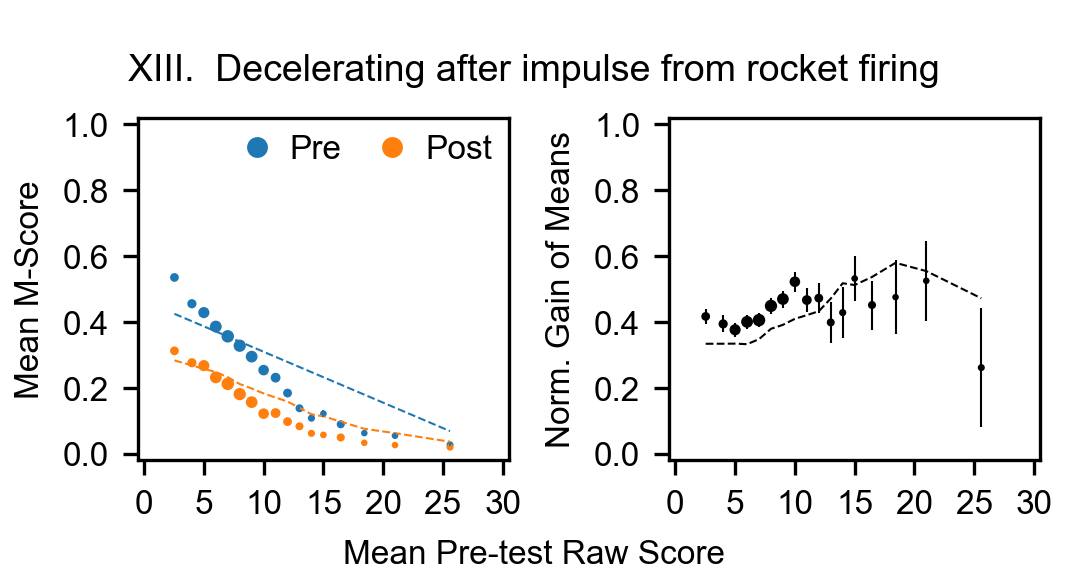}\\
    \includegraphics[width=0.5\textwidth]{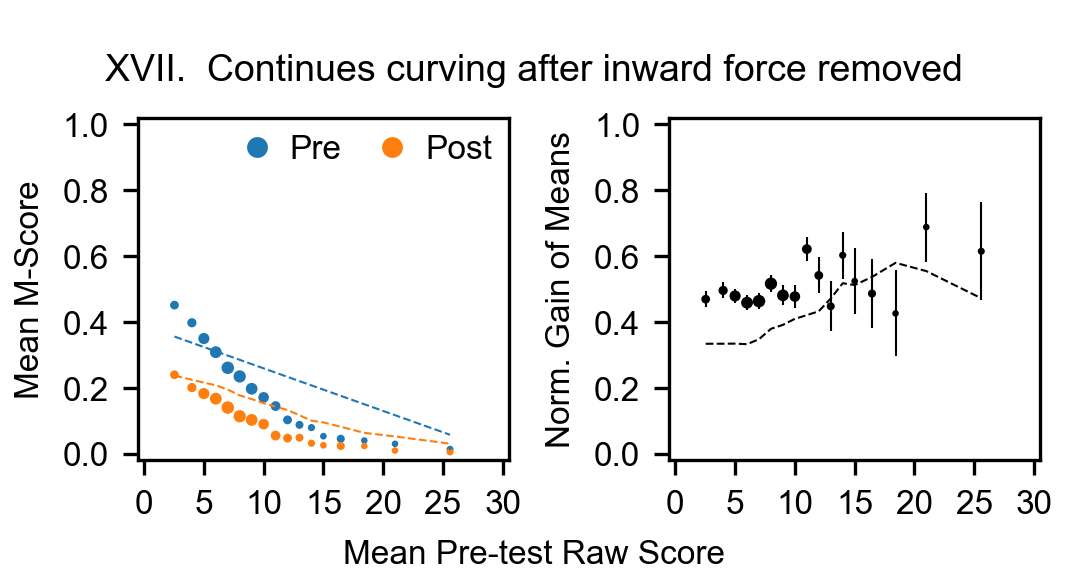}%
    \includegraphics[width=0.5\textwidth]{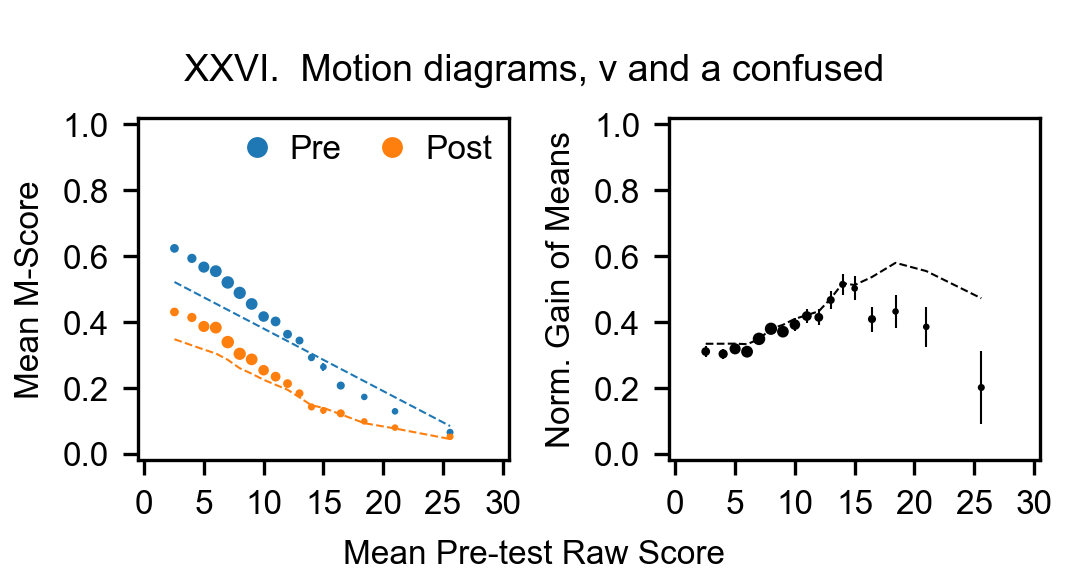}\\
    \caption[\mscoreshortcommoncaptiontext{\emph{medieval}}]{\mscorecommoncaptiontext{\emph{medieval}}}
    \label{fig:mscores-medieval}
\end{figure*}

\begin{figure*}
    \centering
    \includegraphics[width=0.5\textwidth]{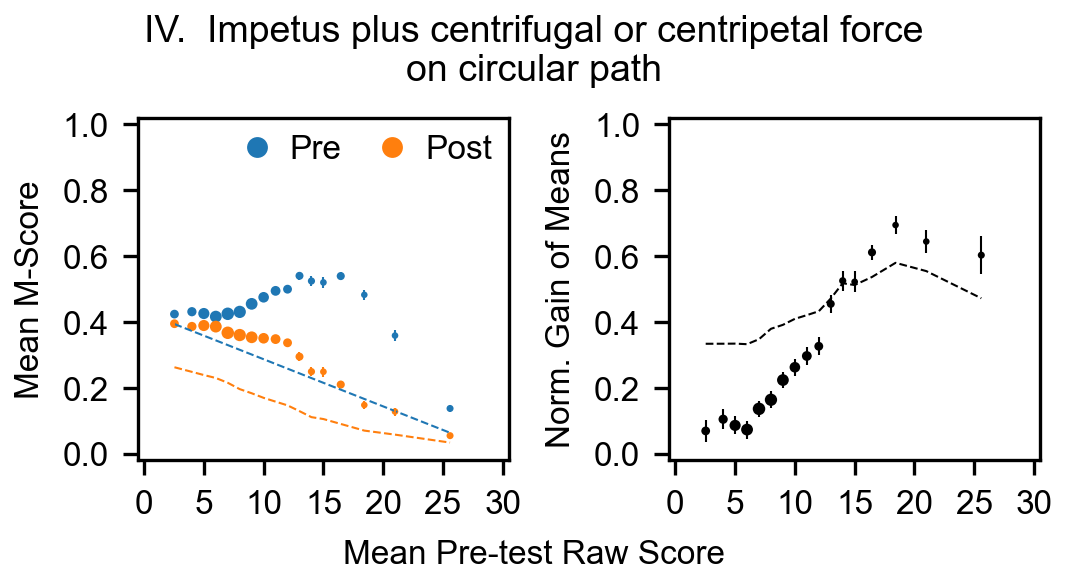}%
    \includegraphics[width=0.5\textwidth]{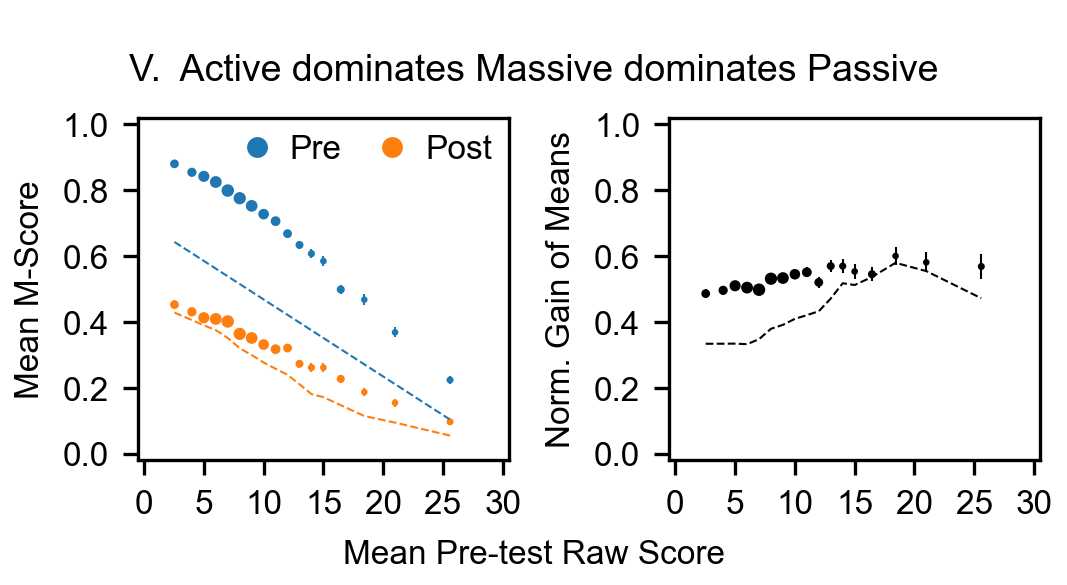}\\
    \includegraphics[width=0.5\textwidth]{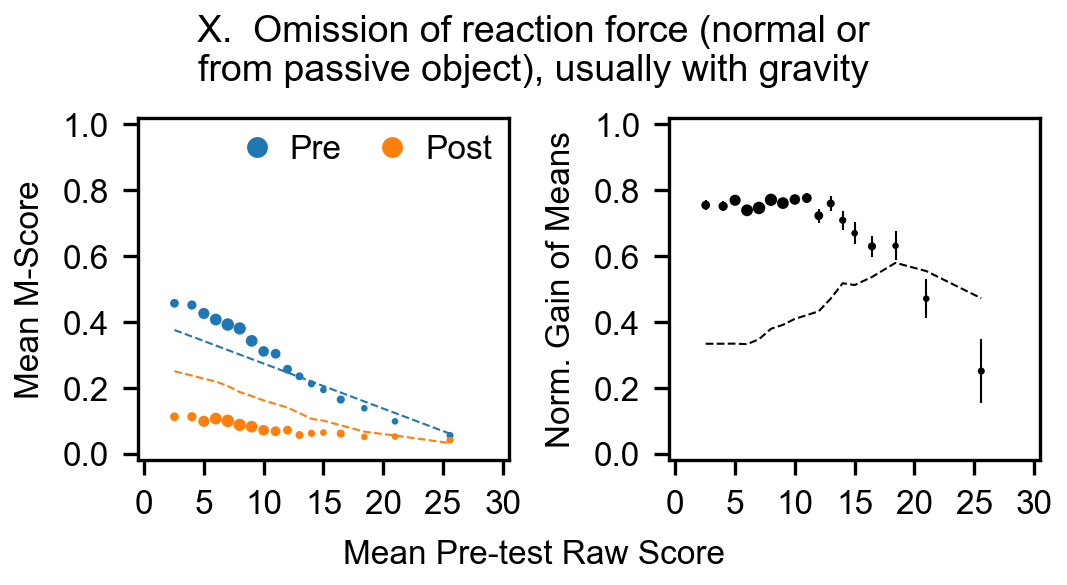}%
    \includegraphics[width=0.5\textwidth]{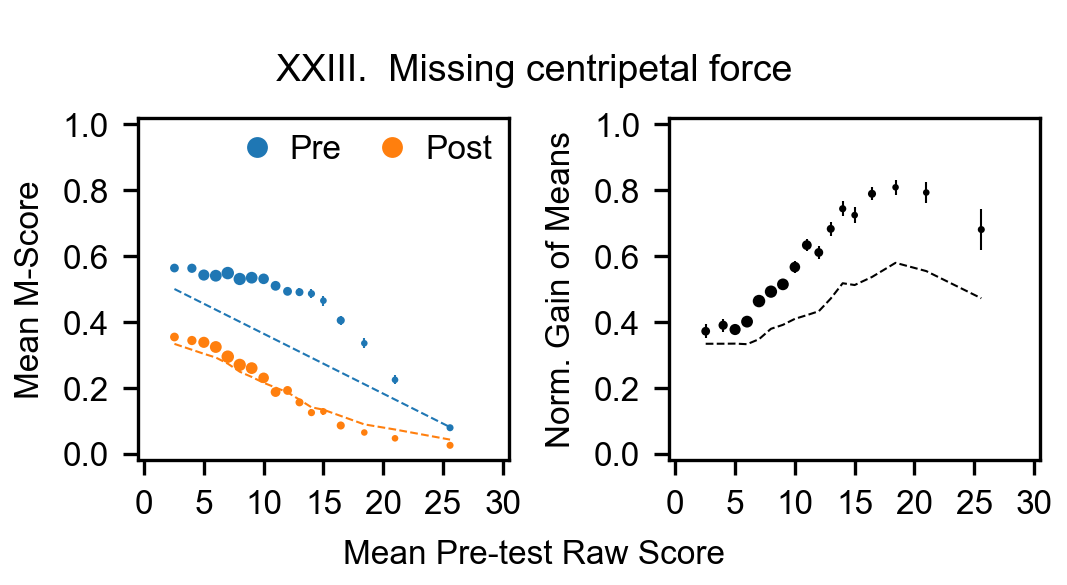}\\
    \includegraphics[width=0.5\textwidth]{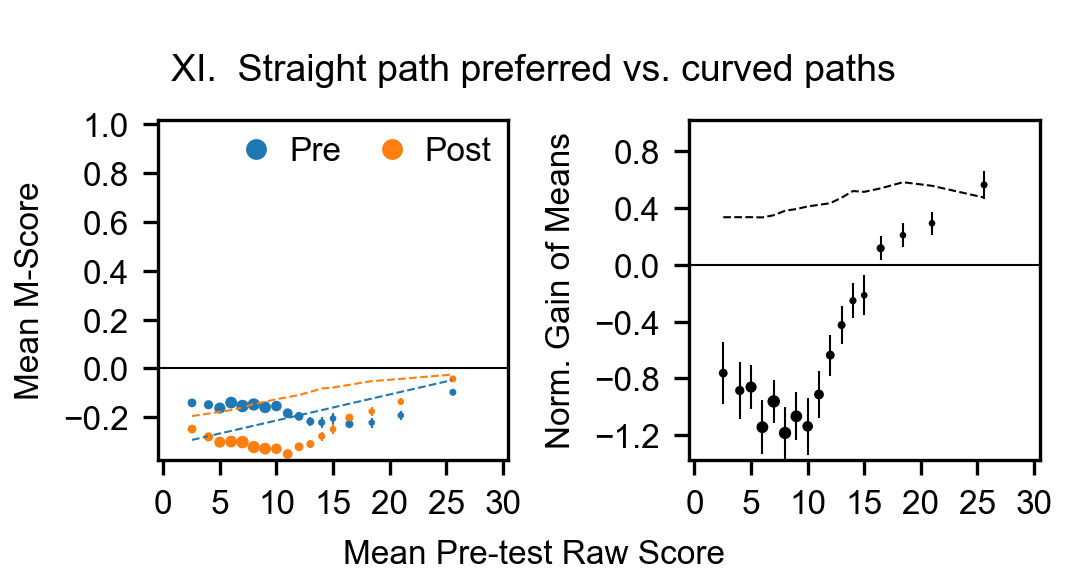}%
    \includegraphics[width=0.5\textwidth]{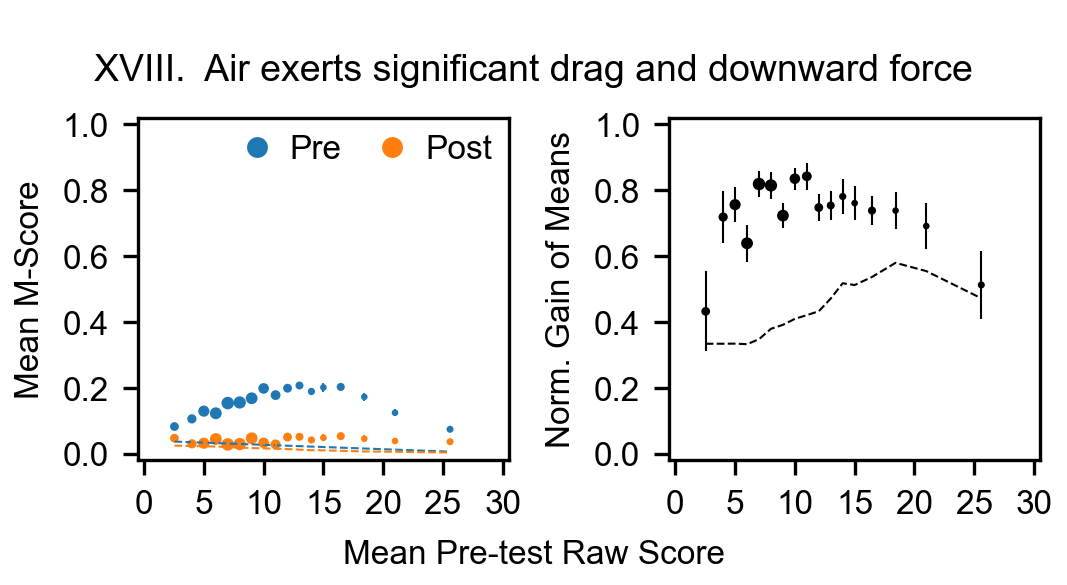}\\
    \includegraphics[width=0.5\textwidth]{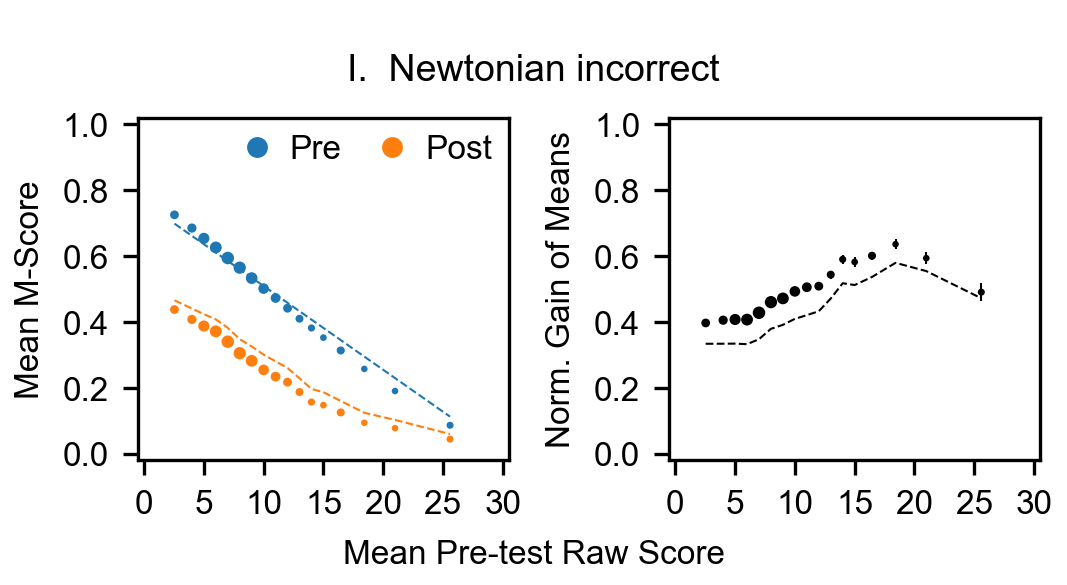}%
    \caption[\mscoreshortcommoncaptiontext{\emph{post-Newtonian} and \emph{novel}}]{\mscorecommoncaptiontext{\emph{post-Newtonian} and \emph{novel}}}
    \label{fig:mscores-post-newtonian-novel}
\end{figure*}

\subsection{Discussion: Misconception Scores and Normalized Gains}

The misconception scores are displayed in three Figures (\ref{fig:mscores-ancient}, \ref{fig:mscores-medieval}, and \ref{fig:mscores-post-newtonian-novel}) for easy tie-in with the above section.  Our discussion however is based on the characteristics of the analyzed curves and transcends the above categorization.

\subsubsection{Pre-instruction Misconception Scores}

The most notable class of the pre-instruction m-scores is \textit{strong}; these typically start above 0.5 at low raw score and are fairly flat (or even rising) up to shoulder around 13 (i.e. significantly above the average), they then decay fairly linearly, often aiming for scores in the range of 20--25 (rather than top score of 30).  Given that the median and mean of the raw pre-test scores are only 9.0 and 9.8 respectively, misconceptions in this class are highly prevalent. This class includes \vecname{II}, all four misconceptions involving impetus, and \vecname{XXIII}.

Most of the remaining pre-instruction m-scores are classified as \textit{declining}, and up to the median raw score they often decline along a tangent that would bring half of them to zero in the raw score range 15-25 (whereas only 1/5 would hit zero beyond 30).  Those with zeros below 20 are \vecnameonly{VIII} (intercept of $\sim$15), \vecnameonly{XXI} ($\sim$17), \vecnameonly{XXVII} ($\sim$15), \vecnameonly{XIII} ($\sim$17), and \vecnameonly{XVII} ($\sim$14). All of these will be referred to as \textit{Naive Misconceptions} later in this section.

\subsubsection{Normalized Gain}

We feel that normalized gain---the \textit{fraction} of what is not known at the start of a course which is learned during the course---is the best single metric of the effectiveness of instruction.  It is often relatively constant across classes of different ability and is known to depend on the nature of instruction \citep{hake1998}.  It is certainly preferable to the amount learned, which generally declines with increasing pre-test score simply because high-scoring students on the pre-test are limited in how much they can learn. 

Except for those misconceptions labeled ``NM'' (for naive misconception) in the \textit{Type} column of Table~\ref{tab:all-misconceptions-table}, the average normalized gain curve starts at a low value of 0.17 and increases to a peak at an average of 0.70, typically at a raw pre-test score of around 20. For more than half of the vectors we found, the normalized gain is at least 0.3 higher at its peak than it is for the lowest ability students (many of these will be classified as \textit{Standard Misconceptions} later). The foremost implication of this is that \textit{current instruction seriously falls to remedy the misconceptions that are not naive misconceptions for students at or below average} although it is effective for well-above-average students. This finding refines the accepted wisdom \citep{Docktor2010c,etkina2005} that misconceptions are highly resistant to standard instruction, by restricting its validity to average and below-average students. 

Turning now to the eight naive misconceptions listed in Table~\ref{tab:all-misconceptions-table}, we note that they are quite dissimilar to the rest of the misconceptions.  Firstly the average normalized gain starts from 0.52 at zero raw score and is nearly flat to 0.51 at a raw pretest score of 15 (well past the average score).  As noted above, the pretest scores for the naive misconceptions tend to decline rapidly with raw score, often initially sloping towards an intercept in the range of 15 to 20.  Above this raw score intercept very few students have non-zero m-score on the pretest and even fewer on the post-text, so that the normalized gain becomes very noisy, reflecting the small size of the sample in that raw score range.

In summary, naive misconceptions are prevalent mainly for average and below-average students and are largely absent among the top students. They are well-remediated by current instruction as evidenced by normalized gains greater than 0.50---even for these low-scoring students (which is in contrast with standard misconceptions).  Inspection of those mentioned above shows that many involve obviously naive views that pertain over a limited range of situations.

\section{Summary and Future} 

The work above makes contributions to both the understanding of misconceptions at a cognitive level and the application of psychometric techniques to studying and evaluating student misconceptions and misunderstandings. We will summarize and discuss these in turn, and then present several steps we hope to take in the near future to continue our work.

\subsection{Cognitive Context} 

This study represents a major advance in the misconceptions model of cognitive processes.  Firstly, this work provides a set of 22 robust misconceptions and misunderstandings determined by a uniform procedure, in contrast to most previous research which studied fewer misconceptions or just one or two in more detail.  Importantly, this work enables measurement of the misconception score on each of these misconceptions for any class (or even a single student, though likely with a great deal of uncertainty). 

We confirm the major features that characterize misconceptions in the literature: our standard misconception dimensions occur across questions with different contexts, involve cognitively similar (but incorrect) reasoning, and they are not well remediated for most students. We show, however, that they are in fact well remediated (normalized gain over 0.5) for those students who are well above average on pretest score.  

More importantly, we show that there are \textit{naive misconceptions} that are well remediated for all students---even those below average in ability.  As the name suggests they tend to be rather narrow mental constructs.  Indeed, several are so narrow that they share elements with ``knowledge in pieces'' and ``resources'' points of view.

While our statistical methods work consistently and we have demonstrated quite robust misconceptions, these methods can only find cognitive states that are built into the distractors on the specific instrument whose results we analyze.  The FCI was carefully designed based on student interviews, prior research, and even experiments on students and is likely quite complete.  We regard it as important to extend our methods to the Force and Motion Concept Inventory which has a more flexible distractor structure with up to nine distractors per question.  The key issue is how similar the misconceptions found by our methods would be to those found here for the FCI and how much the specific misconceptions are dependent on the particular instrument.

With regard to other cognitive explanations for incorrect responses on multiple choice instruments, it is clear that some of our misconceptions involve ontological confusions.  This is clear for \vecname{XXVI}, but a confusion that force is equivalent to motion may lie at the root of some selections of impetus.  The knowledge-in-pieces viewpoint may apply to some of the narrower naive misconceptions like \vecname{XXI}.

\subsection{Psychometrics}

In contrast to most previous studies on the FCI, this work explored a larger and more representative sample of student responses. Using the MNCM allowed us to leverage the considerable information present in students' selection of specific distractors, with our hierarchical Bayesian approach providing the necessary safeguards to prevent overfitting given the large number of free parameters. As a result, we were able to extract a much higher number of dimensions than previously possible, in part because our method could more fully explore our relatively large data set with a total of 34,000 student administrations.

We also took great care in our approach to factor rotation in order to find dimensions that were both robust and interpretable. Following current recommendations in the literature, we explored not only multiple rotation methods but also multiple local minima found by several of these methods. Our evaluation of the various methods and solutions leveraged several quantitative metrics (including consistency across disjoint datasets, simple measures of simplicity, and bootstrap evaluations) as well as our subjective judgment.

Finally, the distractor loadings for each of our interpretable dimensions may provide diagnostic value for both research and classroom use. We hope to incorporate our results into an easy-to-use web-based tool which instructors can use to quantify the prevalence of these misconceptions in their own classes.

\subsection{Next Steps}

Looking to the future, a compelling first task is to examine the correlations between various misconception scores to identify higher-level structures---such as groupings of related misconceptions or hierarchical relationships---in a data-driven manner. Doing so might reinforce some of our existing classification schemas (e.g., historical groupings or naive-vs-standard) or provide new lenses through which to view the identified misconceptions.

In our view, the misconceptions relating to circular motion and impetus deserve particular attention in future work. The correlation analyses described above may prove helpful for examining these in more detail, though these seemed less stable than many other misconceptions even with our very large dataset. We therefore recommend that designers of new concept tests covering force and motion write questions specifically designed to separate the circular motion and impetus misunderstandings, and that interviews with students focus on better understanding their thought processes when answering such questions.  

The same may be said for the \vecname{V} misconception: as currently posed, the questions on the FCI make it difficult to isolate student beliefs about active vs.\ passive agents, heavy vs.\ light agents, and impulsive vs.\ continuous forces as several of these contrasts often exist as once. Future tests might address this by designing questions which probe understanding of Newton's 3rd Law in only one of these contexts at a time as well as in their various combinations.

Finally, we have only scratched the surface of what can be explored with such methods. Applying these to other research-developed instruments would likely prove informative, since misconceptions obviously exist in many other domains besides introductory mechanics. Even without considering other instruments, many opportunities remain for drawing further insight from our FCI dataset, or others that are accompanied by more detailed information about the nature of the instruction or the demographics of the students. Our current dataset already enables us to explore differences in misconception prevalence and instructional effectiveness across different schools. Obtaining additional demographic information would allow us to draw further insights that may improve instruction for students with differing backgrounds. As is often true in research, significant advances give rise to more new questions than they answer.

\section*{Acknowledgments}

We are grateful for support from MIT and the Department of Physics at MIT as well as from the Center for Ultracold Atoms at MIT and Harvard. We acknowledge help in getting data from H. Dedic, M. Dugdale, A. Fuchs, N. Lasry, and S. Osborne Popp, and we thank E. Christman for multiple helpful discussions. This work made use of computational resources provided by subMIT at MIT Physics.

\printbibliography[title={References}]

\clearpage
\appendix

\section*{Supplemental Materials: Robust Distractor Vectors and Accompanying Information}

\graphicspath{{figures/best_vectors}}

\raggedbottom

The following pages contain the set of 27 robust distractor vectors identified using our methods. Each vector is labeled with its roman numeral designation and descriptive name, followed by a visual grid of MNCM slope coefficients with shading highlighting large negative (red) and positive (blue) values. Correct answers---which have slopes fixed to zero per our model identification criteria---are indicated by the symbol `\O{}'.

For each robust distractor vector, two tables with additional information are also included below the raw coefficients. The first table contains several summary statistics for our bootstrap correlations using pre-instruction, post-instruction, and combined datasets. The second contains brief textual descriptions of each distractor in the vector having a rounded slope of magnitude 0.5 or greater, as well as of their associated questions. (The very dense \textit{Newtonian Incorrect} vector is an exception: only the most heavily-loaded distractors are included.)

\clearpage\noindent\begin{minipage}{\textwidth}
\subsection*{I.\hspace{0.5em}Newtonian incorrect}

\vspace{0.5em}
\begin{center}
\includegraphics{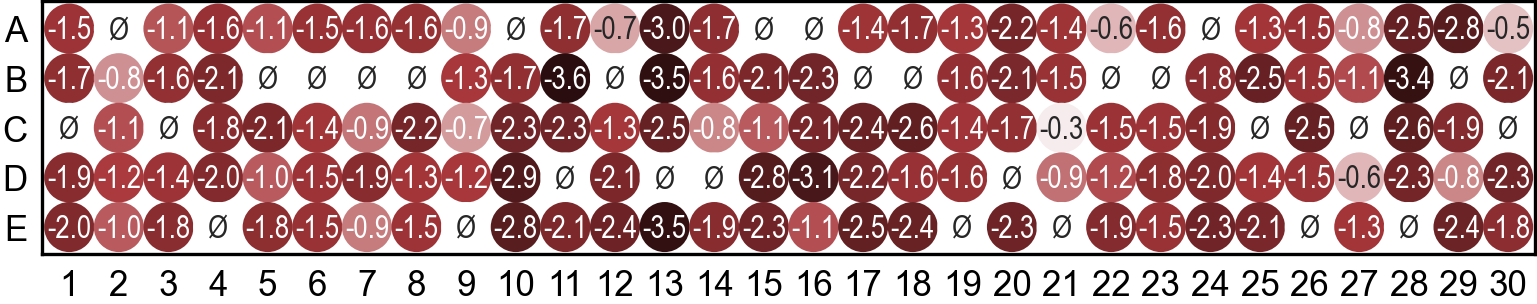}
\end{center}

\begin{center}
\footnotesize\footnotesize\begin{tblr}{
    width = 369pt,
    colspec = {@{}X[l]S[table-format=1.2]S[table-format=1.2]S[table-format=1.2]@{}},
    rows = {rowsep=1pt},
    row{1} = {guard, abovesep+=2pt},
    row{2} = {abovesep+=2pt},
    row{Z} = {belowsep+=1pt},
    hspan = minimal,
}
\toprule
Bootstrap correlation statistic & Pre data & Post data & All data \\
\midrule
Median & 0.99 & 0.99 & 1.00 \\
5th percentile & 0.99 & 0.99 & 1.00 \\
Fraction $\geq$ 0.85 & 1.00 & 1.00 & 1.00 \\
\bottomrule
\end{tblr}
\end{center}\vspace{-0.5em}

\begin{center}
\footnotesize\footnotesize\begin{tblr}{
    width = \textwidth,
    colspec = {@{}lX[l]lllS[table-format=2.1]S[table-format=-1.1]@{}},
    rows = {rowsep=1pt},
    row{1} = {guard, abovesep+=2pt},
    row{2} = {abovesep+=2pt},
    row{Z} = {belowsep+=1pt},
    hspan = minimal,
}
\toprule
\SetCell[c=2]{l} Item/distractor description & & Code & NC-Ex & IRC\textsubscript{pre} & $f$ (\%) & $\widehat{a}$ \\
\midrule
\SetCell[c=4]{l,font=\itshape} After puck on ice is kicked, what forces act \\
11B & downward force of gravity and force in direction of motion & I1, Ob & 1b,5Sa & DW & 22.8 & -3.6 \\
\hborder{abovespace+=3pt, belowspace+=1pt}\hline[dotted]
\SetCell[c=4]{l,font=\itshape} Ball is thrown up, what forces act afterwards \\
13B & decreasing upward force up to top, then increasing gravity down  & I3, G4, G5 & * & DW & 21.7 & -3.5 \\
13E & no force, ball falls back due to natural tendency to rest on surface of earth & G2 & --- & W & 0.7 & -3.5 \\
13A & downward gravity and steadily decreasing upward force & I3 & 2a,2b & INT & 8.2 & -3.0 \\
\hborder{abovespace+=3pt, belowspace+=1pt}\hline[dotted]
\SetCell[c=4]{l,font=\itshape} Larger Boy kicks knee of smaller boy; both on office chairs \\
28B & larger active student exerts force, passive student exerts no force & AF1 & 3b & DW & 9.4 & -3.4 \\
28C & smaller passive student exerts larger force  & --- & 3b & INT & 6.6 & -2.6 \\
\hborder{abovespace+=3pt, belowspace+=1pt}\hline[dotted]
\SetCell[c=4]{l,font=\itshape} Car now pushes truck at constant speed \\
16D & car pushes truck, truck pushed because in way (engine not running so it can't push back on car) & AF1 & 3b & W & 4.5 & -3.1 \\
\hborder{abovespace+=3pt, belowspace+=1pt}\hline[dotted]
\SetCell[c=4]{l,font=\itshape} Along path you picked in 8, what is speed of puck after y-kick? \\
10D & puck speed increases for while after kick, then decreases & I4 & 1b & DW & 18.2 & -2.9 \\
10E & puck speed constant for while, then decreases & --- & 1b & INT & 10.2 & -2.8 \\
\hborder{abovespace+=3pt, belowspace+=1pt}\hline[dotted]
\SetCell[c=4]{l,font=\itshape} Car accelerates while pushing a truck; its force \\
15D & car pushes truck, truck pushed because in way (engine not running so it can't push back on car) & AF1 & 3b & W & 4.6 & -2.8 \\
\hborder{abovespace+=3pt, belowspace+=1pt}\hline[dotted]
\SetCell[c=4]{l,font=\itshape} What forces act on an empty office chair \\
29A & gravity exerts downward force on chair  & Ob & 5SA & DW & 17.3 & -2.8 \\
\hborder{abovespace+=3pt, belowspace+=1pt}\hline[dotted]
\SetCell[c=4]{l,font=\itshape} Forces on a boy swinging on a rope \\
18C & downward gravity, and force in direction of motion & I5f & 2b,2c & DW & 14.7 & -2.6 \\
\bottomrule
\end{tblr}
\end{center}

\end{minipage}

\clearpage\noindent\begin{minipage}{\textwidth}
\subsection*{II.\hspace{0.5em}Applied force exceeds resistance for constant speed}

\vspace{0.5em}
\begin{center}
\includegraphics{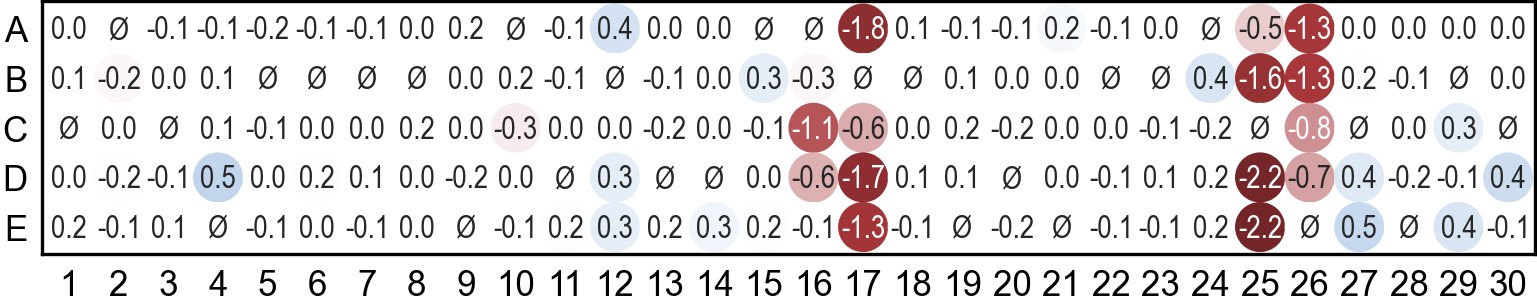}
\end{center}

\begin{center}
\footnotesize\footnotesize\begin{tblr}{
    width = 369pt,
    colspec = {@{}X[l]S[table-format=1.2]S[table-format=1.2]S[table-format=1.2]@{}},
    rows = {rowsep=1pt},
    row{1} = {guard, abovesep+=2pt},
    row{2} = {abovesep+=2pt},
    row{Z} = {belowsep+=1pt},
    hspan = minimal,
}
\toprule
Bootstrap correlation statistic & Pre data & Post data & All data \\
\midrule
Median & 0.95 & 0.99 & 1.00 \\
5th percentile & 0.92 & 0.98 & 0.99 \\
Fraction $\geq$ 0.85 & 1.00 & 1.00 & 1.00 \\
\bottomrule
\end{tblr}
\end{center}\vspace{-0.5em}

\begin{center}
\footnotesize\footnotesize\begin{tblr}{
    width = \textwidth,
    colspec = {@{}lX[l]lllS[table-format=2.1]S[table-format=-1.1]@{}},
    rows = {rowsep=1pt},
    row{1} = {guard, abovesep+=2pt},
    row{2} = {abovesep+=2pt},
    row{Z} = {belowsep+=1pt},
    hspan = minimal,
}
\toprule
\SetCell[c=2]{l} Item/distractor description & & Code & NC-Ex & IRC\textsubscript{pre} & $f$ (\%) & $\widehat{a}$ \\
\midrule
\SetCell[c=4]{l,font=\itshape} Box has v0 due to const force, which \\
25E & is greater than either weight or resistance & CI1 & 1d & INT & 14.9 & -2.2 \\
25D & is greater than resistance & R2 & 1d & INT & 39.6 & -2.2 \\
25B & is greater than weight of box & R2 & 5Sc, 1d & DW & 13.6 & -1.6 \\
25A & has same magnitude as weight of box & R2 & 1d, 5Sc & FLAT & 4.3 & -0.5 \\
\hborder{abovespace+=3pt, belowspace+=1pt}\hline[dotted]
\SetCell[c=4]{l,font=\itshape} Elevator lifted up at constant v by a cable such that \\
17A & upward force of cable exceeds downward force of gravity & CI1 & 1c,1d & INT & 51.7 & -1.8 \\
17D & upward force of cable exceeds gravity and downward force of air & CI1, G1 & 1c,1d & INT & 11.6 & -1.7 \\
17E & no force; elevator goes up because cable gets shorter & AF1 & 1d & DW & 8.8 & -1.3 \\
17C & upward force of cable smaller than force of gravity & --- & 1c,1d & W & 2.4 & -0.6 \\
\hborder{abovespace+=3pt, belowspace+=1pt}\hline[dotted]
\SetCell[c=4]{l,font=\itshape} Woman doubles force that moved box @ v0 \\
26A & speed doubles when force doubled & AF4 & 2b & INT  & 35.8 & -1.3 \\
26B & higher constant speed but not twice as fast & R2, R3 & 2b & INT  & 30.6 & -1.3 \\
26C & initially at higher constant speed, then accelerates faster & I4 & 2b & DW & 4.3 & -0.8 \\
26D & speed increases up to constant v & AF6 & 2b & FLAT & 13.8 & -0.7 \\
\hborder{abovespace+=3pt, belowspace+=1pt}\hline[dotted]
\SetCell[c=4]{l,font=\itshape} Car now pushes truck at constant speed \\
16C & car pushes harder on truck than truck pushes on car & AR2 & 3b & DW & 25.7 & -1.1 \\
16D & car pushes truck, truck pushed because in way (engine not running so it can't push back on car) & AF1 & 3b & W & 4.5 & -0.6 \\
\hborder{abovespace+=3pt, belowspace+=1pt}\hline[dotted]
\SetCell[c=4]{l,font=\itshape} Large truck collides head-on with small compact car \\
4D & truck exerts force on car, but car exerts no force on truck & AR1 & --- & --- & 1.3 & +0.5 \\
\hborder{abovespace+=3pt, belowspace+=1pt}\hline[dotted]
\SetCell[c=4]{l,font=\itshape} Box has v0 due to const force, what happens if force stops? \\
27E & increases speed for while, then slows to stop & I4 & 2b & FLAT & 0.8 & +0.5 \\
\bottomrule
\end{tblr}
\end{center}

\end{minipage}

\clearpage\noindent\begin{minipage}{\textwidth}
\subsection*{III.\hspace{0.5em}Impetus force, mostly on circular path}

\vspace{0.5em}
\begin{center}
\includegraphics{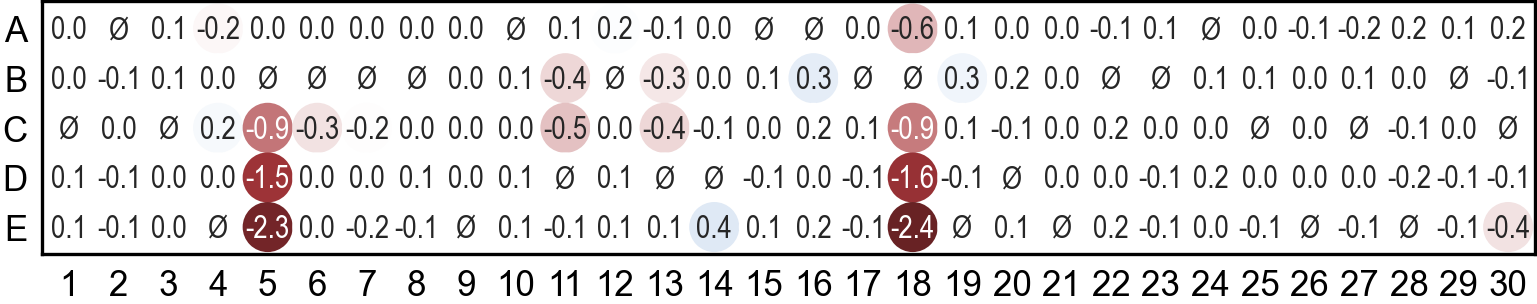}
\end{center}

\begin{center}
\footnotesize\footnotesize\begin{tblr}{
    width = 369pt,
    colspec = {@{}X[l]S[table-format=1.2]S[table-format=1.2]S[table-format=1.2]@{}},
    rows = {rowsep=1pt},
    row{1} = {guard, abovesep+=2pt},
    row{2} = {abovesep+=2pt},
    row{Z} = {belowsep+=1pt},
    hspan = minimal,
}
\toprule
Bootstrap correlation statistic & Pre data & Post data & All data \\
\midrule
Median & 0.91 & 0.99 & 0.99 \\
5th percentile & 0.85 & 0.96 & 0.98 \\
Fraction $\geq$ 0.85 & 0.94 & 1.00 & 1.00 \\
\bottomrule
\end{tblr}
\end{center}\vspace{-0.5em}

\begin{center}
\footnotesize\footnotesize\begin{tblr}{
    width = \textwidth,
    colspec = {@{}lX[l]lllS[table-format=2.1]S[table-format=-1.1]@{}},
    rows = {rowsep=1pt},
    row{1} = {guard, abovesep+=2pt},
    row{2} = {abovesep+=2pt},
    row{Z} = {belowsep+=1pt},
    hspan = minimal,
}
\toprule
\SetCell[c=2]{l} Item/distractor description & & Code & NC-Ex & IRC\textsubscript{pre} & $f$ (\%) & $\widehat{a}$ \\
\midrule
\SetCell[c=4]{l,font=\itshape} Forces on a boy swinging on a rope \\
18E & downward gravity, force in direction of motion, and outward force & CF, I5f & 2b,2c & INT & 24.4 & -2.4 \\
18D & downward gravity, force in direction of motion, and inward force from rope & I5f & 2b & INT & 29.2 & -1.6 \\
18C & downward gravity, and force in direction of motion & I5f & 2b,2c & DW & 14.7 & -0.9 \\
18A & downward force of gravity & AF!, Ob & 2c & FLAT & 3.2 & -0.6 \\
\hborder{abovespace+=3pt, belowspace+=1pt}\hline[dotted]
\SetCell[c=4]{l,font=\itshape} Forces on a ball travelling in a circular track \\
5E & gravity, force in direction of motion, and outward force  & AF2, CF & 2c,2b & INT & 17.5 & -2.3 \\
5D & gravity, force in direction of motion, and inward force  & AF2, I5f & 2b & INT & 20.8 & -1.5 \\
5C & gravity and force in direction of motion  & AF2, I5f & 2c,2b & DW & 29.8 & -0.9 \\
\hborder{abovespace+=3pt, belowspace+=1pt}\hline[dotted]
\SetCell[c=4]{l,font=\itshape} After puck on ice is kicked, what forces act \\
11C & downward gravity, upward by surface, force in direction of motion & I1 & 1b & INT  & 29.2 & -0.5 \\
\bottomrule
\end{tblr}
\end{center}

\end{minipage}

\clearpage\noindent\begin{minipage}{\textwidth}
\subsection*{IV.\hspace{0.5em}Impetus plus centrifugal or centripetal force on circular path}

\vspace{0.5em}
\begin{center}
\includegraphics{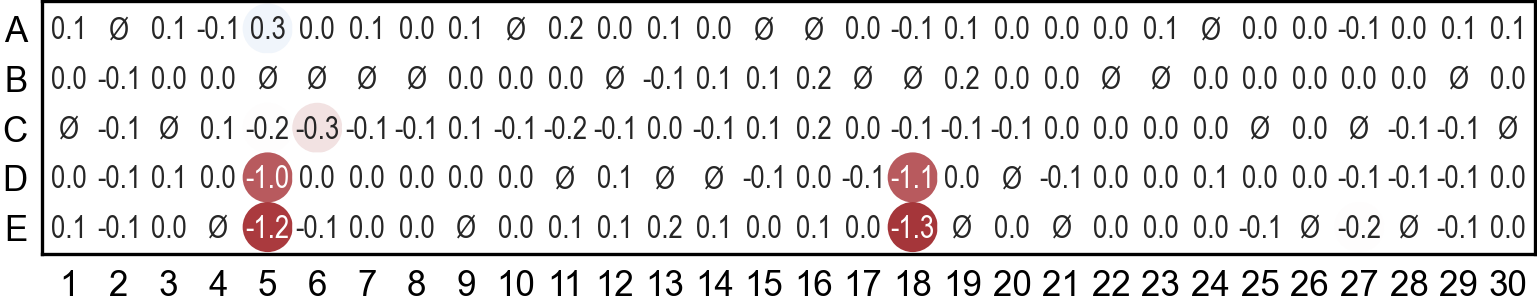}
\end{center}

\begin{center}
\footnotesize\footnotesize\begin{tblr}{
    width = 369pt,
    colspec = {@{}X[l]S[table-format=1.2]S[table-format=1.2]S[table-format=1.2]@{}},
    rows = {rowsep=1pt},
    row{1} = {guard, abovesep+=2pt},
    row{2} = {abovesep+=2pt},
    row{Z} = {belowsep+=1pt},
    hspan = minimal,
}
\toprule
Bootstrap correlation statistic & Pre data & Post data & All data \\
\midrule
Median & 0.88 & 0.96 & 0.99 \\
5th percentile & 0.77 & 0.93 & 0.98 \\
Fraction $\geq$ 0.85 & 0.68 & 1.00 & 1.00 \\
\bottomrule
\end{tblr}
\end{center}\vspace{-0.5em}

\begin{center}
\footnotesize\footnotesize\begin{tblr}{
    width = \textwidth,
    colspec = {@{}lX[l]lllS[table-format=2.1]S[table-format=-1.1]@{}},
    rows = {rowsep=1pt},
    row{1} = {guard, abovesep+=2pt},
    row{2} = {abovesep+=2pt},
    row{Z} = {belowsep+=1pt},
    hspan = minimal,
}
\toprule
\SetCell[c=2]{l} Item/distractor description & & Code & NC-Ex & IRC\textsubscript{pre} & $f$ (\%) & $\widehat{a}$ \\
\midrule
\SetCell[c=4]{l,font=\itshape} Forces on a boy swinging on a rope \\
18E & downward gravity, force in direction of motion, and outward force & CF, I5f & 2b,2c & INT & 24.4 & -1.3 \\
18D & downward gravity, force in direction of motion, and inward force from rope & I5f & 2b & INT & 29.2 & -1.1 \\
\hborder{abovespace+=3pt, belowspace+=1pt}\hline[dotted]
\SetCell[c=4]{l,font=\itshape} Forces on a ball travelling in a circular track \\
5E & gravity, force in direction of motion, and outward force  & AF2, CF & 2c,2b & INT & 17.5 & -1.2 \\
5D & gravity, force in direction of motion, and inward force  & AF2, I5f & 2b & INT & 20.8 & -1.0 \\
\bottomrule
\end{tblr}
\end{center}

\end{minipage}

\clearpage\noindent\begin{minipage}{\textwidth}
\subsection*{V.\hspace{0.5em}Active dominates Massive dominates Passive}

\vspace{0.5em}
\begin{center}
\includegraphics{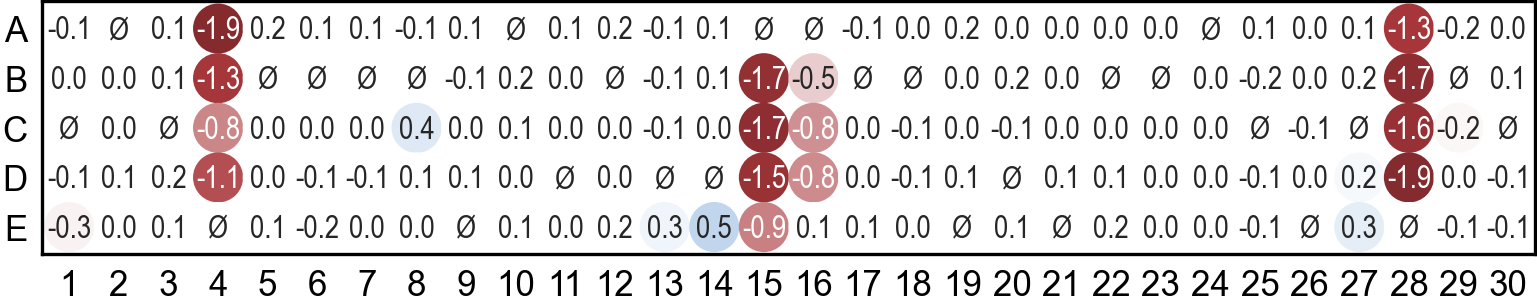}
\end{center}

\begin{center}
\footnotesize\begin{tblr}{
    width = 369pt,
    colspec = {@{}X[l]S[table-format=1.2]S[table-format=1.2]S[table-format=1.2]@{}},
    rows = {rowsep=1pt},
    row{1} = {guard, abovesep+=2pt},
    row{2} = {abovesep+=2pt},
    row{Z} = {belowsep+=1pt},
    hspan = minimal,
}
\toprule
Bootstrap correlation statistic & Pre data & Post data & All data \\
\midrule
Median & 0.97 & 0.99 & 1.00 \\
5th percentile & 0.91 & 0.98 & 0.99 \\
Fraction $\geq$ 0.85 & 1.00 & 1.00 & 1.00 \\
\bottomrule
\end{tblr}
\end{center}\vspace{-0.5em}

\begin{center}
\footnotesize\begin{tblr}{
    width = \textwidth,
    colspec = {@{}lX[l]lllS[table-format=2.1]S[table-format=-1.1]@{}},
    rows = {rowsep=1pt},
    row{1} = {guard, abovesep+=2pt},
    row{2} = {abovesep+=2pt},
    row{Z} = {belowsep+=1pt},
    hspan = minimal,
}
\toprule
\SetCell[c=2]{l} Item/distractor description & & Code & NC-Ex & IRC\textsubscript{pre} & $f$ (\%) & $\widehat{a}$ \\
\midrule
\SetCell[c=4]{l,font=\itshape} Larger Boy kicks knee of smaller boy; both on office chairs \\
28D & larger active student exerts greater force than passive student & AR1, AR2 & 3b & INT & 38.7 & -1.9 \\
28B & larger active student exerts force, passive student exerts no force & AF1 & 3b & DW & 9.4 & -1.7 \\
28C & smaller passive student exerts larger force  & --- & 3b & INT & 6.6 & -1.6 \\
28A & neither student exerts a force on the other & --- & 3b & W & 2.1 & -1.3 \\
\hborder{abovespace+=3pt, belowspace+=1pt}\hline[dotted]
\SetCell[c=4]{l,font=\itshape} Large truck collides head-on with small compact car \\
4A & truck exerts more force than car & AR1 & 3b & DW & 52.7 & -1.9 \\
4B & car exerts greater force on truck & AR1- & 3b & --- & 1.1 & -1.3 \\
4D & truck exerts force on car, but car exerts no force on truck & AR1 & --- & --- & 1.3 & -1.1 \\
4C & no force from either, car smashed because it gets in way & Ob & 3b & --- & 0.7 & -0.8 \\
\hborder{abovespace+=3pt, belowspace+=1pt}\hline[dotted]
\SetCell[c=4]{l,font=\itshape} Car accelerates while pushing a truck; its force \\
15C & accelerating car pushes harder on truck than truck pushes on car & AR2 & 3b & INT & 50.7 & -1.7 \\
15B & car pushes less hard than truck pushes back  & AR1 & 3b & DW & 7.9 & -1.7 \\
15D & car pushes truck, truck pushed because in way (engine not running so it can't push back on car) & AF1 & 3b & W & 4.6 & -1.5 \\
15E & No force from either, truck pushed because in way & Ob & --- & FLAT & 0.4 & -0.9 \\
\hborder{abovespace+=3pt, belowspace+=1pt}\hline[dotted]
\SetCell[c=4]{l,font=\itshape} Car now pushes truck at constant speed \\
16D & car pushes truck, truck pushed because in way (engine not running so it can't push back on car) & AF1 & 3b & W & 4.5 & -0.8 \\
16C & car pushes harder on truck than truck pushes on car & AR2 & 3b & DW & 25.7 & -0.8 \\
16B & car force is smaller than truck force & AR1 & 3b & W & 4.0 & -0.5 \\
\hborder{abovespace+=3pt, belowspace+=1pt}\hline[dotted]
\SetCell[c=4]{l,font=\itshape} What is path of heavy ball dropped from moving plane \\
14E & straight forwards, then sharp curve, then straignt down & I3, G5 & 0c & FLAT & 0.3 & +0.5 \\
\bottomrule
\end{tblr}
\end{center}

\end{minipage}

\clearpage\noindent\begin{minipage}{\textwidth}
\subsection*{VI.\hspace{0.5em}Impetus force, mostly along linear path}

\vspace{0.5em}
\begin{center}
\includegraphics{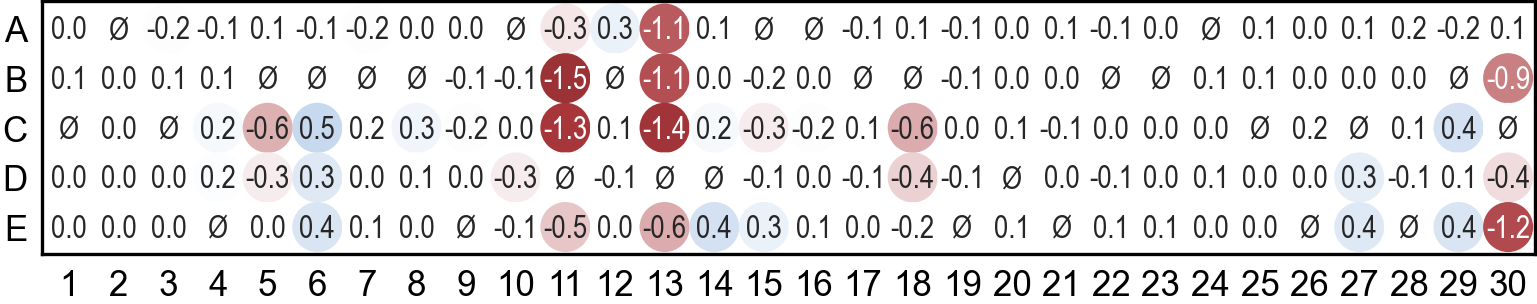}
\end{center}

\begin{center}
\footnotesize\begin{tblr}{
    width = 369pt,
    colspec = {@{}X[l]S[table-format=1.2]S[table-format=1.2]S[table-format=1.2]@{}},
    rows = {rowsep=1pt},
    row{1} = {guard, abovesep+=2pt},
    row{2} = {abovesep+=2pt},
    row{Z} = {belowsep+=1pt},
    hspan = minimal,
}
\toprule
Bootstrap correlation statistic & Pre data & Post data & All data \\
\midrule
Median & 0.92 & 0.95 & 0.99 \\
5th percentile & 0.87 & 0.93 & 0.97 \\
Fraction $\geq$ 0.85 & 0.99 & 1.00 & 1.00 \\
\bottomrule
\end{tblr}
\end{center}\vspace{-0.5em}

\begin{center}
\footnotesize\begin{tblr}{
    width = \textwidth,
    colspec = {@{}lX[l]lllS[table-format=2.1]S[table-format=-1.1]@{}},
    rows = {rowsep=1pt},
    row{1} = {guard, abovesep+=2pt},
    row{2} = {abovesep+=2pt},
    row{Z} = {belowsep+=1pt},
    hspan = minimal,
}
\toprule
\SetCell[c=2]{l} Item/distractor description & & Code & NC-Ex & IRC\textsubscript{pre} & $f$ (\%) & $\widehat{a}$ \\
\midrule
\SetCell[c=4]{l,font=\itshape} After puck on ice is kicked, what forces act \\
11B & downward force of gravity and force in direction of motion & I1, Ob & 1b,5Sa & DW & 22.8 & -1.5 \\
11C & downward gravity, upward by surface, force in direction of motion & I1 & 1b & INT  & 29.2 & -1.3 \\
11E & no forces on puck & G2 & 5Sa,1d & FLAT & 5.6 & -0.5 \\
\hborder{abovespace+=3pt, belowspace+=1pt}\hline[dotted]
\SetCell[c=4]{l,font=\itshape} Ball is thrown up, what forces act afterwards \\
13C &  $\sim$constant gravity and decreasing upward force on way up, then const grav down & I3 & 2a & INT & 35.4 & -1.4 \\
13B & decreasing upward force up to top, then increasing gravity down  & I3, G4, G5 & * & DW & 21.7 & -1.1 \\
13A & downward gravity and steadily decreasing upward force & I3 & 2a,2b & INT & 8.2 & -1.1 \\
13E & no force, ball falls back due to natural tendency to rest on surface of earth & G2 & --- & W & 0.7 & -0.6 \\
\hborder{abovespace+=3pt, belowspace+=1pt}\hline[dotted]
\SetCell[c=4]{l,font=\itshape} Tennis ball hit into a strong wind, what forces act \\
30E & force from gravity, force by hit, and force by air  & I1l & 2a & INT & 54.2 & -1.2 \\
30B & downward force of gravity and force by hit & I1l & 5FA, 2a & W & 8.1 & -0.9 \\
\hborder{abovespace+=3pt, belowspace+=1pt}\hline[dotted]
\SetCell[c=4]{l,font=\itshape} Forces on a boy swinging on a rope \\
18C & downward gravity, and force in direction of motion & I5f & 2b,2c & DW & 14.7 & -0.6 \\
\hborder{abovespace+=3pt, belowspace+=1pt}\hline[dotted]
\SetCell[c=4]{l,font=\itshape} Forces on a ball travelling in a circular track \\
5C & gravity and force in direction of motion  & AF2, I5f & 2c,2b & DW & 29.8 & -0.6 \\
\hborder{abovespace+=3pt, belowspace+=1pt}\hline[dotted]
\SetCell[c=4]{l,font=\itshape} Path of Ball after Exiting Curved Track \\
6C & curves outward somewhat & CF & 1b & W & 3.5 & +0.5 \\
\bottomrule
\end{tblr}
\end{center}

\end{minipage}

\clearpage\noindent\begin{minipage}{\textwidth}
\subsection*{VII.\hspace{0.5em}Impetus force on both linear and circular paths}

\vspace{0.5em}
\begin{center}
\includegraphics{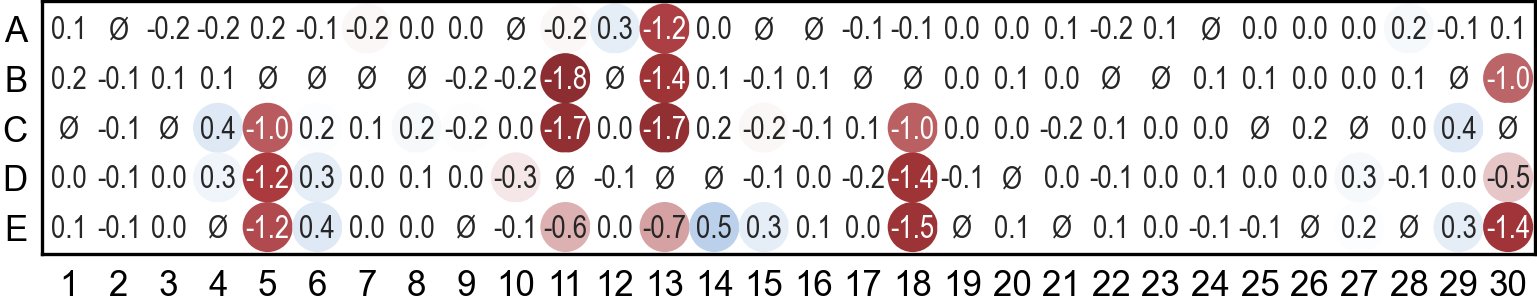}
\end{center}

\begin{center}
\footnotesize\begin{tblr}{
    width = 369pt,
    colspec = {@{}X[l]S[table-format=1.2]S[table-format=1.2]S[table-format=1.2]@{}},
    rows = {rowsep=1pt},
    row{1} = {guard, abovesep+=2pt},
    row{2} = {abovesep+=2pt},
    row{Z} = {belowsep+=1pt},
    hspan = minimal,
}
\toprule
Bootstrap correlation statistic & Pre data & Post data & All data \\
\midrule
Median & 0.98 & 0.98 & 0.99 \\
5th percentile & 0.97 & 0.97 & 0.99 \\
Fraction $\geq$ 0.85 & 1.00 & 1.00 & 1.00 \\
\bottomrule
\end{tblr}
\end{center}\vspace{-0.5em}

\begin{center}
\footnotesize\begin{tblr}{
    width = \textwidth,
    colspec = {@{}lX[l]lllS[table-format=2.1]S[table-format=-1.1]@{}},
    rows = {rowsep=1pt},
    row{1} = {guard, abovesep+=2pt},
    row{2} = {abovesep+=2pt},
    row{Z} = {belowsep+=1pt},
    hspan = minimal,
}
\toprule
\SetCell[c=2]{l} Item/distractor description & & Code & NC-Ex & IRC\textsubscript{pre} & $f$ (\%) & $\widehat{a}$ \\
\midrule
\SetCell[c=4]{l,font=\itshape} After puck on ice is kicked, what forces act \\
11B & downward force of gravity and force in direction of motion & I1, Ob & 1b,5Sa & DW & 22.8 & -1.8 \\
11C & downward gravity, upward by surface, force in direction of motion & I1 & 1b & INT  & 29.2 & -1.7 \\
11E & no forces on puck & G2 & 5Sa,1d & FLAT & 5.6 & -0.6 \\
\hborder{abovespace+=3pt, belowspace+=1pt}\hline[dotted]
\SetCell[c=4]{l,font=\itshape} Ball is thrown up, what forces act afterwards \\
13C &  $\sim$constant gravity and decreasing upward force on way up, then const grav down & I3 & 2a & INT & 35.4 & -1.7 \\
13B & decreasing upward force up to top, then increasing gravity down  & I3, G4, G5 & * & DW & 21.7 & -1.4 \\
13A & downward gravity and steadily decreasing upward force & I3 & 2a,2b & INT & 8.2 & -1.2 \\
13E & no force, ball falls back due to natural tendency to rest on surface of earth & G2 & --- & W & 0.7 & -0.7 \\
\hborder{abovespace+=3pt, belowspace+=1pt}\hline[dotted]
\SetCell[c=4]{l,font=\itshape} Forces on a boy swinging on a rope \\
18E & downward gravity, force in direction of motion, and outward force & CF, I5f & 2b,2c & INT & 24.4 & -1.5 \\
18D & downward gravity, force in direction of motion, and inward force from rope & I5f & 2b & INT & 29.2 & -1.4 \\
18C & downward gravity, and force in direction of motion & I5f & 2b,2c & DW & 14.7 & -1.0 \\
\hborder{abovespace+=3pt, belowspace+=1pt}\hline[dotted]
\SetCell[c=4]{l,font=\itshape} Tennis ball hit into a strong wind, what forces act \\
30E & force from gravity, force by hit, and force by air  & I1l & 2a & INT & 54.2 & -1.4 \\
30B & downward force of gravity and force by hit & I1l & 5FA, 2a & W & 8.1 & -1.0 \\
30D & force by hit, and  force by air & I1l & 5G,2a & DW & 3.4 & -0.5 \\
\hborder{abovespace+=3pt, belowspace+=1pt}\hline[dotted]
\SetCell[c=4]{l,font=\itshape} Forces on a ball travelling in a circular track \\
5D & gravity, force in direction of motion, and inward force  & AF2, I5f & 2b & INT & 20.8 & -1.2 \\
5E & gravity, force in direction of motion, and outward force  & AF2, CF & 2c,2b & INT & 17.5 & -1.2 \\
5C & gravity and force in direction of motion  & AF2, I5f & 2c,2b & DW & 29.8 & -1.0 \\
\hborder{abovespace+=3pt, belowspace+=1pt}\hline[dotted]
\SetCell[c=4]{l,font=\itshape} What is path of heavy ball dropped from moving plane \\
14E & straight forwards, then sharp curve, then straignt down & I3, G5 & 0c & FLAT & 0.3 & +0.5 \\
\bottomrule
\end{tblr}
\end{center}

\end{minipage}

\clearpage\noindent\begin{minipage}{\textwidth}
\subsection*{VIII.\hspace{0.5em}Passive object pushed “because in the way”}

\vspace{0.5em}
\begin{center}
\includegraphics{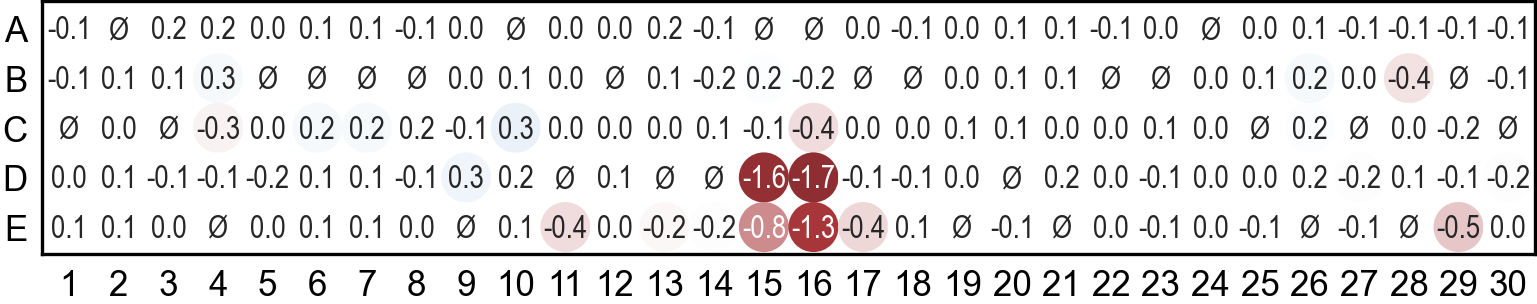}
\end{center}

\begin{center}
\footnotesize\begin{tblr}{
    width = 369pt,
    colspec = {@{}X[l]S[table-format=1.2]S[table-format=1.2]S[table-format=1.2]@{}},
    rows = {rowsep=1pt},
    row{1} = {guard, abovesep+=2pt},
    row{2} = {abovesep+=2pt},
    row{Z} = {belowsep+=1pt},
    hspan = minimal,
}
\toprule
Bootstrap correlation statistic & Pre data & Post data & All data \\
\midrule
Median & 0.96 & 0.71 & 0.99 \\
5th percentile & 0.89 & 0.59 & 0.98 \\
Fraction $\geq$ 0.85 & 1.00 & 0.00 & 1.00 \\
\bottomrule
\end{tblr}
\end{center}\vspace{-0.5em}

\begin{center}
\footnotesize\begin{tblr}{
    width = \textwidth,
    colspec = {@{}lX[l]lllS[table-format=2.1]S[table-format=-1.1]@{}},
    rows = {rowsep=1pt},
    row{1} = {guard, abovesep+=2pt},
    row{2} = {abovesep+=2pt},
    row{Z} = {belowsep+=1pt},
    hspan = minimal,
}
\toprule
\SetCell[c=2]{l} Item/distractor description & & Code & NC-Ex & IRC\textsubscript{pre} & $f$ (\%) & $\widehat{a}$ \\
\midrule
\SetCell[c=4]{l,font=\itshape} Car now pushes truck at constant speed \\
16D & car pushes truck, truck pushed because in way (engine not running so it can't push back on car) & AF1 & 3b & W & 4.5 & -1.7 \\
16E & no force from either, truck pushed because in way & Ob & --- & FLAT & 2.2 & -1.3 \\
\hborder{abovespace+=3pt, belowspace+=1pt}\hline[dotted]
\SetCell[c=4]{l,font=\itshape} Car accelerates while pushing a truck; its force \\
15D & car pushes truck, truck pushed because in way (engine not running so it can't push back on car) & AF1 & 3b & W & 4.6 & -1.6 \\
15E & No force from either, truck pushed because in way & Ob & --- & FLAT & 0.4 & -0.8 \\
\hborder{abovespace+=3pt, belowspace+=1pt}\hline[dotted]
\SetCell[c=4]{l,font=\itshape} What forces act on an empty office chair \\
29E & no forces (since chair is at rest) & AF3 & 5SA or 5G & W & 3.2 & -0.5 \\
\bottomrule
\end{tblr}
\end{center}

\end{minipage}

\clearpage\noindent\begin{minipage}{\textwidth}
\subsection*{IX.\hspace{0.5em}After rocket starts or stops, goes in direction of latest force or earlier motion}

\vspace{0.5em}
\begin{center}
\includegraphics{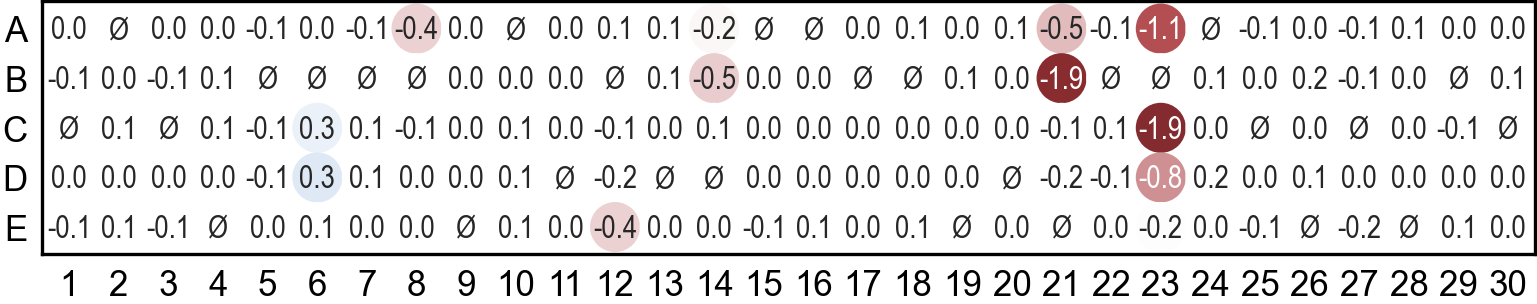}
\end{center}

\begin{center}
\footnotesize\begin{tblr}{
    width = 369pt,
    colspec = {@{}X[l]S[table-format=1.2]S[table-format=1.2]S[table-format=1.2]@{}},
    rows = {rowsep=1pt},
    row{1} = {guard, abovesep+=2pt},
    row{2} = {abovesep+=2pt},
    row{Z} = {belowsep+=1pt},
    hspan = minimal,
}
\toprule
Bootstrap correlation statistic & Pre data & Post data & All data \\
\midrule
Median & 0.98 & 0.98 & 1.00 \\
5th percentile & 0.97 & 0.96 & 0.99 \\
Fraction $\geq$ 0.85 & 1.00 & 1.00 & 1.00 \\
\bottomrule
\end{tblr}
\end{center}\vspace{-0.5em}

\begin{center}
\footnotesize\begin{tblr}{
    width = \textwidth,
    colspec = {@{}lX[l]lllS[table-format=2.1]S[table-format=-1.1]@{}},
    rows = {rowsep=1pt},
    row{1} = {guard, abovesep+=2pt},
    row{2} = {abovesep+=2pt},
    row{Z} = {belowsep+=1pt},
    hspan = minimal,
}
\toprule
\SetCell[c=2]{l} Item/distractor description & & Code & NC-Ex & IRC\textsubscript{pre} & $f$ (\%) & $\widehat{a}$ \\
\midrule
\SetCell[c=4]{l,font=\itshape} Rocket engine now turned off, what is new path? \\
23C & straight in direction of rocket's force & CI3 & 1b & INT & 20.7 & -1.9 \\
23A & immediately reverts to direction prior to rocket firing & I2 & 1b & W & 13.3 & -1.1 \\
23D & starts along thrust, curves back to initial direction (along x) & I2, I3 & 1b & W & 20.3 & -0.8 \\
\hborder{abovespace+=3pt, belowspace+=1pt}\hline[dotted]
\SetCell[c=4]{l,font=\itshape} Rocket going along x pointing along y starts firing, it \\
21B & immediately goes straight  in direction of thrust & CI3 & 0c & DW & 17.3 & -1.9 \\
21A & immediately goes in direction of thrust, then reverts to initial direction & I2 & 0c & W & 8.2 & -0.5 \\
\hborder{abovespace+=3pt, belowspace+=1pt}\hline[dotted]
\SetCell[c=4]{l,font=\itshape} What is path of heavy ball dropped from moving plane \\
14B & ball falls straight down  & K4 & 1a & W & 16.3 & -0.5 \\
\bottomrule
\end{tblr}
\end{center}

\end{minipage}

\clearpage\noindent\begin{minipage}{\textwidth}
\subsection*{X.\hspace{0.5em}Omission of reaction force (normal or from passive object), usually with gravity}

\vspace{0.5em}
\begin{center}
\includegraphics{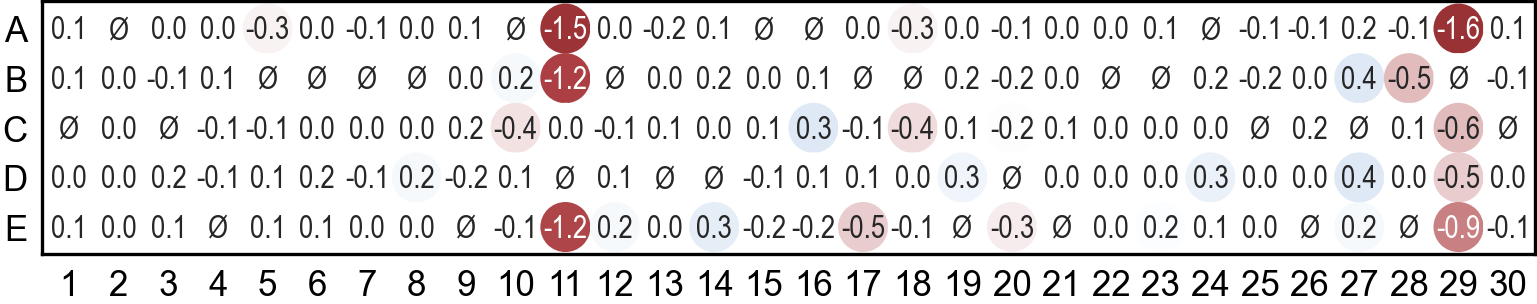}
\end{center}

\begin{center}
\footnotesize\begin{tblr}{
    width = 369pt,
    colspec = {@{}X[l]S[table-format=1.2]S[table-format=1.2]S[table-format=1.2]@{}},
    rows = {rowsep=1pt},
    row{1} = {guard, abovesep+=2pt},
    row{2} = {abovesep+=2pt},
    row{Z} = {belowsep+=1pt},
    hspan = minimal,
}
\toprule
Bootstrap correlation statistic & Pre data & Post data & All data \\
\midrule
Median & 0.96 & 0.79 & 0.99 \\
5th percentile & 0.93 & 0.59 & 0.97 \\
Fraction $\geq$ 0.85 & 1.00 & 0.16 & 1.00 \\
\bottomrule
\end{tblr}
\end{center}\vspace{-0.5em}

\begin{center}
\footnotesize\begin{tblr}{
    width = \textwidth,
    colspec = {@{}lX[l]lllS[table-format=2.1]S[table-format=-1.1]@{}},
    rows = {rowsep=1pt},
    row{1} = {guard, abovesep+=2pt},
    row{2} = {abovesep+=2pt},
    row{Z} = {belowsep+=1pt},
    hspan = minimal,
}
\toprule
\SetCell[c=2]{l} Item/distractor description & & Code & NC-Ex & IRC\textsubscript{pre} & $f$ (\%) & $\widehat{a}$ \\
\midrule
\SetCell[c=4]{l,font=\itshape} What forces act on an empty office chair \\
29A & gravity exerts downward force on chair  & Ob & 5SA & DW & 17.3 & -1.6 \\
29E & no forces (since chair is at rest) & AF3 & 5SA or 5G & W & 3.2 & -0.9 \\
29C & upward force by floor and downward force by air & G1, G2 & 5G & W & 2.5 & -0.6 \\
29D & downward force from gravity, upward force by floor, and downward force from air & G1 & 5Fb- & INT & 21.9 & -0.5 \\
\hborder{abovespace+=3pt, belowspace+=1pt}\hline[dotted]
\SetCell[c=4]{l,font=\itshape} After puck on ice is kicked, what forces act \\
11A & downward force of gravity & Ob, G1 & 5Sa,2b,1d & INT  & 9.6 & -1.5 \\
11B & downward force of gravity and force in direction of motion & I1, Ob & 1b,5Sa & DW & 22.8 & -1.2 \\
11E & no forces on puck & G2 & 5Sa,1d & FLAT & 5.6 & -1.2 \\
\hborder{abovespace+=3pt, belowspace+=1pt}\hline[dotted]
\SetCell[c=4]{l,font=\itshape} Larger Boy kicks knee of smaller boy; both on office chairs \\
28B & larger active student exerts force, passive student exerts no force & AF1 & 3b & DW & 9.4 & -0.5 \\
\hborder{abovespace+=3pt, belowspace+=1pt}\hline[dotted]
\SetCell[c=4]{l,font=\itshape} Elevator lifted up at constant v by a cable such that \\
17E & no force; elevator goes up because cable gets shorter & AF1 & 1d & DW & 8.8 & -0.5 \\
\bottomrule
\end{tblr}
\end{center}

\end{minipage}

\clearpage\noindent\begin{minipage}{\textwidth}
\subsection*{XI.\hspace{0.5em}Straight path preferred vs. curved paths}

\vspace{0.5em}
\begin{center}
\includegraphics{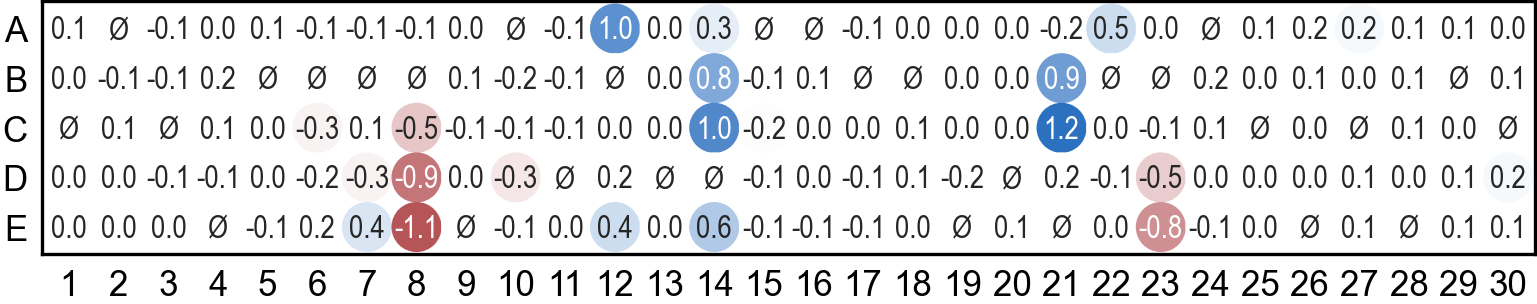}
\end{center}

\begin{center}
\footnotesize\begin{tblr}{
    width = 369pt,
    colspec = {@{}X[l]S[table-format=1.2]S[table-format=1.2]S[table-format=1.2]@{}},
    rows = {rowsep=1pt},
    row{1} = {guard, abovesep+=2pt},
    row{2} = {abovesep+=2pt},
    row{Z} = {belowsep+=1pt},
    hspan = minimal,
}
\toprule
Bootstrap correlation statistic & Pre data & Post data & All data \\
\midrule
Median & 0.91 & 0.98 & 0.99 \\
5th percentile & 0.87 & 0.97 & 0.98 \\
Fraction $\geq$ 0.85 & 0.98 & 1.00 & 1.00 \\
\bottomrule
\end{tblr}
\end{center}\vspace{-0.5em}

\begin{center}
\footnotesize\begin{tblr}{
    width = \textwidth,
    colspec = {@{}lX[l]lllS[table-format=2.1]S[table-format=-1.1]@{}},
    rows = {rowsep=1pt},
    row{1} = {guard, abovesep+=2pt},
    row{2} = {abovesep+=2pt},
    row{Z} = {belowsep+=1pt},
    hspan = minimal,
}
\toprule
\SetCell[c=2]{l} Item/distractor description & & Code & NC-Ex & IRC\textsubscript{pre} & $f$ (\%) & $\widehat{a}$ \\
\midrule
\SetCell[c=4]{l,font=\itshape} Rocket going along x pointing along y starts firing, it \\
21C & straight diagonal at 45 degrees up and forward & CI2 & 0c & INT & 27.0 & +1.2 \\
21B & immediately goes straight  in direction of thrust & CI3 & 0c & DW & 17.3 & +0.9 \\
\hborder{abovespace+=3pt, belowspace+=1pt}\hline[dotted]
\SetCell[c=4]{l,font=\itshape} Puck moving along x @ vx given kick along y: new path? \\
8E & path nearly along kick, then curves back toward former direction & I2 & 1b & DW & 17.3 & -1.1 \\
8D & starts at slight angle then curving in direction of kick & I4 & 1b & W & 10.2 & -0.9 \\
8C & starts along direction of kick, then sharp curve to old direction & I2 & 1b & W & 1.1 & -0.5 \\
\hborder{abovespace+=3pt, belowspace+=1pt}\hline[dotted]
\SetCell[c=4]{l,font=\itshape} What is path of heavy ball dropped from moving plane \\
14C & straight diagonal down and forward & CI2 & 0c & INT & 9.4 & +1.0 \\
14B & ball falls straight down  & K4 & 1a & W & 16.3 & +0.8 \\
14E & straight forwards, then sharp curve, then straignt down & I3, G5 & 0c & FLAT & 0.3 & +0.6 \\
\hborder{abovespace+=3pt, belowspace+=1pt}\hline[dotted]
\SetCell[c=4]{l,font=\itshape} Path of cannonball shot horizontally from cliff \\
12A & straight diagonal down and forward & CI2 & 0c & FLAT & 0.9 & +1.0 \\
\hborder{abovespace+=3pt, belowspace+=1pt}\hline[dotted]
\SetCell[c=4]{l,font=\itshape} Rocket engine now turned off, what is new path? \\
23E & starts straight along diagonal, curves to direction of rocket thrust & I4 & 1b & W & 5.3 & -0.8 \\
23D & starts along thrust, curves back to initial direction (along x) & I2, I3 & 1b & W & 20.3 & -0.5 \\
\hborder{abovespace+=3pt, belowspace+=1pt}\hline[dotted]
\SetCell[c=4]{l,font=\itshape} Rocket moving along x turns on y rocket, its speed \\
22A & speed is constant & AF4 & 2b & INT & 30.3 & +0.5 \\
\bottomrule
\end{tblr}
\end{center}

\end{minipage}

\clearpage\noindent\begin{minipage}{\textwidth}
\subsection*{XII.\hspace{0.5em}Eliminated—similar to XI but weaker}

\vspace{0.5em}
\begin{center}
\includegraphics{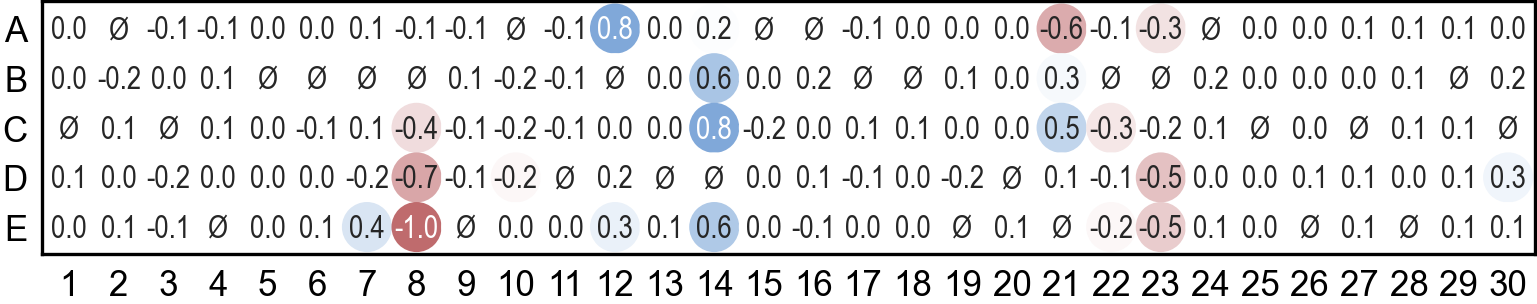}
\end{center}

\begin{center}
\footnotesize\begin{tblr}{
    width = 369pt,
    colspec = {@{}X[l]S[table-format=1.2]S[table-format=1.2]S[table-format=1.2]@{}},
    rows = {rowsep=1pt},
    row{1} = {guard, abovesep+=2pt},
    row{2} = {abovesep+=2pt},
    row{Z} = {belowsep+=1pt},
    hspan = minimal,
}
\toprule
Bootstrap correlation statistic & Pre data & Post data & All data \\
\midrule
Median & 0.74 & 0.94 & 0.97 \\
5th percentile & 0.61 & 0.90 & 0.93 \\
Fraction $\geq$ 0.85 & 0.05 & 0.99 & 1.00 \\
\bottomrule
\end{tblr}
\end{center}\vspace{-0.5em}

\begin{center}
\footnotesize\begin{tblr}{
    width = \textwidth,
    colspec = {@{}lX[l]lllS[table-format=2.1]S[table-format=-1.1]@{}},
    rows = {rowsep=1pt},
    row{1} = {guard, abovesep+=2pt},
    row{2} = {abovesep+=2pt},
    row{Z} = {belowsep+=1pt},
    hspan = minimal,
}
\toprule
\SetCell[c=2]{l} Item/distractor description & & Code & NC-Ex & IRC\textsubscript{pre} & $f$ (\%) & $\widehat{a}$ \\
\midrule
\SetCell[c=4]{l,font=\itshape} Puck moving along x @ vx given kick along y: new path? \\
8E & path nearly along kick, then curves back toward former direction & I2 & 1b & DW & 17.3 & -1.0 \\
8D & starts at slight angle then curving in direction of kick & I4 & 1b & W & 10.2 & -0.7 \\
\hborder{abovespace+=3pt, belowspace+=1pt}\hline[dotted]
\SetCell[c=4]{l,font=\itshape} What is path of heavy ball dropped from moving plane \\
14C & straight diagonal down and forward & CI2 & 0c & INT & 9.4 & +0.8 \\
14B & ball falls straight down  & K4 & 1a & W & 16.3 & +0.6 \\
14E & straight forwards, then sharp curve, then straignt down & I3, G5 & 0c & FLAT & 0.3 & +0.6 \\
\hborder{abovespace+=3pt, belowspace+=1pt}\hline[dotted]
\SetCell[c=4]{l,font=\itshape} Path of cannonball shot horizontally from cliff \\
12A & straight diagonal down and forward & CI2 & 0c & FLAT & 0.9 & +0.8 \\
\hborder{abovespace+=3pt, belowspace+=1pt}\hline[dotted]
\SetCell[c=4]{l,font=\itshape} Rocket going along x pointing along y starts firing, it \\
21A & immediately goes in direction of thrust, then reverts to initial direction & I2 & 0c & W & 8.2 & -0.6 \\
21C & straight diagonal at 45 degrees up and forward & CI2 & 0c & INT & 27.0 & +0.5 \\
\hborder{abovespace+=3pt, belowspace+=1pt}\hline[dotted]
\SetCell[c=4]{l,font=\itshape} Rocket engine now turned off, what is new path? \\
23D & starts along thrust, curves back to initial direction (along x) & I2, I3 & 1b & W & 20.3 & -0.5 \\
23E & starts straight along diagonal, curves to direction of rocket thrust & I4 & 1b & W & 5.3 & -0.5 \\
\bottomrule
\end{tblr}
\end{center}

\end{minipage}

\clearpage\noindent\begin{minipage}{\textwidth}
\subsection*{XIII.\hspace{0.5em}Decelerating after impulse from rocket firing}

\vspace{0.5em}
\begin{center}
\includegraphics{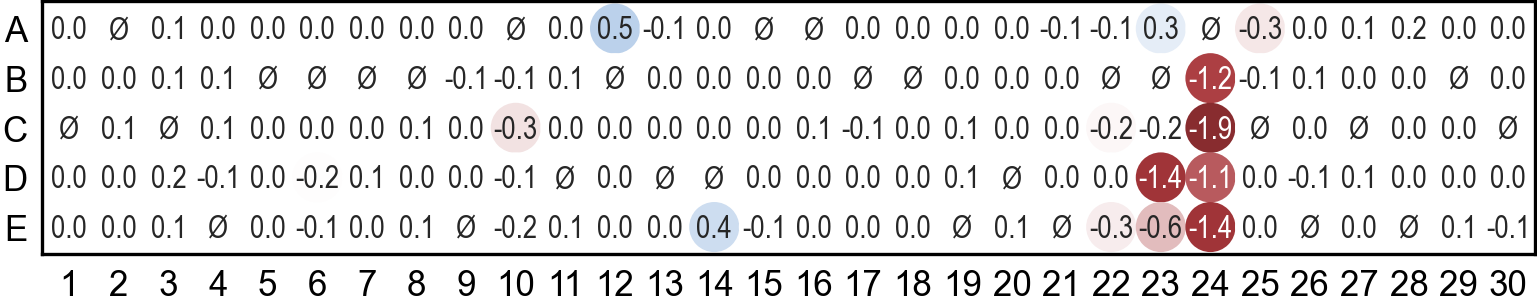}
\end{center}

\begin{center}
\footnotesize\begin{tblr}{
    width = 369pt,
    colspec = {@{}X[l]S[table-format=1.2]S[table-format=1.2]S[table-format=1.2]@{}},
    rows = {rowsep=1pt},
    row{1} = {guard, abovesep+=2pt},
    row{2} = {abovesep+=2pt},
    row{Z} = {belowsep+=1pt},
    hspan = minimal,
}
\toprule
Bootstrap correlation statistic & Pre data & Post data & All data \\
\midrule
Median & 0.99 & 0.98 & 1.00 \\
5th percentile & 0.99 & 0.97 & 0.99 \\
Fraction $\geq$ 0.85 & 1.00 & 1.00 & 1.00 \\
\bottomrule
\end{tblr}
\end{center}\vspace{-0.5em}

\begin{center}
\footnotesize\begin{tblr}{
    width = \textwidth,
    colspec = {@{}lX[l]lllS[table-format=2.1]S[table-format=-1.1]@{}},
    rows = {rowsep=1pt},
    row{1} = {guard, abovesep+=2pt},
    row{2} = {abovesep+=2pt},
    row{Z} = {belowsep+=1pt},
    hspan = minimal,
}
\toprule
\SetCell[c=2]{l} Item/distractor description & & Code & NC-Ex & IRC\textsubscript{pre} & $f$ (\%) & $\widehat{a}$ \\
\midrule
\SetCell[c=4]{l,font=\itshape} Rocket Pointing Sideways Turns off \\
24C & continuously decreasing & I3 & 1c & DW & 20.2 & -1.9 \\
24E & constant for while, then decreasing thereafter & I3 & 1c & DW & 5.7 & -1.4 \\
24B & continuously increasing & --- & 1c & W & 2.7 & -1.2 \\
24D & increasing for while then constant thereafter & --- & 1c & W & 2.4 & -1.1 \\
\hborder{abovespace+=3pt, belowspace+=1pt}\hline[dotted]
\SetCell[c=4]{l,font=\itshape} Rocket engine now turned off, what is new path? \\
23D & starts along thrust, curves back to initial direction (along x) & I2, I3 & 1b & W & 20.3 & -1.4 \\
23E & starts straight along diagonal, curves to direction of rocket thrust & I4 & 1b & W & 5.3 & -0.6 \\
\hborder{abovespace+=3pt, belowspace+=1pt}\hline[dotted]
\SetCell[c=4]{l,font=\itshape} Path of cannonball shot horizontally from cliff \\
12A & straight diagonal down and forward & CI2 & 0c & FLAT & 0.9 & +0.5 \\
\bottomrule
\end{tblr}
\end{center}

\end{minipage}

\clearpage\noindent\begin{minipage}{\textwidth}
\subsection*{XIV.\hspace{0.5em}2M and M balls differ by factor of 2}

\vspace{0.5em}
\begin{center}
\includegraphics{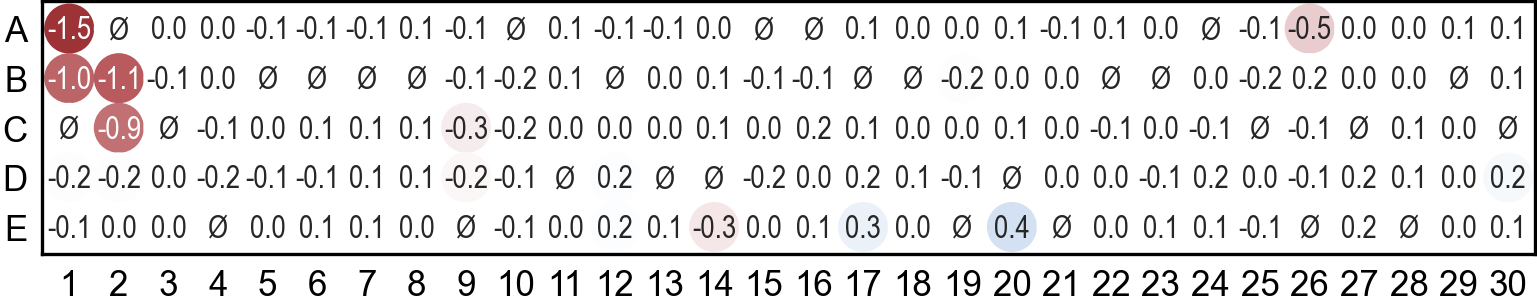}
\end{center}

\begin{center}
\footnotesize\begin{tblr}{
    width = 369pt,
    colspec = {@{}X[l]S[table-format=1.2]S[table-format=1.2]S[table-format=1.2]@{}},
    rows = {rowsep=1pt},
    row{1} = {guard, abovesep+=2pt},
    row{2} = {abovesep+=2pt},
    row{Z} = {belowsep+=1pt},
    hspan = minimal,
}
\toprule
Bootstrap correlation statistic & Pre data & Post data & All data \\
\midrule
Median & 0.98 & 0.96 & 0.99 \\
5th percentile & 0.96 & 0.92 & 0.99 \\
Fraction $\geq$ 0.85 & 1.00 & 1.00 & 1.00 \\
\bottomrule
\end{tblr}
\end{center}\vspace{-0.5em}

\begin{center}
\footnotesize\begin{tblr}{
    width = \textwidth,
    colspec = {@{}lX[l]lllS[table-format=2.1]S[table-format=-1.1]@{}},
    rows = {rowsep=1pt},
    row{1} = {guard, abovesep+=2pt},
    row{2} = {abovesep+=2pt},
    row{Z} = {belowsep+=1pt},
    hspan = minimal,
}
\toprule
\SetCell[c=2]{l} Item/distractor description & & Code & NC-Ex & IRC\textsubscript{pre} & $f$ (\%) & $\widehat{a}$ \\
\midrule
\SetCell[c=4]{l,font=\itshape} Two metal balls dropped: time to earth's surface \\
1A & $\sim$1/2 the time for 2M as M & G3 & 2 + 5Gm & W & 7.0 & -1.5 \\
1B & $\sim$1/2 the time for M as for 2M & G3- & 2 + 5Gm & W & 4.2 & -1.0 \\
\hborder{abovespace+=3pt, belowspace+=1pt}\hline[dotted]
\SetCell[c=4]{l,font=\itshape} Two metal balls roll off table with equal speeds \\
2B & 2M hits floor about 1/2 distance of M & G3 & 2 + 5Gm & INT & 18.9 & -1.1 \\
2C & M hits floor about 1/2 distance of 2M & G3- & 2 + 5Gm & FLAT & 5.5 & -0.9 \\
\hborder{abovespace+=3pt, belowspace+=1pt}\hline[dotted]
\SetCell[c=4]{l,font=\itshape} Woman doubles force that moved box @ v0 \\
26A & speed doubles when force doubled & AF4 & 2b & INT  & 35.8 & -0.5 \\
\bottomrule
\end{tblr}
\end{center}

\end{minipage}

\clearpage\noindent\begin{minipage}{\textwidth}
\subsection*{XV.\hspace{0.5em}Latest force dominates after sudden force}

\vspace{0.5em}
\begin{center}
\includegraphics{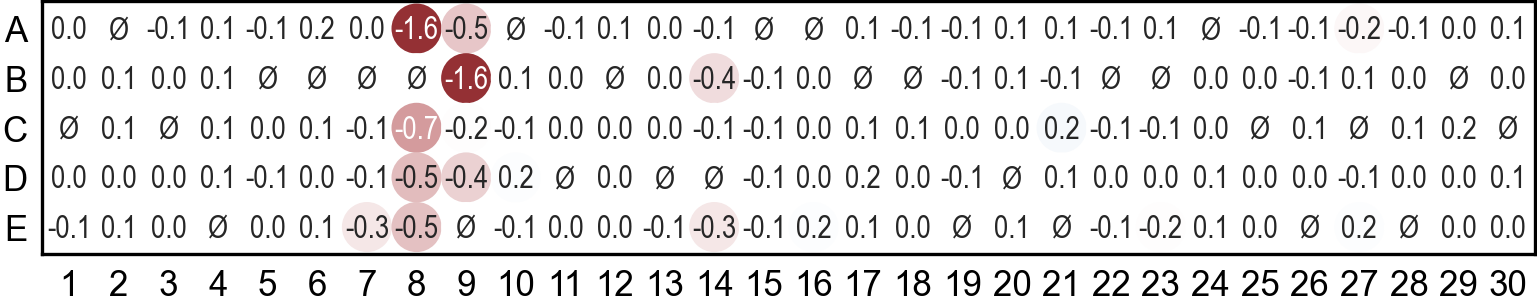}
\end{center}

\begin{center}
\footnotesize\begin{tblr}{
    width = 369pt,
    colspec = {@{}X[l]S[table-format=1.2]S[table-format=1.2]S[table-format=1.2]@{}},
    rows = {rowsep=1pt},
    row{1} = {guard, abovesep+=2pt},
    row{2} = {abovesep+=2pt},
    row{Z} = {belowsep+=1pt},
    hspan = minimal,
}
\toprule
Bootstrap correlation statistic & Pre data & Post data & All data \\
\midrule
Median & 0.98 & 0.98 & 0.99 \\
5th percentile & 0.97 & 0.96 & 0.99 \\
Fraction $\geq$ 0.85 & 1.00 & 1.00 & 1.00 \\
\bottomrule
\end{tblr}
\end{center}\vspace{-0.5em}

\begin{center}
\footnotesize\begin{tblr}{
    width = \textwidth,
    colspec = {@{}lX[l]lllS[table-format=2.1]S[table-format=-1.1]@{}},
    rows = {rowsep=1pt},
    row{1} = {guard, abovesep+=2pt},
    row{2} = {abovesep+=2pt},
    row{Z} = {belowsep+=1pt},
    hspan = minimal,
}
\toprule
\SetCell[c=2]{l} Item/distractor description & & Code & NC-Ex & IRC\textsubscript{pre} & $f$ (\%) & $\widehat{a}$ \\
\midrule
\SetCell[c=4]{l,font=\itshape} Puck moving along x @ vx given kick along y: new path? \\
8A & straight path in direction of kick  & CI3 & 0e,2a & DW & 17.9 & -1.6 \\
8C & starts along direction of kick, then sharp curve to old direction & I2 & 1b & W & 1.1 & -0.7 \\
8D & starts at slight angle then curving in direction of kick & I4 & 1b & W & 10.2 & -0.5 \\
8E & path nearly along kick, then curves back toward former direction & I2 & 1b & DW & 17.3 & -0.5 \\
\hborder{abovespace+=3pt, belowspace+=1pt}\hline[dotted]
\SetCell[c=4]{l,font=\itshape} The speed of puck just after kick is \\
9B & speed equal to vkick, independent of v0 & CI3 & 0e,2a & DW & 22.6 & -1.6 \\
9A & speed equals v0  & --- & 0e,2a & FLAT & 4.7 & -0.5 \\
\bottomrule
\end{tblr}
\end{center}

\end{minipage}

\clearpage\noindent\begin{minipage}{\textwidth}
\subsection*{XVI.\hspace{0.5em}Force change increases speed, sometimes with initial acceleration}

\vspace{0.5em}
\begin{center}
\includegraphics{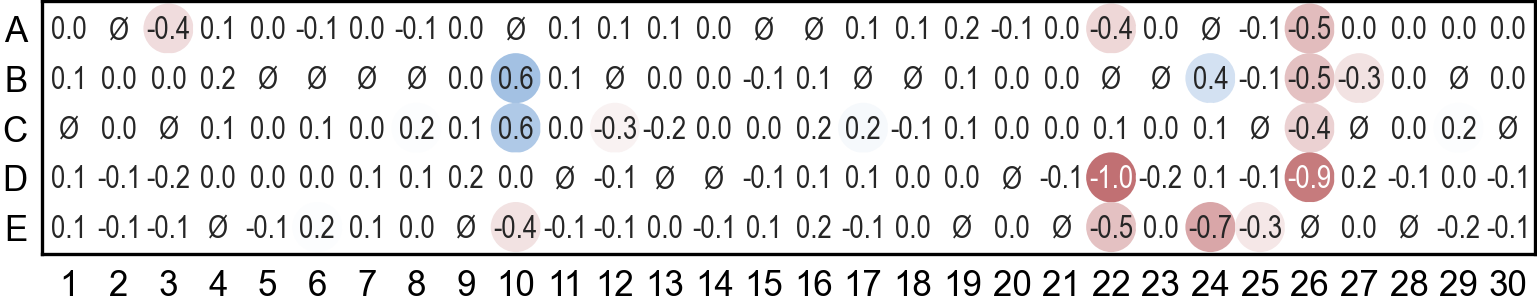}
\end{center}

\begin{center}
\footnotesize\begin{tblr}{
    width = 369pt,
    colspec = {@{}X[l]S[table-format=1.2]S[table-format=1.2]S[table-format=1.2]@{}},
    rows = {rowsep=1pt},
    row{1} = {guard, abovesep+=2pt},
    row{2} = {abovesep+=2pt},
    row{Z} = {belowsep+=1pt},
    hspan = minimal,
}
\toprule
Bootstrap correlation statistic & Pre data & Post data & All data \\
\midrule
Median & 0.92 & 0.92 & 0.98 \\
5th percentile & 0.86 & 0.86 & 0.96 \\
Fraction $\geq$ 0.85 & 0.97 & 0.97 & 1.00 \\
\bottomrule
\end{tblr}
\end{center}\vspace{-0.5em}

\begin{center}
\footnotesize\begin{tblr}{
    width = \textwidth,
    colspec = {@{}lX[l]lllS[table-format=2.1]S[table-format=-1.1]@{}},
    rows = {rowsep=1pt},
    row{1} = {guard, abovesep+=2pt},
    row{2} = {abovesep+=2pt},
    row{Z} = {belowsep+=1pt},
    hspan = minimal,
}
\toprule
\SetCell[c=2]{l} Item/distractor description & & Code & NC-Ex & IRC\textsubscript{pre} & $f$ (\%) & $\widehat{a}$ \\
\midrule
\SetCell[c=4]{l,font=\itshape} Rocket moving along x turns on y rocket, its speed \\
22D & increases for a while, then constant & AF6 & 2b & DW & 22.7 & -1.0 \\
22E & constant for a while, then decreases & AF7 & 2b & W & 2.9 & -0.5 \\
\hborder{abovespace+=3pt, belowspace+=1pt}\hline[dotted]
\SetCell[c=4]{l,font=\itshape} Woman doubles force that moved box @ v0 \\
26D & speed increases up to constant v & AF6 & 2b & FLAT & 13.8 & -0.9 \\
26A & speed doubles when force doubled & AF4 & 2b & INT  & 35.8 & -0.5 \\
26B & higher constant speed but not twice as fast & R2, R3 & 2b & INT  & 30.6 & -0.5 \\
\hborder{abovespace+=3pt, belowspace+=1pt}\hline[dotted]
\SetCell[c=4]{l,font=\itshape} Rocket Pointing Sideways Turns off \\
24E & constant for while, then decreasing thereafter & I3 & 1c & DW & 5.7 & -0.7 \\
\hborder{abovespace+=3pt, belowspace+=1pt}\hline[dotted]
\SetCell[c=4]{l,font=\itshape} Along path you picked in 8, what is speed of puck after y-kick? \\
10B & continuously increases & I4 & 1b & FLAT & 2.9 & +0.6 \\
10C & continuously decreases & --- & 1b & W & 11.4 & +0.6 \\
\bottomrule
\end{tblr}
\end{center}

\end{minipage}

\clearpage\noindent\begin{minipage}{\textwidth}
\subsection*{XVII.\hspace{0.5em}Continues curving after inward force removed}

\vspace{0.5em}
\begin{center}
\includegraphics{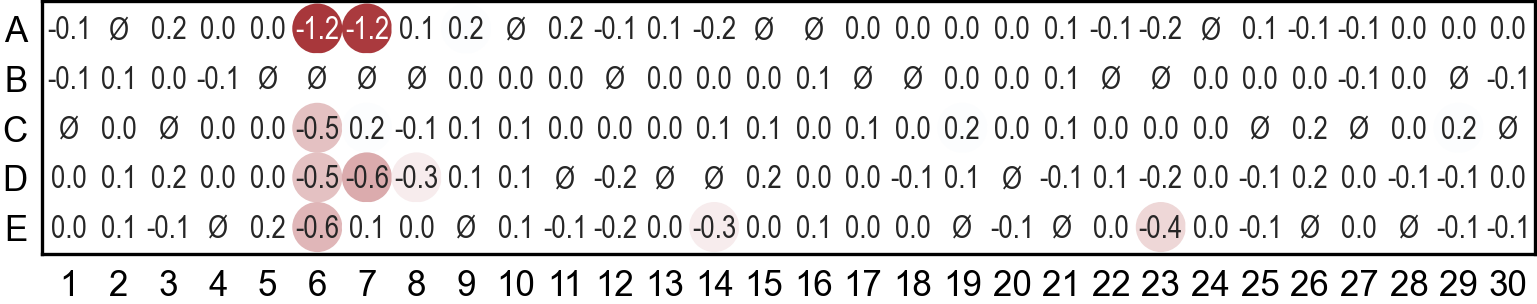}
\end{center}

\begin{center}
\footnotesize\begin{tblr}{
    width = 369pt,
    colspec = {@{}X[l]S[table-format=1.2]S[table-format=1.2]S[table-format=1.2]@{}},
    rows = {rowsep=1pt},
    row{1} = {guard, abovesep+=2pt},
    row{2} = {abovesep+=2pt},
    row{Z} = {belowsep+=1pt},
    hspan = minimal,
}
\toprule
Bootstrap correlation statistic & Pre data & Post data & All data \\
\midrule
Median & 0.94 & 0.97 & 0.99 \\
5th percentile & 0.84 & 0.96 & 0.97 \\
Fraction $\geq$ 0.85 & 0.94 & 1.00 & 1.00 \\
\bottomrule
\end{tblr}
\end{center}\vspace{-0.5em}

\begin{center}
\footnotesize\begin{tblr}{
    width = \textwidth,
    colspec = {@{}lX[l]lllS[table-format=2.1]S[table-format=-1.1]@{}},
    rows = {rowsep=1pt},
    row{1} = {guard, abovesep+=2pt},
    row{2} = {abovesep+=2pt},
    row{Z} = {belowsep+=1pt},
    hspan = minimal,
}
\toprule
\SetCell[c=2]{l} Item/distractor description & & Code & NC-Ex & IRC\textsubscript{pre} & $f$ (\%) & $\widehat{a}$ \\
\midrule
\SetCell[c=4]{l,font=\itshape} Path of Ball after Exiting Curved Track \\
6A & curved  inward path after exiting curved channel & I5p & 1b & DW & 19.2 & -1.2 \\
6E & goes almost straight out from center & CF & 1b & --- & 0.7 & -0.6 \\
6D & curves out significantly & CI2, CF & 1b & --- & 1.1 & -0.5 \\
6C & curves outward somewhat & CF & 1b & W & 3.5 & -0.5 \\
\hborder{abovespace+=3pt, belowspace+=1pt}\hline[dotted]
\SetCell[c=4]{l,font=\itshape} Steel Ball circles on string; path after string breaks? \\
7A & curved inwards path after string breaks & I5p & 1b & DW & 13.3 & -1.2 \\
7D & starts straight out, then curves to old tangential direction & I2, CF,I5p & 1b & W & 4.3 & -0.6 \\
\bottomrule
\end{tblr}
\end{center}

\end{minipage}

\clearpage
\clearpage\noindent\begin{minipage}{\textwidth}
\subsection*{XVIII.\hspace{0.5em}Air exerts significant drag and downward force}

\vspace{0.5em}
\begin{center}
\includegraphics{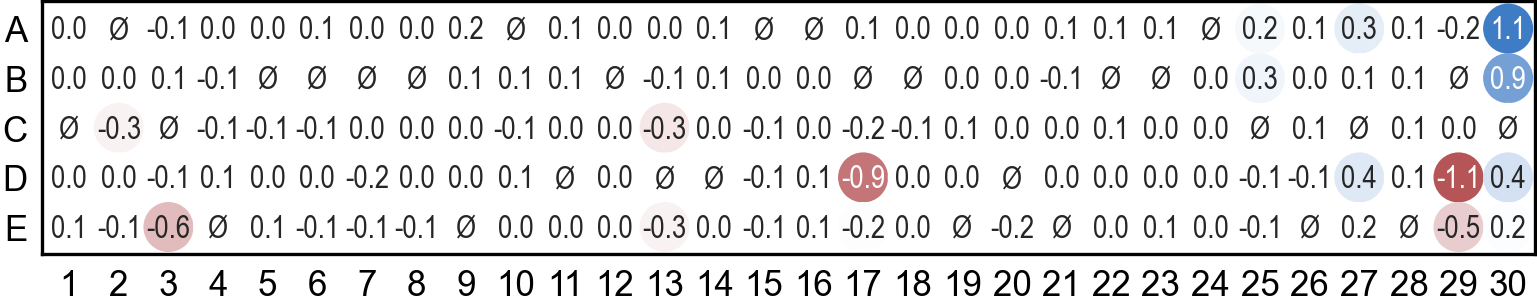}
\end{center}

\begin{center}
\footnotesize\begin{tblr}{
    width = 369pt,
    colspec = {@{}X[l]S[table-format=1.2]S[table-format=1.2]S[table-format=1.2]@{}},
    rows = {rowsep=1pt},
    row{1} = {guard, abovesep+=2pt},
    row{2} = {abovesep+=2pt},
    row{Z} = {belowsep+=1pt},
    hspan = minimal,
}
\toprule
Bootstrap correlation statistic & Pre data & Post data & All data \\
\midrule
Median & 0.93 & 0.95 & 0.99 \\
5th percentile & 0.87 & 0.92 & 0.98 \\
Fraction $\geq$ 0.85 & 0.98 & 0.99 & 1.00 \\
\bottomrule
\end{tblr}
\end{center}\vspace{-0.5em}

\begin{center}
\footnotesize\begin{tblr}{
    width = \textwidth,
    colspec = {@{}lX[l]lllS[table-format=2.1]S[table-format=-1.1]@{}},
    rows = {rowsep=1pt},
    row{1} = {guard, abovesep+=2pt},
    row{2} = {abovesep+=2pt},
    row{Z} = {belowsep+=1pt},
    hspan = minimal,
}
\toprule
\SetCell[c=2]{l} Item/distractor description & & Code & NC-Ex & IRC\textsubscript{pre} & $f$ (\%) & $\widehat{a}$ \\
\midrule
\SetCell[c=4]{l,font=\itshape} Tennis ball hit into a strong wind, what forces act \\
30A & downward force of gravity & AF1 & 5FA & W & 4.4 & +1.1 \\
30B & downward force of gravity and force by hit & I1l & 5FA, 2a & W & 8.1 & +0.9 \\
\hborder{abovespace+=3pt, belowspace+=1pt}\hline[dotted]
\SetCell[c=4]{l,font=\itshape} What forces act on an empty office chair \\
29D & downward force from gravity, upward force by floor, and downward force from air & G1 & 5Fb- & INT & 21.9 & -1.1 \\
29E & no forces (since chair is at rest) & AF3 & 5SA or 5G & W & 3.2 & -0.5 \\
\hborder{abovespace+=3pt, belowspace+=1pt}\hline[dotted]
\SetCell[c=4]{l,font=\itshape} Elevator lifted up at constant v by a cable such that \\
17D & upward force of cable exceeds gravity and downward force of air & CI1, G1 & 1c,1d & INT & 11.6 & -0.9 \\
\hborder{abovespace+=3pt, belowspace+=1pt}\hline[dotted]
\SetCell[c=4]{l,font=\itshape} Stone Dropped from Roof of Single Story Building \\
3E & falls because of combined gravity and air pushing it down & G1 & 5Fb- & W & 5.7 & -0.6 \\
\bottomrule
\end{tblr}
\end{center}

\end{minipage}

\clearpage\noindent\begin{minipage}{\textwidth}
\subsection*{XIX.\hspace{0.5em}No acceleration while rocket firing, then speed decreases}

\vspace{0.5em}
\begin{center}
\includegraphics{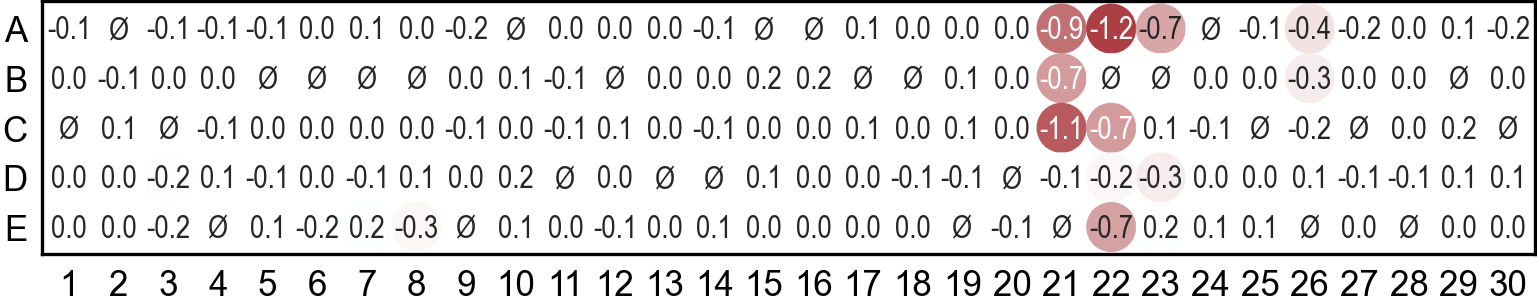}
\end{center}

\begin{center}
\footnotesize\begin{tblr}{
    width = 369pt,
    colspec = {@{}X[l]S[table-format=1.2]S[table-format=1.2]S[table-format=1.2]@{}},
    rows = {rowsep=1pt},
    row{1} = {guard, abovesep+=2pt},
    row{2} = {abovesep+=2pt},
    row{Z} = {belowsep+=1pt},
    hspan = minimal,
}
\toprule
Bootstrap correlation statistic & Pre data & Post data & All data \\
\midrule
Median & 0.88 & 0.98 & 0.99 \\
5th percentile & 0.80 & 0.96 & 0.97 \\
Fraction $\geq$ 0.85 & 0.77 & 1.00 & 1.00 \\
\bottomrule
\end{tblr}
\end{center}\vspace{-0.5em}

\begin{center}
\footnotesize\begin{tblr}{
    width = \textwidth,
    colspec = {@{}lX[l]lllS[table-format=2.1]S[table-format=-1.1]@{}},
    rows = {rowsep=1pt},
    row{1} = {guard, abovesep+=2pt},
    row{2} = {abovesep+=2pt},
    row{Z} = {belowsep+=1pt},
    hspan = minimal,
}
\toprule
\SetCell[c=2]{l} Item/distractor description & & Code & NC-Ex & IRC\textsubscript{pre} & $f$ (\%) & $\widehat{a}$ \\
\midrule
\SetCell[c=4]{l,font=\itshape} Rocket moving along x turns on y rocket, its speed \\
22A & speed is constant & AF4 & 2b & INT & 30.3 & -1.2 \\
22C & rocket speed continuously decreases  & AF7 & 2b & W & 3.2 & -0.7 \\
22E & constant for a while, then decreases & AF7 & 2b & W & 2.9 & -0.7 \\
\hborder{abovespace+=3pt, belowspace+=1pt}\hline[dotted]
\SetCell[c=4]{l,font=\itshape} Rocket going along x pointing along y starts firing, it \\
21C & straight diagonal at 45 degrees up and forward & CI2 & 0c & INT & 27.0 & -1.1 \\
21A & immediately goes in direction of thrust, then reverts to initial direction & I2 & 0c & W & 8.2 & -0.9 \\
21B & immediately goes straight  in direction of thrust & CI3 & 0c & DW & 17.3 & -0.7 \\
\hborder{abovespace+=3pt, belowspace+=1pt}\hline[dotted]
\SetCell[c=4]{l,font=\itshape} Rocket engine now turned off, what is new path? \\
23A & immediately reverts to direction prior to rocket firing & I2 & 1b & W & 13.3 & -0.7 \\
\bottomrule
\end{tblr}
\end{center}

\end{minipage}

\clearpage\noindent\begin{minipage}{\textwidth}
\subsection*{XX.\hspace{0.5em}Eliminated—similar to XIX but less sparse}

\vspace{0.5em}
\begin{center}
\includegraphics{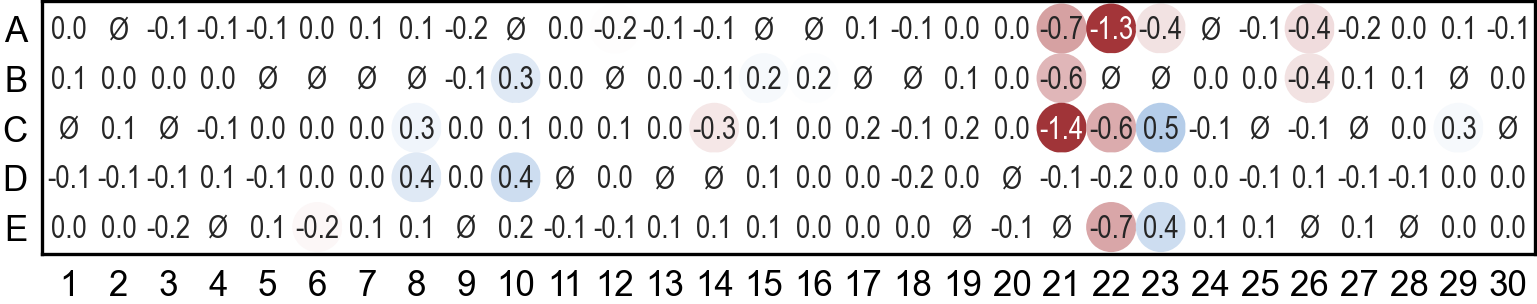}
\end{center}

\begin{center}
\footnotesize\begin{tblr}{
    width = 369pt,
    colspec = {@{}X[l]S[table-format=1.2]S[table-format=1.2]S[table-format=1.2]@{}},
    rows = {rowsep=1pt},
    row{1} = {guard, abovesep+=2pt},
    row{2} = {abovesep+=2pt},
    row{Z} = {belowsep+=1pt},
    hspan = minimal,
}
\toprule
Bootstrap correlation statistic & Pre data & Post data & All data \\
\midrule
Median & 0.90 & 0.97 & 0.98 \\
5th percentile & 0.84 & 0.95 & 0.96 \\
Fraction $\geq$ 0.85 & 0.91 & 1.00 & 1.00 \\
\bottomrule
\end{tblr}
\end{center}\vspace{-0.5em}

\begin{center}
\footnotesize\begin{tblr}{
    width = \textwidth,
    colspec = {@{}lX[l]lllS[table-format=2.1]S[table-format=-1.1]@{}},
    rows = {rowsep=1pt},
    row{1} = {guard, abovesep+=2pt},
    row{2} = {abovesep+=2pt},
    row{Z} = {belowsep+=1pt},
    hspan = minimal,
}
\toprule
\SetCell[c=2]{l} Item/distractor description & & Code & NC-Ex & IRC\textsubscript{pre} & $f$ (\%) & $\widehat{a}$ \\
\midrule
\SetCell[c=4]{l,font=\itshape} Rocket going along x pointing along y starts firing, it \\
21C & straight diagonal at 45 degrees up and forward & CI2 & 0c & INT & 27.0 & -1.4 \\
21A & immediately goes in direction of thrust, then reverts to initial direction & I2 & 0c & W & 8.2 & -0.7 \\
21B & immediately goes straight  in direction of thrust & CI3 & 0c & DW & 17.3 & -0.6 \\
\hborder{abovespace+=3pt, belowspace+=1pt}\hline[dotted]
\SetCell[c=4]{l,font=\itshape} Rocket moving along x turns on y rocket, its speed \\
22A & speed is constant & AF4 & 2b & INT & 30.3 & -1.3 \\
22E & constant for a while, then decreases & AF7 & 2b & W & 2.9 & -0.7 \\
22C & rocket speed continuously decreases  & AF7 & 2b & W & 3.2 & -0.6 \\
\hborder{abovespace+=3pt, belowspace+=1pt}\hline[dotted]
\SetCell[c=4]{l,font=\itshape} Rocket engine now turned off, what is new path? \\
23C & straight in direction of rocket's force & CI3 & 1b & INT & 20.7 & +0.5 \\
\bottomrule
\end{tblr}
\end{center}

\end{minipage}

\clearpage\noindent\begin{minipage}{\textwidth}
\subsection*{XXI.\hspace{0.5em}2M ball falls significantly faster, but not by factor of 2}

\vspace{0.5em}
\begin{center}
\includegraphics{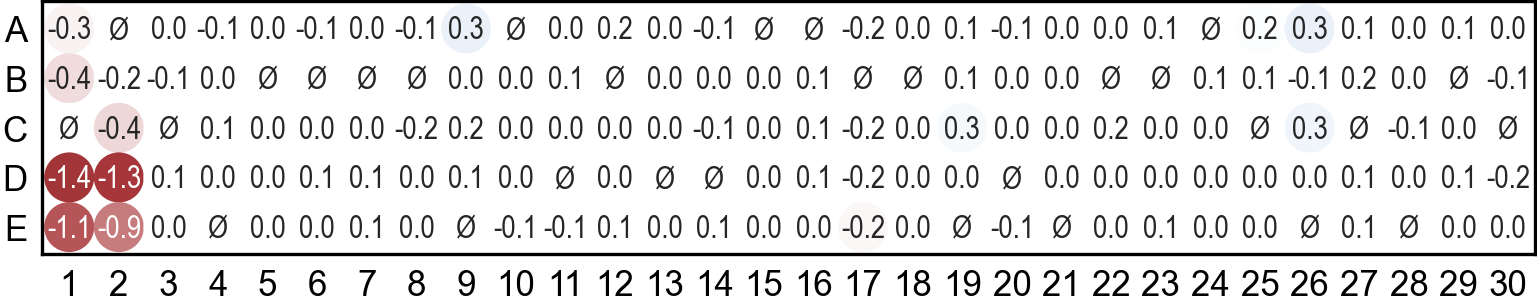}
\end{center}

\begin{center}
\footnotesize\begin{tblr}{
    width = 369pt,
    colspec = {@{}X[l]S[table-format=1.2]S[table-format=1.2]S[table-format=1.2]@{}},
    rows = {rowsep=1pt},
    row{1} = {guard, abovesep+=2pt},
    row{2} = {abovesep+=2pt},
    row{Z} = {belowsep+=1pt},
    hspan = minimal,
}
\toprule
Bootstrap correlation statistic & Pre data & Post data & All data \\
\midrule
Median & 0.99 & 0.95 & 0.99 \\
5th percentile & 0.97 & 0.91 & 0.98 \\
Fraction $\geq$ 0.85 & 1.00 & 0.99 & 1.00 \\
\bottomrule
\end{tblr}
\end{center}\vspace{-0.5em}

\begin{center}
\footnotesize\begin{tblr}{
    width = \textwidth,
    colspec = {@{}lX[l]lllS[table-format=2.1]S[table-format=-1.1]@{}},
    rows = {rowsep=1pt},
    row{1} = {guard, abovesep+=2pt},
    row{2} = {abovesep+=2pt},
    row{Z} = {belowsep+=1pt},
    hspan = minimal,
}
\toprule
\SetCell[c=2]{l} Item/distractor description & & Code & NC-Ex & IRC\textsubscript{pre} & $f$ (\%) & $\widehat{a}$ \\
\midrule
\SetCell[c=4]{l,font=\itshape} Two metal balls dropped: time to earth's surface \\
1D & considerably less for 2M, but not 1/2 & G3 & 2 + 5Gm & DW & 11.1 & -1.4 \\
1E & considerably less for M, but not 1/2 & G3- & 2 + 5Gm & W & 2.9 & -1.1 \\
\hborder{abovespace+=3pt, belowspace+=1pt}\hline[dotted]
\SetCell[c=4]{l,font=\itshape} Two metal balls roll off table with equal speeds \\
2D & 2M  hits floor closer, but not 1/2 of M  & G3 & 2 + 5Gm & DW & 29.0 & -1.3 \\
2E & M hits floor closer, but not 1/2 distance of 2M & G3- & 2 + 5Gm & FLAT & 5.3 & -0.9 \\
\bottomrule
\end{tblr}
\end{center}

\end{minipage}

\clearpage\noindent\begin{minipage}{\textwidth}
\subsection*{XXII.\hspace{0.5em}Eliminated—low bootstrap correlations}

\vspace{0.5em}
\begin{center}
\includegraphics{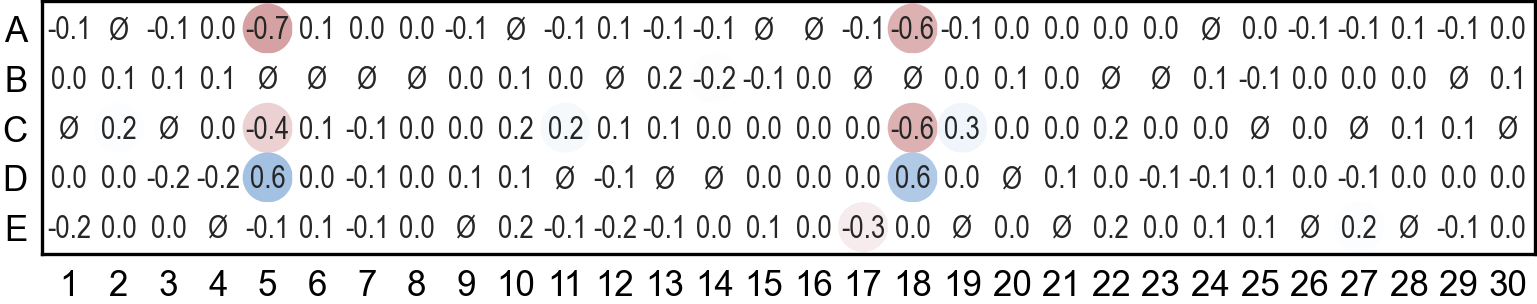}
\end{center}

\begin{center}
\footnotesize\begin{tblr}{
    width = 369pt,
    colspec = {@{}X[l]S[table-format=1.2]S[table-format=1.2]S[table-format=1.2]@{}},
    rows = {rowsep=1pt},
    row{1} = {guard, abovesep+=2pt},
    row{2} = {abovesep+=2pt},
    row{Z} = {belowsep+=1pt},
    hspan = minimal,
}
\toprule
Bootstrap correlation statistic & Pre data & Post data & All data \\
\midrule
Median & 0.44 & 0.85 & 0.95 \\
5th percentile & 0.22 & 0.75 & 0.76 \\
Fraction $\geq$ 0.85 & 0.00 & 0.46 & 0.84 \\
\bottomrule
\end{tblr}
\end{center}\vspace{-0.5em}

\begin{center}
\footnotesize\begin{tblr}{
    width = \textwidth,
    colspec = {@{}lX[l]lllS[table-format=2.1]S[table-format=-1.1]@{}},
    rows = {rowsep=1pt},
    row{1} = {guard, abovesep+=2pt},
    row{2} = {abovesep+=2pt},
    row{Z} = {belowsep+=1pt},
    hspan = minimal,
}
\toprule
\SetCell[c=2]{l} Item/distractor description & & Code & NC-Ex & IRC\textsubscript{pre} & $f$ (\%) & $\widehat{a}$ \\
\midrule
\SetCell[c=4]{l,font=\itshape} Forces on a ball travelling in a circular track \\
5A & gravity only force  & Ob & 2c & FLAT & 6.4 & -0.7 \\
5D & gravity, force in direction of motion, and inward force  & AF2, I5f & 2b & INT & 20.8 & +0.6 \\
\hborder{abovespace+=3pt, belowspace+=1pt}\hline[dotted]
\SetCell[c=4]{l,font=\itshape} Forces on a boy swinging on a rope \\
18A & downward force of gravity & AF!, Ob & 2c & FLAT & 3.2 & -0.6 \\
18C & downward gravity, and force in direction of motion & I5f & 2b,2c & DW & 14.7 & -0.6 \\
18D & downward gravity, force in direction of motion, and inward force from rope & I5f & 2b & INT & 29.2 & +0.6 \\
\bottomrule
\end{tblr}
\end{center}

\end{minipage}

\clearpage\noindent\begin{minipage}{\textwidth}
\subsection*{XXIII.\hspace{0.5em}Missing centripetal force}

\vspace{0.5em}
\begin{center}
\includegraphics{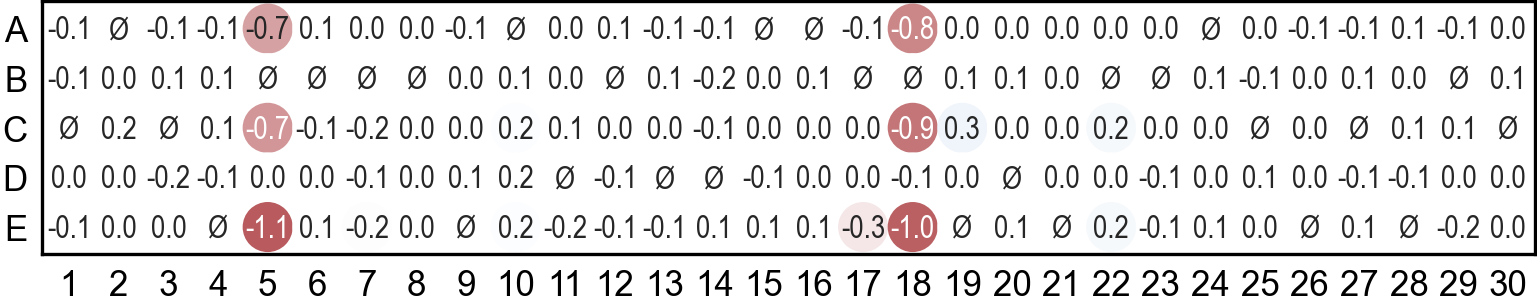}
\end{center}

\begin{center}
\footnotesize\begin{tblr}{
    width = 369pt,
    colspec = {@{}X[l]S[table-format=1.2]S[table-format=1.2]S[table-format=1.2]@{}},
    rows = {rowsep=1pt},
    row{1} = {guard, abovesep+=2pt},
    row{2} = {abovesep+=2pt},
    row{Z} = {belowsep+=1pt},
    hspan = minimal,
}
\toprule
Bootstrap correlation statistic & Pre data & Post data & All data \\
\midrule
Median & 0.74 & 0.98 & 0.98 \\
5th percentile & 0.63 & 0.96 & 0.87 \\
Fraction $\geq$ 0.85 & 0.01 & 1.00 & 0.96 \\
\bottomrule
\end{tblr}
\end{center}\vspace{-0.5em}

\begin{center}
\footnotesize\begin{tblr}{
    width = \textwidth,
    colspec = {@{}lX[l]lllS[table-format=2.1]S[table-format=-1.1]@{}},
    rows = {rowsep=1pt},
    row{1} = {guard, abovesep+=2pt},
    row{2} = {abovesep+=2pt},
    row{Z} = {belowsep+=1pt},
    hspan = minimal,
}
\toprule
\SetCell[c=2]{l} Item/distractor description & & Code & NC-Ex & IRC\textsubscript{pre} & $f$ (\%) & $\widehat{a}$ \\
\midrule
\SetCell[c=4]{l,font=\itshape} Forces on a ball travelling in a circular track \\
5E & gravity, force in direction of motion, and outward force  & AF2, CF & 2c,2b & INT & 17.5 & -1.1 \\
5C & gravity and force in direction of motion  & AF2, I5f & 2c,2b & DW & 29.8 & -0.7 \\
5A & gravity only force  & Ob & 2c & FLAT & 6.4 & -0.7 \\
\hborder{abovespace+=3pt, belowspace+=1pt}\hline[dotted]
\SetCell[c=4]{l,font=\itshape} Forces on a boy swinging on a rope \\
18E & downward gravity, force in direction of motion, and outward force & CF, I5f & 2b,2c & INT & 24.4 & -1.0 \\
18C & downward gravity, and force in direction of motion & I5f & 2b,2c & DW & 14.7 & -0.9 \\
18A & downward force of gravity & AF!, Ob & 2c & FLAT & 3.2 & -0.8 \\
\bottomrule
\end{tblr}
\end{center}

\end{minipage}

\clearpage\noindent\begin{minipage}{\textwidth}
\subsection*{XXIV.\hspace{0.5em}Eliminated—appears to be a residual}

\vspace{0.5em}
\begin{center}
\includegraphics{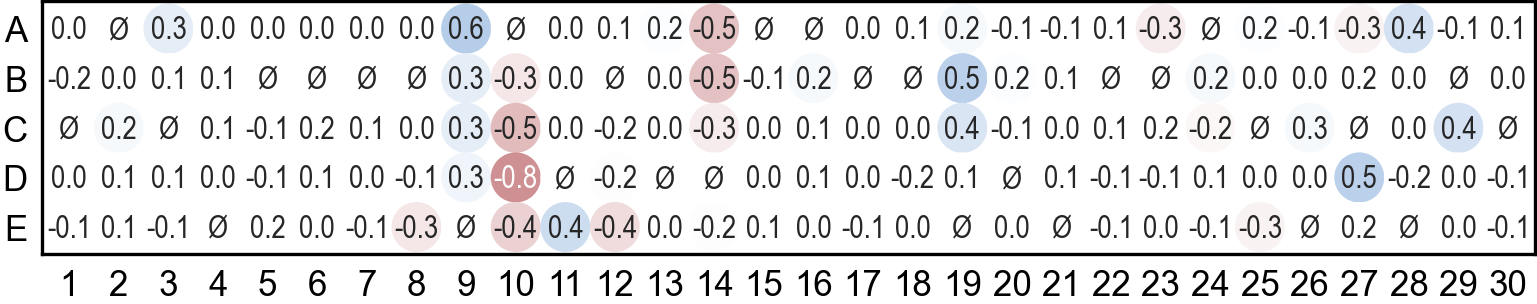}
\end{center}

\begin{center}
\footnotesize\begin{tblr}{
    width = 369pt,
    colspec = {@{}X[l]S[table-format=1.2]S[table-format=1.2]S[table-format=1.2]@{}},
    rows = {rowsep=1pt},
    row{1} = {guard, abovesep+=2pt},
    row{2} = {abovesep+=2pt},
    row{Z} = {belowsep+=1pt},
    hspan = minimal,
}
\toprule
Bootstrap correlation statistic & Pre data & Post data & All data \\
\midrule
Median & 0.82 & 0.61 & 0.95 \\
5th percentile & 0.59 & 0.49 & 0.89 \\
Fraction $\geq$ 0.85 & 0.33 & 0.00 & 0.98 \\
\bottomrule
\end{tblr}
\end{center}\vspace{-0.5em}

\begin{center}
\footnotesize\begin{tblr}{
    width = \textwidth,
    colspec = {@{}lX[l]lllS[table-format=2.1]S[table-format=-1.1]@{}},
    rows = {rowsep=1pt},
    row{1} = {guard, abovesep+=2pt},
    row{2} = {abovesep+=2pt},
    row{Z} = {belowsep+=1pt},
    hspan = minimal,
}
\toprule
\SetCell[c=2]{l} Item/distractor description & & Code & NC-Ex & IRC\textsubscript{pre} & $f$ (\%) & $\widehat{a}$ \\
\midrule
\SetCell[c=4]{l,font=\itshape} Along path you picked in 8, what is speed of puck after y-kick? \\
10D & puck speed increases for while after kick, then decreases & I4 & 1b & DW & 18.2 & -0.8 \\
10C & continuously decreases & --- & 1b & W & 11.4 & -0.5 \\
\hborder{abovespace+=3pt, belowspace+=1pt}\hline[dotted]
\SetCell[c=4]{l,font=\itshape} The speed of puck just after kick is \\
9A & speed equals v0  & --- & 0e,2a & FLAT & 4.7 & +0.6 \\
\hborder{abovespace+=3pt, belowspace+=1pt}\hline[dotted]
\SetCell[c=4]{l,font=\itshape} What is path of heavy ball dropped from moving plane \\
14B & ball falls straight down  & K4 & 1a & W & 16.3 & -0.5 \\
14A & ball falls backwards along parabola  & K4 & 1a & DW & 32.7 & -0.5 \\
\hborder{abovespace+=3pt, belowspace+=1pt}\hline[dotted]
\SetCell[c=4]{l,font=\itshape} Given motion diagrams showing x(t), do two blocks ever have same speed \\
19B & yes, at first time it has same position & K1 & 0a & W & 3.6 & +0.5 \\
\hborder{abovespace+=3pt, belowspace+=1pt}\hline[dotted]
\SetCell[c=4]{l,font=\itshape} Box has v0 due to const force, what happens if force stops? \\
27D & continues at a constant speed & I1 & 2b & FLAT & 2.3 & +0.5 \\
\bottomrule
\end{tblr}
\end{center}

\end{minipage}

\clearpage\noindent\begin{minipage}{\textwidth}
\subsection*{XXV.\hspace{0.5em}Eliminated—low bootstrap correlations}

\vspace{0.5em}
\begin{center}
\includegraphics{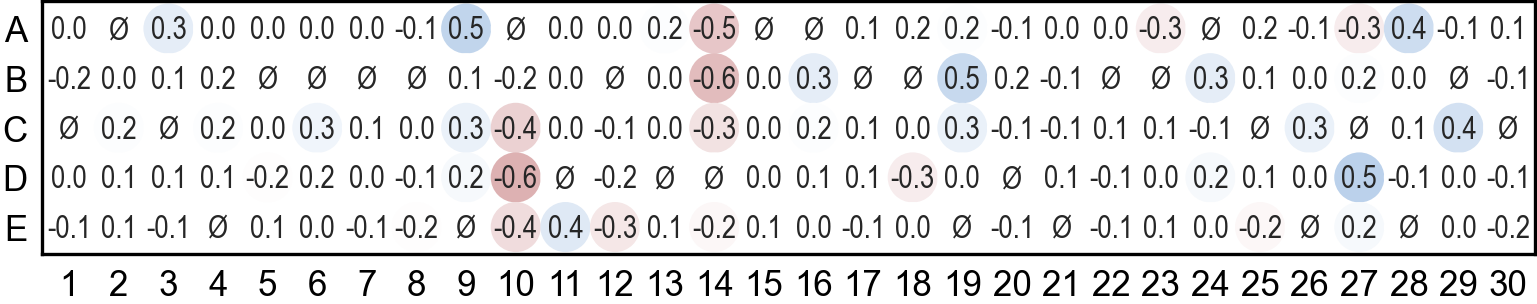}
\end{center}

\begin{center}
\footnotesize\begin{tblr}{
    width = 369pt,
    colspec = {@{}X[l]S[table-format=1.2]S[table-format=1.2]S[table-format=1.2]@{}},
    rows = {rowsep=1pt},
    row{1} = {guard, abovesep+=2pt},
    row{2} = {abovesep+=2pt},
    row{Z} = {belowsep+=1pt},
    hspan = minimal,
}
\toprule
Bootstrap correlation statistic & Pre data & Post data & All data \\
\midrule
Median & 0.81 & 0.51 & 0.93 \\
5th percentile & 0.58 & 0.39 & 0.88 \\
Fraction $\geq$ 0.85 & 0.24 & 0.00 & 0.98 \\
\bottomrule
\end{tblr}
\end{center}\vspace{-0.5em}

\begin{center}
\footnotesize\begin{tblr}{
    width = \textwidth,
    colspec = {@{}lX[l]lllS[table-format=2.1]S[table-format=-1.1]@{}},
    rows = {rowsep=1pt},
    row{1} = {guard, abovesep+=2pt},
    row{2} = {abovesep+=2pt},
    row{Z} = {belowsep+=1pt},
    hspan = minimal,
}
\toprule
\SetCell[c=2]{l} Item/distractor description & & Code & NC-Ex & IRC\textsubscript{pre} & $f$ (\%) & $\widehat{a}$ \\
\midrule
\SetCell[c=4]{l,font=\itshape} Along path you picked in 8, what is speed of puck after y-kick? \\
10D & puck speed increases for while after kick, then decreases & I4 & 1b & DW & 18.2 & -0.6 \\
\hborder{abovespace+=3pt, belowspace+=1pt}\hline[dotted]
\SetCell[c=4]{l,font=\itshape} What is path of heavy ball dropped from moving plane \\
14B & ball falls straight down  & K4 & 1a & W & 16.3 & -0.6 \\
14A & ball falls backwards along parabola  & K4 & 1a & DW & 32.7 & -0.5 \\
\hborder{abovespace+=3pt, belowspace+=1pt}\hline[dotted]
\SetCell[c=4]{l,font=\itshape} Box has v0 due to const force, what happens if force stops? \\
27D & continues at a constant speed & I1 & 2b & FLAT & 2.3 & +0.5 \\
\hborder{abovespace+=3pt, belowspace+=1pt}\hline[dotted]
\SetCell[c=4]{l,font=\itshape} The speed of puck just after kick is \\
9A & speed equals v0  & --- & 0e,2a & FLAT & 4.7 & +0.5 \\
\hborder{abovespace+=3pt, belowspace+=1pt}\hline[dotted]
\SetCell[c=4]{l,font=\itshape} Given motion diagrams showing x(t), do two blocks ever have same speed \\
19B & yes, at first time it has same position & K1 & 0a & W & 3.6 & +0.5 \\
\bottomrule
\end{tblr}
\end{center}

\end{minipage}

\clearpage\noindent\begin{minipage}{\textwidth}
\subsection*{XXVI.\hspace{0.5em}Motion diagrams, v and a confused}

\vspace{0.5em}
\begin{center}
\includegraphics{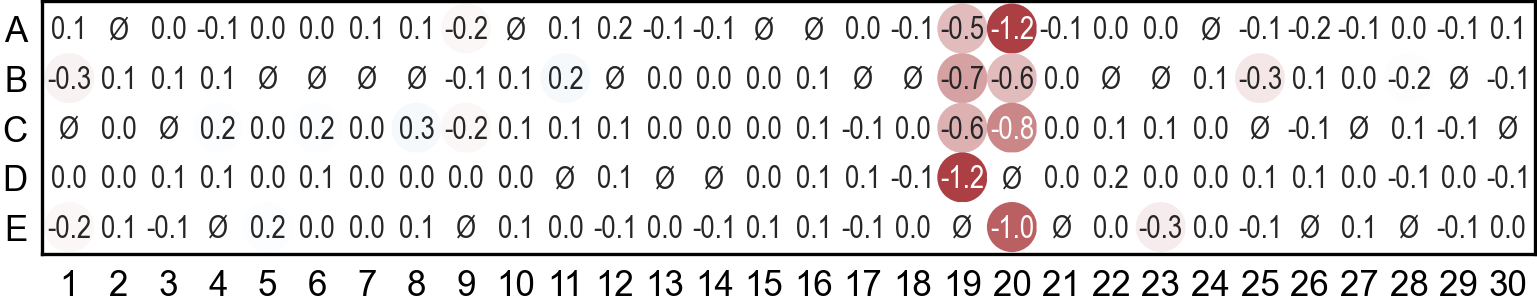}
\end{center}

\begin{center}
\footnotesize\begin{tblr}{
    width = 369pt,
    colspec = {@{}X[l]S[table-format=1.2]S[table-format=1.2]S[table-format=1.2]@{}},
    rows = {rowsep=1pt},
    row{1} = {guard, abovesep+=2pt},
    row{2} = {abovesep+=2pt},
    row{Z} = {belowsep+=1pt},
    hspan = minimal,
}
\toprule
Bootstrap correlation statistic & Pre data & Post data & All data \\
\midrule
Median & 0.98 & 0.95 & 0.99 \\
5th percentile & 0.96 & 0.89 & 0.98 \\
Fraction $\geq$ 0.85 & 1.00 & 0.99 & 1.00 \\
\bottomrule
\end{tblr}
\end{center}\vspace{-0.5em}

\begin{center}
\footnotesize\begin{tblr}{
    width = \textwidth,
    colspec = {@{}lX[l]lllS[table-format=2.1]S[table-format=-1.1]@{}},
    rows = {rowsep=1pt},
    row{1} = {guard, abovesep+=2pt},
    row{2} = {abovesep+=2pt},
    row{Z} = {belowsep+=1pt},
    hspan = minimal,
}
\toprule
\SetCell[c=2]{l} Item/distractor description & & Code & NC-Ex & IRC\textsubscript{pre} & $f$ (\%) & $\widehat{a}$ \\
\midrule
\SetCell[c=4]{l,font=\itshape} Motion diagram of two blocks moving at different constant speeds, their accelerations \\
20A & acceleration of slower block is greater & --- & 0b & DW & 16.6 & -1.2 \\
20E & not enough information & --- & 0f & W & 7.6 & -1.0 \\
20C & acceleration of faster block is greater & K2 & 0b & INT & 31.8 & -0.8 \\
20B & accelerations are equal and both are greater than zero & K2 & 0b & INT & 5.2 & -0.6 \\
\hborder{abovespace+=3pt, belowspace+=1pt}\hline[dotted]
\SetCell[c=4]{l,font=\itshape} Given motion diagrams showing x(t), do two blocks ever have same speed \\
19D & yes, at both times it has same position & K1 & 0a & DW & 26.8 & -1.2 \\
19B & yes, at first time it has same position & K1 & 0a & W & 3.6 & -0.7 \\
19C & yes, at second time it has same position & K1 & 0a & W & 4.5 & -0.6 \\
19A & No & K2 & --- & W & 14.0 & -0.5 \\
\bottomrule
\end{tblr}
\end{center}

\end{minipage}

\clearpage\noindent\begin{minipage}{\textwidth}
\subsection*{XXVII.\hspace{0.5em}Gravity stronger closer to ground}

\vspace{0.5em}
\begin{center}
\includegraphics{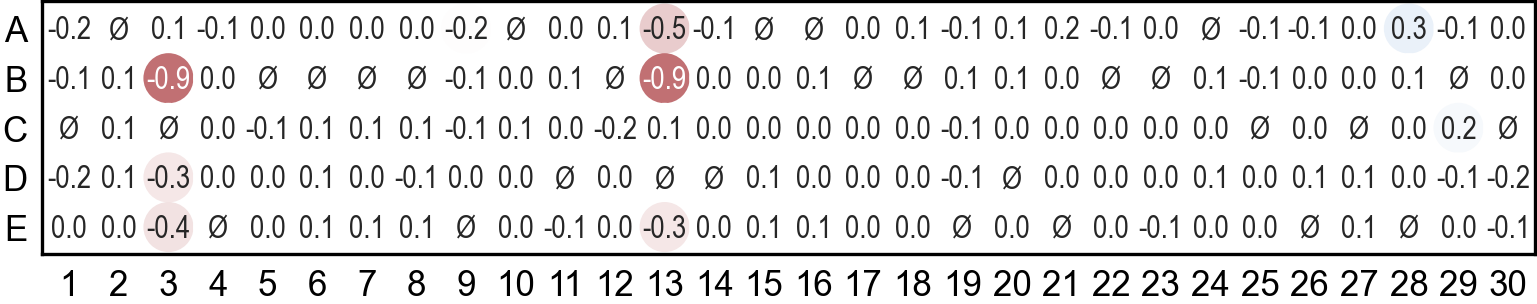}
\end{center}

\begin{center}
\footnotesize\begin{tblr}{
    width = 369pt,
    colspec = {@{}X[l]S[table-format=1.2]S[table-format=1.2]S[table-format=1.2]@{}},
    rows = {rowsep=1pt},
    row{1} = {guard, abovesep+=2pt},
    row{2} = {abovesep+=2pt},
    row{Z} = {belowsep+=1pt},
    hspan = minimal,
}
\toprule
Bootstrap correlation statistic & Pre data & Post data & All data \\
\midrule
Median & 0.96 & 0.71 & 0.98 \\
5th percentile & 0.91 & 0.52 & 0.96 \\
Fraction $\geq$ 0.85 & 0.99 & 0.04 & 1.00 \\
\bottomrule
\end{tblr}
\end{center}\vspace{-0.5em}

\begin{center}
\footnotesize\begin{tblr}{
    width = \textwidth,
    colspec = {@{}lX[l]lllS[table-format=2.1]S[table-format=-1.1]@{}},
    rows = {rowsep=1pt},
    row{1} = {guard, abovesep+=2pt},
    row{2} = {abovesep+=2pt},
    row{Z} = {belowsep+=1pt},
    hspan = minimal,
}
\toprule
\SetCell[c=2]{l} Item/distractor description & & Code & NC-Ex & IRC\textsubscript{pre} & $f$ (\%) & $\widehat{a}$ \\
\midrule
\SetCell[c=4]{l,font=\itshape} Ball is thrown up, what forces act afterwards \\
13B & decreasing upward force up to top, then increasing gravity down  & I3, G4, G5 & * & DW & 21.7 & -0.9 \\
13A & downward gravity and steadily decreasing upward force & I3 & 2a,2b & INT & 8.2 & -0.5 \\
\hborder{abovespace+=3pt, belowspace+=1pt}\hline[dotted]
\SetCell[c=4]{l,font=\itshape} Stone Dropped from Roof of Single Story Building \\
3B & falling body speeds up because gravity increases closer to earth & AF5,G4 & 5Gc & DW & 18.7 & -0.9 \\
\bottomrule
\end{tblr}
\end{center}

\end{minipage}

\end{document}